\documentclass{aa}  

\usepackage{siunitx}
\usepackage{graphicx}
\usepackage{txfonts}

\usepackage{lineno}
\AtBeginDocument{\nolinenumbers\renewcommand{\linenumbers}{}}

\DeclareSymbolFont{cmletters}{OML}{cmm}{m}{it}
\DeclareMathSymbol{v}{\mathalpha}{cmletters}{`v}

\usepackage{lipsum}
\usepackage{subcaption}         
\usepackage{lscape}             
\usepackage{placeins}           
                                
\begin{document}

   \title{All-sky spectroscopic survey of bright $\delta$~Scuti stars from TESS}

   \subtitle{Projected rotational velocities and pulsation characteristics}

%
%
%

   \author{M.~Zummer\inst{1,2}\corrauth{michalmiskozummer@gmail.com}     
        \and M.~Skarka\inst{2}\email{skarka@asu.cas.cz}
        \and D.~Kor\v{c}\'{a}kov\'{a}\inst{1}\email{kor@sirrah.troja.mff.cuni.cz}
        \and M.~Va\v{n}ko\inst{3}\email{vanko@ta3.sk}
        \and T.~Pribulla\inst{3}\email{pribulla@ta3.sk}
        \and J.~\v{Z}\'{a}k\inst{2}\email{zak@asu.cas.cz}
        \and S.~Murphy\inst{4}\email{Simon.Murphy@unisq.edu.au}
        \and L.~Antonucci\inst{5}\email{leantonucci@uc.cl}
        \and R.~Brahm\inst{6,7}\email{rafael.brahm@uai.cl}
        \and P.~Gajdo\v{s}\inst{8}\email{pavol.gajdos@upjs.sk}
        \and J.~Jan\'{i}k\inst{9}\email{honza@physics.muni.cz}
        \and J.~Ne\v{c}\'{a}sek\inst{10}\email{jakub.necasek@tul.cz}
        \and L.~Vanzi\inst{5}\email{lvanzi@uc.cl} 
        }

   \institute{Astronomical Institute of Charles University,
   Faculty of Mathematics and Physics,\hfill\break
   V~Hole\v{s}ovi\v{c}k\'{a}ch~2, CZ-180 00 Praha~8 - Troja, Czech Republic
   \and Astronomical Institute of the Czech Academy of Sciences, Fri\v{c}ova 298, 25165 Ond\v{r}ejov, Czech Republic\and Astronomical Institute of the Slovak Academy of Sciences, Tatransk\'a Lomnica, SK-059~60 Vysok\'e Tatry, Slovakia \and Centre for Astrophysics, University of Southern Queensland, Toowoomba QLD 4350, Australia
   \and Center of Astro Engineering, Pontificia Universidad Cat\'{o}lica de Chile, Av. Vicu\~{n}a Mackenna 4860, 782-043 Santiago, Chile
   \and Facultad de Ingenier\'{i}a y Ciencias, Universidad Adolfo Ib\'{a}\~{n}ez, Av. Diagonal las Torres 2640, Pe\~{n}alol\'{e}n, Santiago, Chile
   \and Millennium Institute for Astrophysics, Av. Vicuna Mackenna 4860, 782-0436 Macul, Santiago, Chile
   \and Institute of Physics, Faculty of Science, Pavol Jozef \v{S}af\'{a}rik University, Park Angelinum 9, 040 01 Ko\v{s}ice, Slovakia
   \and Department of Theoretical Physics and Astrophysics, Masaryk University, Kotl\'a\v{r}sk\'a 2, CZ-611 37 Brno, Czech Republic
   \and Institute of Plasma Physics of the Czech Academy of Sciences, Research Centre for Special Optics and Optoelectronic Systems TOPTEC, U Slovanky 2525/1a 182 00, Praha 8, Czech Republic}
   \date{Received \today}
 
  \abstract
   {The second most numerous class of pulsating stars -- $\delta$~Scuti ($\delta$~Sct) -- and 
   their spectroscopic properties have not been investigated on a large scale since the launch of the space missions \textit{Kepler} and TESS, which may limit the interpretation of data. The lack of a homogeneous spectroscopic sample of $\delta$~Sct prohibits a deeper understanding of the influence of rotation on stellar pulsations and limits bridging the gap between observations and pulsation theory.}
   {To determine the influence of rotation on the excited pulsation modes of $\delta$ Sct stars, we aim to analyse their projected rotational velocities from high-resolution spectroscopy and combine it with 2-minute TESS photometry. We intend to investigate empirical relationships between stellar rotation and pulsation parameters.} 
   {We conduct spectroscopic observations with three high-resolution échelle spectrographs  and combine them with two spectral libraries to gather a homogeneous all-sky high-resolution ($R = \SI{25000}{}$ to $\SI{115000}{}$) spectroscopic sample of 262 $\delta$~Sct stars. We derive projected rotational velocities $v\sin i$ using three independent methods and extract pulsation frequencies from TESS photometry. We analyse projected rotation $v\sin i$ together with extracted pulsation frequencies and amplitudes.}
   {The distribution of projected rotational velocities $v\sin i$ for $\delta$~Sct stars is unimodal, with a mean value of \SI{120.3}{\kilo\meter\per\second} and peak probability at \SI{113.8}{\kilo\meter\per\second}. We find that the minimum frequency $\nu_\textrm{min}$ and other frequency metrics (dominant $\nu_\textrm{dom}$, amplitude-weighted $\nu_\textrm{AW}$, and power-weighted frequency $\nu_\textrm{PW}$) \textit{decrease} with $v\sin i/v_\textrm{crit}$, whereas the maximum frequency $\nu_\textrm{max}$ does not. We show that fast rotating stars have a much more limited range of photometric and pulsation amplitudes compared to stars with low $v\sin i$. We find that the lower mean stellar density $\bar{\rho}$ of rapid rotators cannot fully explain the observed decrease in pulsation frequencies. We calibrate the Gaia \texttt{vbroad} broadening parameter with our $v\sin i$ measurements, which removes the systematic offset and enables the use of Gaia measurements for a much larger sample of $\delta$~Sct stars. Analysing multiple spectra per star from different instruments, we were able to compare different methods for determining $v\sin i$, with the biggest differences occurring for fast rotators with $v\sin i>\SI{200}{\kilo\meter\per\second}$. We also show that different spectrographs may yield different values of the $v\sin i$, sometimes by up to \SI{20}{\kilo\meter\per\second}.}
   {We present the largest high-resolution sample of projected rotational velocities $v\sin i$ for bright $\delta$~Sct stars and place it in the context of stellar pulsations from TESS. We conclude that deriving projected rotational velocities $v\sin i$ for pulsating stars requires particular care, as standard methods may not perform equally well for all types of stars, instruments, or rotational velocity ranges.}

   \maketitle

\section{Introduction}
The pulsating stars of $\delta$ Scuti ($\delta$~Sct) are situated where the classical cepheid instability strip crosses the main sequence; their cores are therefore in the stage of burning hydrogen. This population of pulsating stars of early A- to late F-type has masses in a range of \SI{1.5}{\mathcal{M}_\odot} to \SI{2.3}{\mathcal{M}_\odot} \citep{Murphy2026_09}, temperatures in a range of \SI{6000}{\kelvin} to \SI{9000}{\kelvin} \citep{Uytterhoeven2011, Kurtz2022}, luminosities of 2 to \SI{50}{\mathcal{L}_\odot}, and frequencies between $\SI{4}{\day^{-1}}$ \citep{Breger2000, Aerts2010_Astroseismology} and $\SI{70}{\day^{-1}}$ \citep{Bedding2020}. The frequency of \SI{4}{\per\day} is commonly used as a distinction between $\delta$~Sct ($>\SI{4}{\per\day}$) and $\gamma$~Doradus ($\gamma$~Dor) stars ($<\SI{4}{\per\day}$), although many authors adopt \SI{5}{\per\day} as a dividing threshold \citep{Handler2002, Grigahcene2010, Balona2011, Uytterhoeven2011, Skarka2022, Skarka2024}.

Generally, both radial and non-radial pressure (p) mode pulsations exist among $\delta$~Sct stars. Some of these stars can have hundreds of frequencies that require p, gravity (g), and Rossby (r) modes \citep{Saio2018, Henriksen2023}, rotational splitting, nonlinear combinations of frequencies, as well as additional complexities  such as variable amplitudes on longer timescales \citep{Kurtz2022}. Alternatively, they can have ridge-like frequency clusters, as studied by \citet{Antoci2026}. The complexity of their pulsation spectra makes the identification and
interpretation of individual modes challenging, particularly in hybrid
pulsators \citep{Grigahcene2010, Kurtz2014, Yang2021}. A large fraction of $\delta$~Sct stars have frequencies also in the $\gamma$~Dor range \citep{Uytterhoeven2011,Balona2014a, Balona2015, Skarka2022}. Pulsation modes are primarily excited by the classical $\kappa$ mechanism operating in He\,\textsc{ii} ionisation zones \citep{Cox1963,Chavalier1971,Aerts2010_Astroseismology}. \citet{Antoci2014} showed that the opacity mechanism alone cannot explain oscillations, and turbulent pressure in the hydrogen ionisation zone also contributes to the excitation of pulsations, mainly moderate radial p modes.

The last two decades have seen enormous developments in asteroseismology, thanks to the launch of space missions CoRoT \citep{Auvergne2009corot}, the \textit{Kepler} space telescope \citep{KeplerBorucki2010}, and most recently, the Transiting Exoplanet Survey Satellite (TESS, \citeauthor{TESSRicker2015}~\citeyear{TESSRicker2015}). Space photometry provides precision about 100 times better than ground-based photometry, with nearly continuous, evenly sampled data \citep{Kurtz2022}. This precision and the sheer number of objects observed have increased the number of known pulsating stars and allowed us to probe their interiors. Building on the \textit{Kepler} era, TESS brought the possibility of ensemble studies of $\delta$~Sct stars across the whole sky, see e. g. \citet{Antoci2019, Cunha2019, Bedding2020, Barac2022, Gootkin2024}. 
Another advantage of TESS is its long-term observational coverage, with
observations spanning multiple sectors and, for some targets, several
years. Two-minute cadence data are also available for many stars. Additionally, TESS all-sky coverage includes many early-type hotter stars not observed by \textit{Kepler}, along with bright stars that can be observed by smaller telescopes.  

Because of the overabundance of space photometry compared to spectroscopy, most of the recent studies of $\delta$~Sct stars rely purely on space photometry, and only a handful of stars have any spectroscopic information. Before the advent of space photometry, \citet{Solano1997} conducted a survey of the rotational velocities of 68 $\delta$~Sct stars and found that stars with low $v\sin i$ tend to have higher pulsation amplitudes, often being part of a distinct high-amplitude $\delta$~Sct (HADS) subclass. The fact that fast rotators pulsate with lower amplitudes was already pointed out by \citet{Danziger1972}. No other dependence on projected rotation has been found \citep{Kahraman2017}. Using more than \SI{4000}{} $\delta$~Sct and $\gamma$~Dor stars observed with the medium-resolution LAMOST data $(R\approx7500)$, it was discovered that both $\delta$~Sct and $\gamma$~Dor stars have similar $v\sin i$ distributions \citep{Wang2025}. 

Recently, multiple studies have looked into the role of stellar rotation in the populations of A- and F-type stars \citep{Gootkin2024, Murphy2024, Murphy2026_08, Murphy2026_09, Berry2026}. Even with the abundance of the Gaia data \citep{GaiaDR2, GaiaDR3, Fremat2023}, the effect of projected rotation on pulsation characteristics cannot be fully understood without proper spectroscopic follow-up, as the Gaia \texttt{vbroad} spectral line broadening parameters are systematically biassed \citep{Gootkin2024, Murphy2024, Wang2025}. 

In light of these recent discoveries and the huge amount of precise photometric data, we decided to carry out a comprehensive study of bright, manually vetted $\delta$~Sct stars with high-resolution spectroscopy. We aim to understand the influence of projected rotation $v\sin i$ on stellar oscillations and excited pulsation modes. 
We also revisit the results of previous studies in the context of high precision photometry from short cadence 2-minute TESS data and search for scaling relations. We observed 262 bright ($V$ mag < 7) stars and gathered 307 high-resolution échelle spectra.

The article is structured as follows: we define the sample and observations in Sect.~\ref{Sec:Spectroscopic_Observations}; in Sect.~\ref{Sec:Measurements_of_vsini}, we measure projected rotational velocities $v\sin i$ using 3 methods; in Sect.~\ref{Sec:TESS_photometry}, we analyse TESS data and extract pulsation frequencies; and we conclude in Sect.~\ref{Sec:Conclusions}.

\section{Spectroscopic observations}
\label{Sec:Spectroscopic_Observations}
\subsection{Sample definition}
Our studied sample of $\delta$~Sct stars is based on the TESS Input Catalog (TIC, \citeauthor{Stassun2019TIC}~\citeyear{Stassun2019TIC}), which incorporates more than 1.5 billion stars from Gaia DR2 \citep{GaiaDR2}, available publicly at the Mikulski Archive for Space Telescopes\footnote{\url{https://archive.stsci.edu/missions-and-data/tess}}. We have selected stars with $V$\,mag~<~7 and a temperature range of \SI{6500}{\kelvin} to \SI{9000}{\kelvin}. Then, we manually inspected the selected objects using a \texttt{Python} script for visual representation based on \citet{Skarka2022, Skarka2024}. We visually inspected the candidates and selected stars showing
pulsation in the frequency range above $\SI{5}{d^{-1}}$, while ignoring $\gamma$~Dor/$\delta$~Sct hybrids with significant peaks below $\SI{5}{d^{-1}}$. We have also excluded stars with temperatures above \SI{9000}{\kelvin} that show $\delta$~Sct-like pulsations above $>\SI{5}{d^{-1}}$ to avoid crossing the blue boundary of the instability strip, even though, for example, \citet{Balona2020} use \SI{10000}{\kelvin} as a boundary between $\delta$~Sct and Maia variables.

\begin{figure}[h!]
\resizebox{\hsize}{!}{\includegraphics[angle=0]{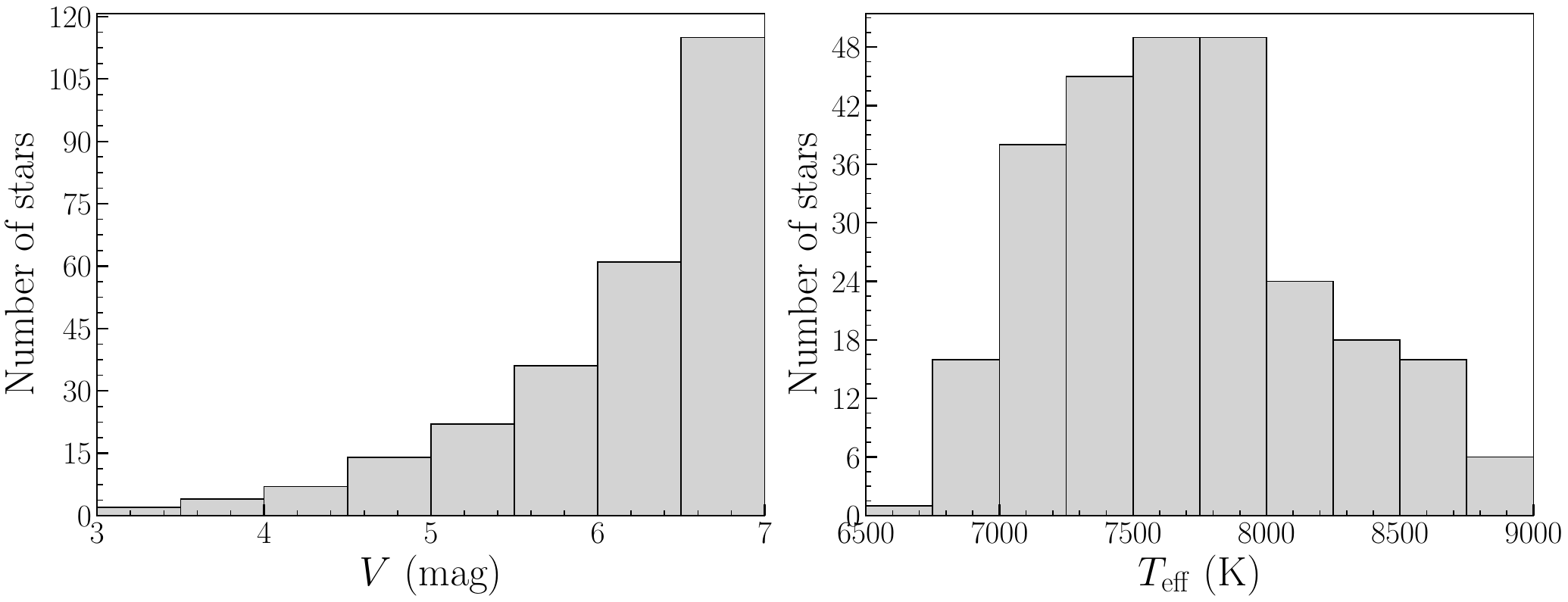}}
\caption{Distribution of $V$ magnitudes and effective temperatures $T_\textrm{eff}$ from TIC of observed $\delta$~Scuti stars.}
\label{fig:observed_distribution}
\end{figure}

\begin{figure}[h!]
\resizebox{\hsize}{!}{\includegraphics[angle=0]{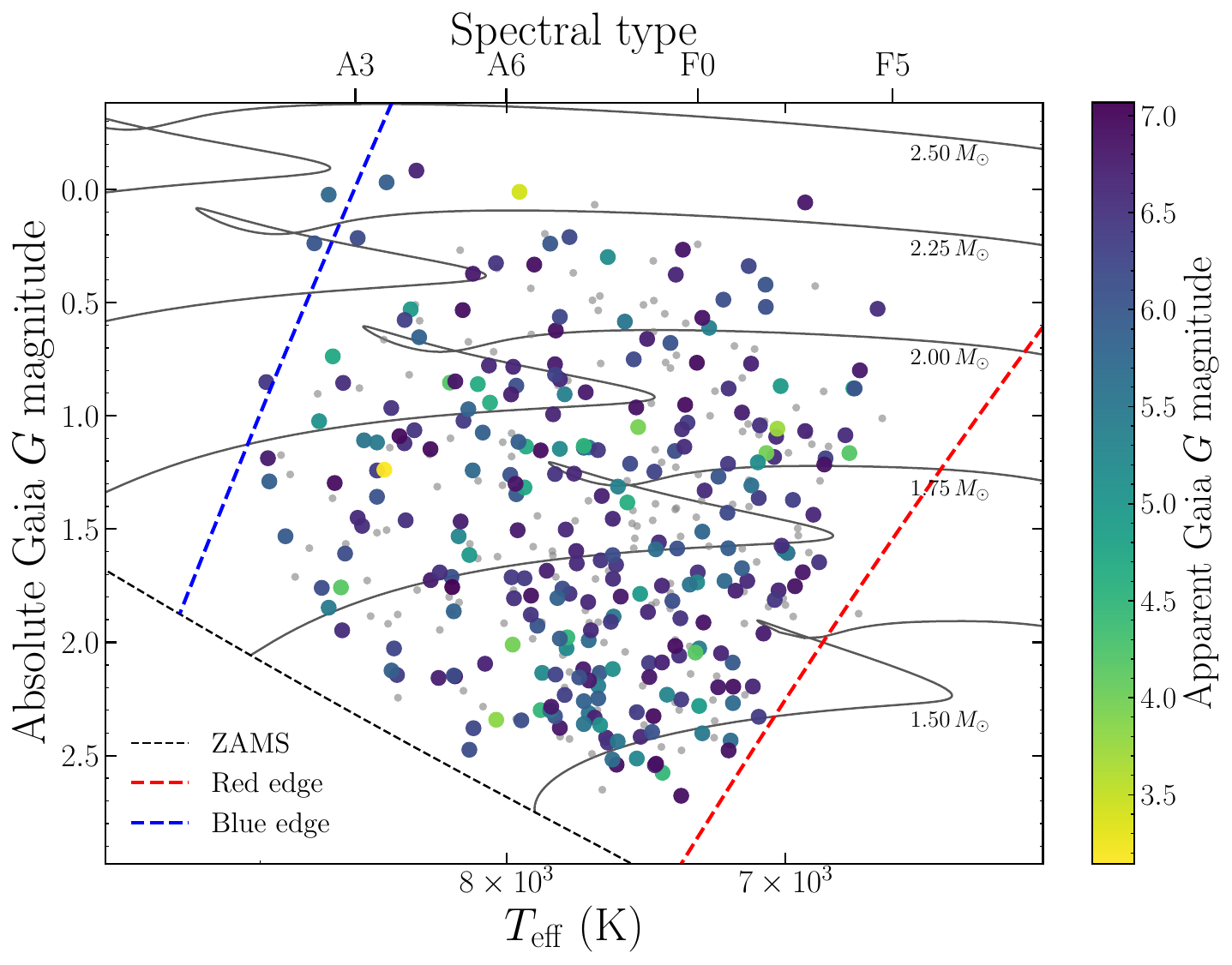}}
\caption{Gaia Hertzsprung–Russell diagram of the initial sample (gray) and observed sample (colored, 262 stars) of $\delta$~Scuti stars. The dashed line is the ZAMS, the black lines are the evolutionary tracks for stars with different masses, the blue and red line show the empirical boundaries of the instability strip all derived by \citet{Murphy2019}.}
\label{fig:HRdiagram}
\end{figure}

\begin{figure*}[ht!]
    \centering
     {\includegraphics[width=\linewidth]{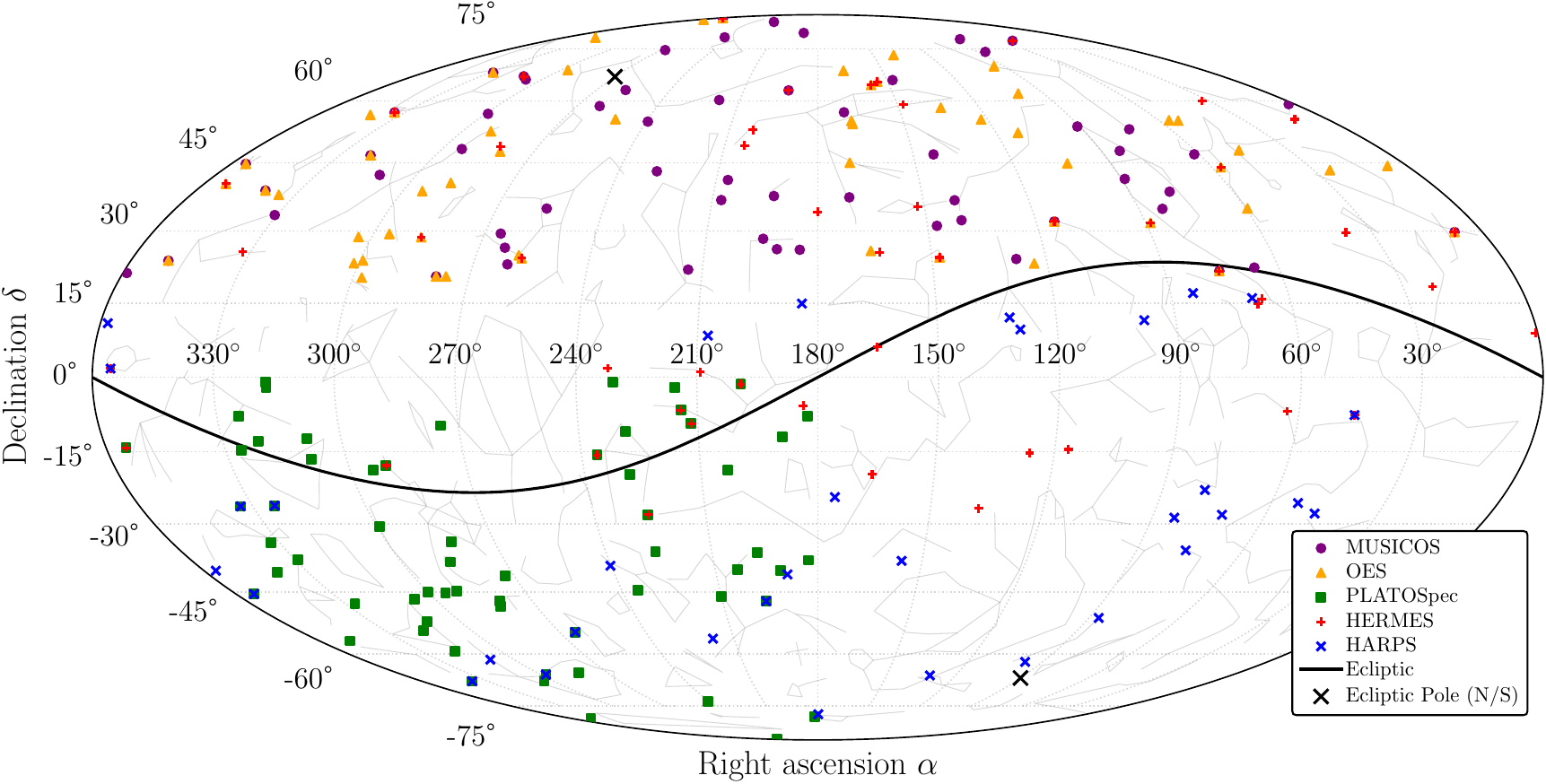}}
     \caption{$\delta$~Scuti stars for which at least one spectrum was obtained.}
      \label{fig_skymap}
\end{figure*}

After visual inspection, we have obtained 435 objects that form our initial sample. Out of this sample, 262 were observed by at least one of the spectrographs at our disposal. Altogether, we observed 307 high-resolution échelle spectra. The distribution of $V$~magnitudes and effective temperatures $T_\textrm{eff}$ of the observed sample can be seen in Fig.~\ref{fig:observed_distribution} and the Gaia Hertzsprung–Russell diagram in Fig.~\ref{fig:HRdiagram}. The sky map of all observed objects can be seen in Fig.~\ref{fig_skymap}.

\subsection{Instruments}
We used three échelle spectrographs for our observations: MUSICOS, OES, and PLATOSpec, together with MELCHIORS and HARPS/ESO spectral libraries.

MUSICOS \citep{Pribulla2024} is an échelle spectrograph at the \SI{1.3}{\meter} telescope at Skalnaté Pleso Observatory ($20^\circ 14^\prime 02^{\prime\prime}\,\mathrm{E},\ 49^\circ 11^\prime 22^{\prime\prime}\,\mathrm{N}$), \SI{1786}{\meter}  above sea level. It has a resolving power of $R\approx\SI{25000}{}-\SI{38500}{}$ and a spectral range of \SI{4250}{}~to~\SI{7375}{\AA} covering 65 échelle orders.

Ondřejov Echelle Spectrograph (OES) \citep{Koubsky2004, Kabath2020} is an échelle spectrograph installed on a \SI{2}{\meter} Perek telescope in Ondřejov $(14^\circ 47^\prime 01.1^{\prime\prime}\,\mathrm{E},\ 49^\circ 54^\prime 38^{\prime\prime}\,\mathrm{N})$, \SI{528}{\meter} above sea level, \SI{30}{\kilo\meter} southeast of Prague. It has a resolving power of $R\approx\SI{50000}{}$ and covers a spectral range \SI{3750}{}\ to \SI{9200}{\AA}. 

PLATOSpec \citep{Kabath2026} is a high-resolution, blue sensitive spectrograph with a resolving power of $R\approx\SI{70000}{}$ and a spectral range of \SI{3800}{} to \SI{7000}{\AA}, installed on the refurbished ESO \SI{1.52}{m} telescope (E152)\footnote{\url{https://www.eso.org/public/teles-instr/lasilla/152metre/}} at La Silla observatory $(70^\circ 44^\prime 00^{\prime\prime}\,\mathrm{W},\ 29^\circ 15^\prime 00^{\prime\prime}\,\mathrm{S})$, \SI{2400}{\meter} above sea level. It was built to screen and validate exoplanetary candidates and study stellar variability as a support facility for the PLATO mission \citep{rauer2025plato}.

We have also searched the Mercator Library of High Resolution Stellar Spectroscopy (MELCHIORS, \citeauthor{Royer2024MELCHIORS}~\citeyear{Royer2024MELCHIORS}) and HARPS spectral databases. The list of spectra is summarised in Table~\ref{table:Joural of spectra}. 

HERMES \citep{Raskin2011HERMES} is a high-resolution spectrograph installed on \SI{1.2}{\meter} Mercator telescope at the Roque de Los Muchachos Observatory on the island of La Palma, Spain $(17^\circ 52^\prime 43^{\prime\prime}\,\mathrm{W},\ 28^\circ 45^\prime 45^{\prime\prime}\,\mathrm{N})$, \SI{2331}{\meter} above sea level. It has a resolving power of $R\approx\SI{85000}{}$ and a spectral range of \SI{3770}{}~to~\SI{9000}{\AA}.

HARPS \citep{HARPS} is a high-resolution spectrograph on the ESO \SI{3.6}{\m} telescope at the La Silla observatory, with a primary aim of high precision spectroscopy measurements of exoplanets. It has a spectral range of \SI{3800}{}~to~\SI{6800}{\AA}. 
We have searched the ESO Science Archive and found that 37 objects were also observed by HARPS. We have selected and downloaded one spectrum with the highest $S/N$ ratio if multiple spectra were present. All spectra have a high resolving power of either $R\approx\SI{80000}{}$ or $\SI{115000}{}$. 

Spectra from MUSICOS and OES were reduced and normalised by the standard custom-made observatory pipelines. Observations from PLATOSpec were reduced by the \texttt{Ceres} pipeline \citep{Brahm2017}. Spectra from HERMES were downloaded in an already normalised form, and for HARPS spectra normalisation, we have used an automated neural network normalisation tool \texttt{SUPPNet} \citep{Rozanski2022SUPPnet} that is trained to predict a pseudo-continuum, which should lead to results similar to careful manual normalisation.

\begin{table}[ht]
\caption[]{\label{table:Joural of spectra}Journal of spectra.}
\begin{tabular}{lccc}
 \hline \hline
 Instrument&Resolution $R$&Range (\SI{}{\AA})& No. of spectra\\
\hline
MUSICOS&25000--38500&4250--7375&72\\
OES&50000&3750--9200&75\\
PLATOSpec&70000&3800--7000&71\\
HERMES&85000&3770--9000&52\\
HARPS&80000/115000&3800--6800&37\\
\hline
\end{tabular}
\end{table}

\section{Measurements of $v\sin i$}
\label{Sec:Measurements_of_vsini}
There are multiple methods to determine projected rotational velocities $v\sin{i}$ of stars, and none of these, as pointed out by \citet{Gray1992}, are standard. Most simple methods measure the direct rotational broadening of a spectral line \citep{Slettebak1954, Slettebak1955, Slettebak1956}. \citet{Slettebak1975} developed a calibration of rotational velocity based on the full width at half depth of the absorption lines in stellar spectra. In the first study of a homogeneous sample of $\delta$~Sct stars, \citet{Solano1997} determined the rotational velocities of 68 stars using the standard rotational profile by \citet{Gray1992}. The introduction of the Fourier transform (FT) on line profiles allows for much deeper analysis. There have been multiple spectroscopic surveys of A and F spectral type stars by \citet{Royer2002_1, Royer2002_2, Royer2007} that used FT of single spectral lines, as it allows for better identification of the rotation profile. All of the previous studies rely heavily on the calibration of single strong lines. Identification of suitable spectral lines is a difficult task, as summarised by \cite{Diaz2011}, since one has to find 
\begin{enumerate}
    \item lines that have no blends from surrounding lines,
    \item lines broadened primarily by the effect of rotation, e.g. no Balmer lines as they show a significant Stark effect,
    \item lines that are present for stars from low to critical rotators,  
\end{enumerate}
to which we add
\begin{enumerate}
    \item [4.] lines that are available across a wide temperature range,
    \item [5.] lines that are not in close proximity to telluric lines.
\end{enumerate}
Because $\delta$~Sct stars are often characterised as mid-to-rapid rotators \citep{Breger2000, Aerts2010_Astroseismology}, estimating their rotational velocities is a challenging task. With high $v\sin{i}$, it is increasingly difficult to find multiple lines that satisfy the conditions mentioned above. We have therefore carried out a threefold measurement of $v\sin i$ using the following methods
\begin{itemize}
    \item \textit{method\,1} fitting of the single line Fe\,\textsc{i}\,4957.613\,\AA\ by the convolution of the Voigt profile and the standard rotational kernel by \citep{Gray1992},
    \item \textit{method\,2} fitting the spectral region using \texttt{iSpec} \citep{Blanco2014iSpec, Blanco2019iSpec},
    \item \textit{method\,3} using the Fourier transform of the cross correlation function (CCF) of the observed and non-rotation template spectrum developed by \citet{Diaz2011}. 
\end{itemize}

Examples of the results for all spectrographs and all three methods can be seen in Appendix~\ref{Apendix:A}. For \textit{method\,2} and \textit{method\,3} we use the same spectral range of \SI{5000}{\AA} to \SI{5800}{\AA} as this region is well covered among all used spectrographs. Results are presented in Table~2 only available at the Strasbourg Astronomical Data Centre (CDS).

\subsection{Method\,1 - Single spectral line}
\label{Subsection:Single_spectral_line}
We fitted the spectral line Fe\,\textsc{i}\,4957.613\,\AA\ with the convolution of a Voigt profile and a rotational kernel in the standard form by \citet{Gray1992}. A similar method was employed by \cite{Solano1997}, who fitted only the rotational kernel but used multiple spectral lines. Given the different origins of the spectra, we decided to use a single
spectral line that is consistently present in all spectra and provides
stable results compared to other surveys or different spectral lines. We have used the standard linear limb darkening law \citep{Schwarzschild1906}
\begin{equation}
    I(\mu) = I(1)(1-\epsilon(1-\mu))\,,
\end{equation}
where $I(1)$ is the stellar intensity at the centre of the disc, $\mu=\cos\theta$ and $\theta$ being the angle between the line of sight and the direction of the emerging flux. We have used $\epsilon=0.6$ to be consistent with previous studies, e.g. \citet{Solano1997}. After the fit, we inspected each fitted line and discarded those that clearly did not fit. The advantage of \textit{method~1} compared to \textit{method~2} and \textit{method~3} is that it allows us to derive $v\sin i$ even for very fast rotators, while the others can sometimes fail for $v\sin i>\SI{250}{\kilo\meter\per\second}$.

\subsection{Method\,2 - Fitting spectral regions}
\label{Subsection:iSpec}
The quality of our spectra allows us to derive projected rotational velocities by fitting large segments of the spectra. We have used \texttt{iSpec} \citep{Blanco2014iSpec, Blanco2019iSpec}, which compares observed and synthetic spectra, employing $\chi^2$ minimisation by executing a nonlinear least-squares method \citep{Markwardt2009}. The code searches the parameter space and performs multiple iterations until $\chi^2$ is lower than a given threshold. We have processed and corrected the spectrum for systemic velocity using the built-in \texttt{iSpec} templates and renormalised it to one. After this, we used the effective temperature $T_\textrm{eff}$ and $\log g$ from Gaia DR3 \citep{GaiaDR3} and iterated for all the available parameters except for $\log g$, which was kept fixed. We have decided not to iterate $\log g$ as it often leads to degeneracies in the model of the spectra, since the spectrum is not strongly dependent on  $\log g$ compared to other parameters, and it speeds up our computations. Synthetic spectra were computed with the LTE radiative transfer code MOOG
\citep{Sneden1973, Sneden2012} using ATLAS9 model atmospheres
\citep{Castelli2003} and solar abundances from \citet{Grevesse1998}. The linear limb darkening coefficient was fixed at $\epsilon=0.6$ to ensure consistency with the \textit{method\,1} and previous studies. This procedure was performed individually for each spectrum over the wavelength range of 5000 to \SI{5800}{\AA}, as the automatic processing of the spectra would not be reliable for both slow and fast rotators. Examples of the spectra for different $v\sin{i}$ with their models are in Figs~\ref{fig:Apendix:iSpec1}~and~\ref{fig:Apendix:iSpec2}.

\begin{figure}[h!]
  \centering
  \includegraphics[width=\columnwidth]{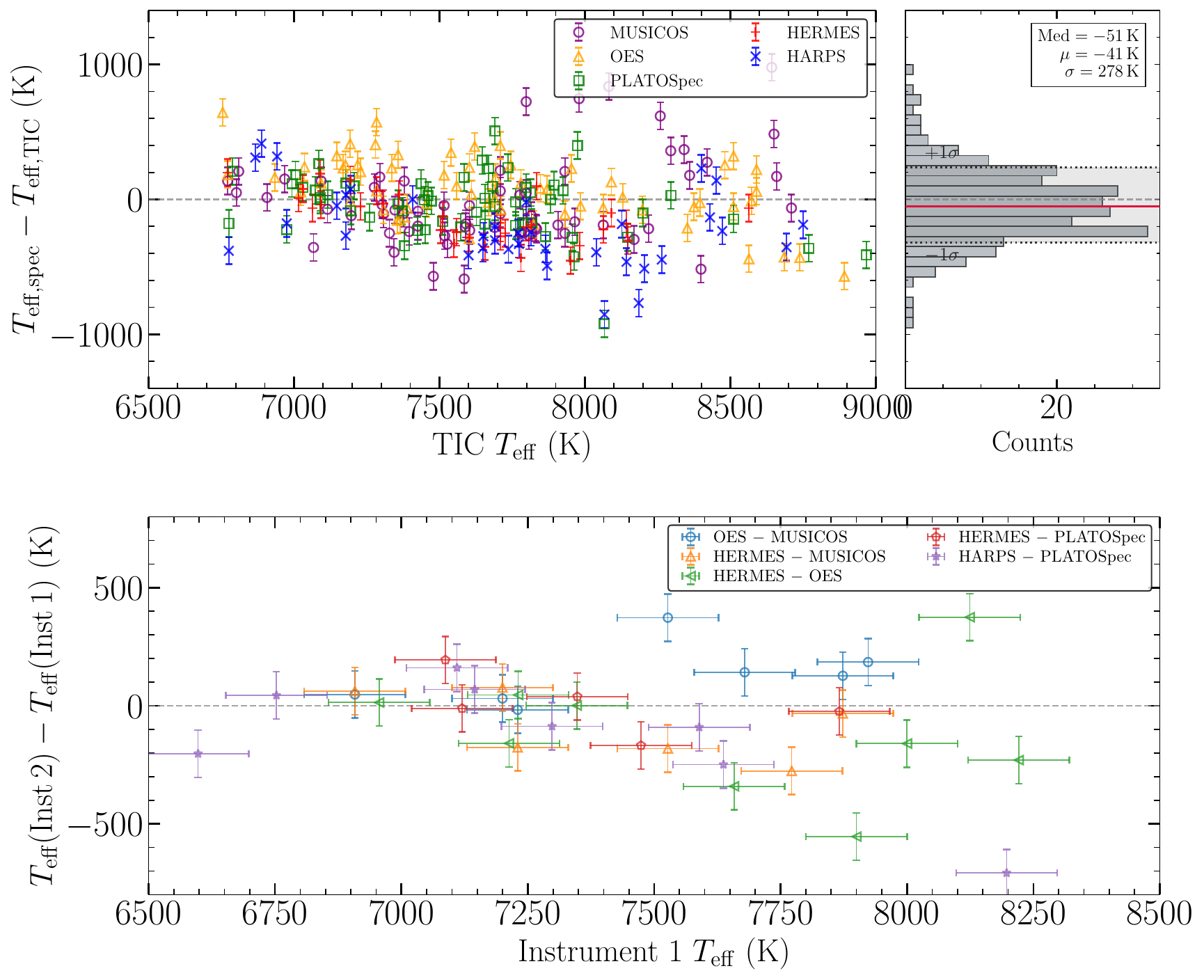}
  \caption{Top: Comparison of the effective temperature $T_\textrm{eff, TIC}$ from TIC catalogue and values derived by \texttt{iSpec} $T_\textrm{eff, spec}$. Dashed gray is the 1:1 line. Right: histogram of differences with mean $\mu = -\SI{41}{\kelvin}$ (red line) and standard deviation $\sigma = \SI{278}{\kelvin}$ (dashed gray lines). Bottom: Comparison of effective temperatures $T_\textrm{eff}$ derived by different spectrographs. Dashed gray line is the 1:1 line.}
  \label{fig:iSpec_Teff_vs_TIC_Teff}
\end{figure}

Derived values of the $v\sin i$ and the effective temperature $T_\textrm{eff}$ can be found in Table~2, which is only available at the Strasbourg Astronomical Data Centre (CDS). Obtaining the error of the \texttt{iSpec} model requires uncertainties of spectral flux due to $\chi^2$ minimisation, which are often either not present or are underestimated or overestimated. As the largest contribution to the error is a coarse grid of synthetic models that do not include gravity darkening, we determined the error using a Monte Carlo method by perturbing basic atmospheric and stellar parameters according to a Gaussian distribution from their expected values. The errors are then estimated as the standard deviation from the statistical dispersion across 100 iterations. We have performed a Monte Carlo analysis for a small subset of spectra from each spectrograph and calculated the mean uncertainty. This statistical error is close to \SI{100}{\kelvin}, and we use it for all spectra. However, as seen in the bottom of Fig. \ref{fig:iSpec_Teff_vs_TIC_Teff}, different spectrographs can give different values for the effective temperature. The standard deviation for different instruments is $\approx\SI{300}{\kelvin}$, which might be closer to the actual estimate of the temperature error. The difference in temperature from different spectrographs may come from poor normalisation, but the actual change in the stellar photosphere may also play a role since we are dealing with pulsating stars. The biggest outlier, TIC\,405410684, is also a fast rotator, where finding a best fit may be uncertain.

\subsection{Method\,3 - Fourier method for spectral region}
\citet{Diaz2011} developed a method for finding $v\sin{i}$ in Fourier space using the CCF of observed spectra and a non-rotation template, which has the advantage of not requiring single spectral lines. We derived $v\sin{i}$ using a version of this method implemented in \texttt{Python}, which requires an object and a template spectrum. We used template spectra that have been specifically created on a star-by-star basis by using the derived parameters in \texttt{iSpec} \citep{Blanco2014iSpec, Blanco2019iSpec} from Sect.~\ref{Subsection:iSpec}, while assuming slow rotation of \SI{1}{\kilo\meter\per\second}. The whole method can be summarised in the following stages 
\begin{itemize}
    \item calculation of the no rotation template spectrum for each object,
    \item calculation of the object$\times$template and template$\times$template CCF,
    \item removal of the CCF maximum from the template$\times$template CCF by linear interpolation,
    \item decontamination of the object$\times$template CCF to obtain the cleaned rotational broadening profile,
    \item measurement of the first zero of the FT,
    \item converting the first zero $\sigma_1$ of the FT into $v\sin{i}$ using the formula $v\sin i = \frac{k_n(\epsilon)}{\sigma_n}\,,$ where $k_n(\epsilon)$ can be approximated by the formula $k_n(\epsilon) =0.60975 +0.0639\epsilon+0.0205\epsilon^2 +0.021\epsilon^3\,,$ and $\sigma_n=\sigma_1$ is the first zero of the FT \citep{Diaz2011}.
\end{itemize}
This was repeated for multiple overlapping wavelength regions with a length of about \SI{220}{\AA} in the interest region of \SI{5000}{} to \SI{5800}{\AA}. The final value of $v\sin{i}$ is the mean of the $v\sin{i}$ of selected regions. The estimate of its error is taken from the standard deviation of $v\sin{i}$ of selected regions. We have also assumed a fixed limb darkening coefficient $\epsilon=0.6$, consistent with \textit{method\,1} and \textit{method\,2}, \citet{Solano1997} and \citet{Royer2002_1, Royer2002_2, Royer2007} and excluded the region 516 to \SI{519}{\nano\meter}, due to blended lines and the Mg\,\textsc{i}\,b triplet. We have also assumed that the rotational profile is symmetric, which might not always be the case and neglected the impact of non-radial pulsations on the spectral line profiles \citep{Aerts2014}.

\subsection{Comparisons}
\begin{figure*}[ht!]
    \centering
     \includegraphics[width=\linewidth]{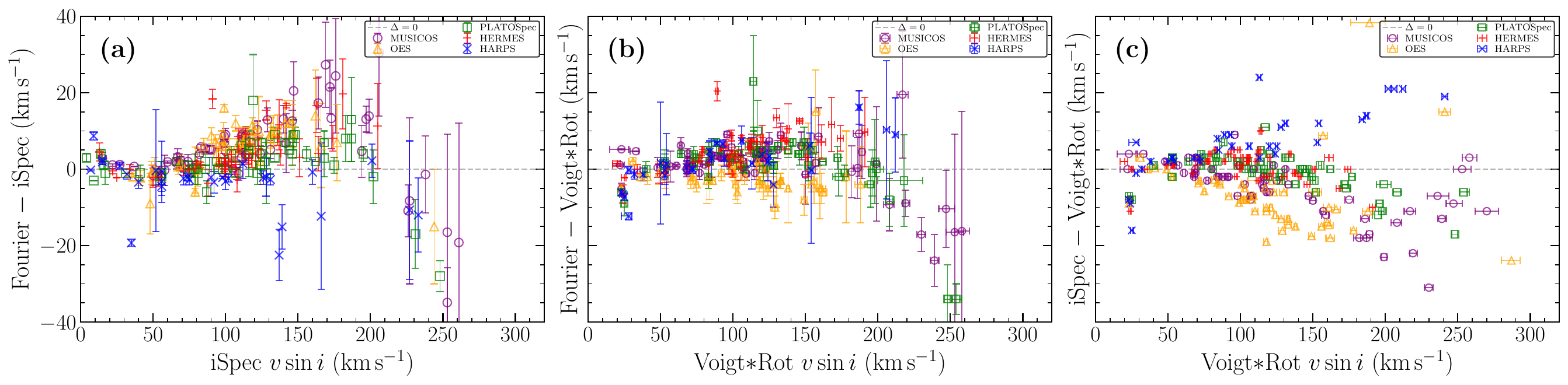}
     \caption{Left: Comparison of $v\sin i$ measured with Fourier method and derived by fitting spectra in \texttt{iSpec}
     Middle: Comparison of $v\sin i$ measured with Fourier method and by the standard rotational profile. Right: Comparison of $v\sin i$ derived by fitting spectra in \texttt{iSpec} and single line fitting of the standard rotation profile. Dashed gray is the 1:1 line. Error bars in the $y$ axes are for the first method in the difference.}
      \label{fig:vsini_comparison}
\end{figure*}

The comparison of all three methods is shown in Fig.~\ref{fig:vsini_comparison}, subplots (a), (b) and (c). Methods are compatible in the low to medium range, but they start to differ at high $v\sin i$ ($>\SI{200}{\kilo\meter\per\second}$). Deviation at high $v\sin i$ cannot be explained by the spectral resolution, as both high and medium-resolution spectrographs deviate in the same direction in subplots (a) and (b). The differences between spectrographs can also be seen, for example, OES (yellow points) drifts the most in all 3 subplots. In subplots (b) and (c), OES drifts in the opposite direction $\Delta = 0$ compared to HARPS, probably caused by different resolution, instrumental profile, continuum normalisation, or the handling of microturbulence. HARPS, which has the highest resolution, deviates already in the range of \SI{100}{} to \SI{150}{\kilo\meter\per\second} in subplot (c) while in subplot (a) it seems stable compared to other spectrographs.

It has been a long established fact that gravity darkening can underestimate the value of $v\sin i$, e.g. \citet{Fremat2005, Zorec2016, Zorec2017}, and has been also studied in detail for the rapidly rotating Be stars \citep{Townsend2004}. In some cases, due to the quality of the spectra, poor normalisation, and high effective stellar temperature, high or low stellar rotation, not all methods could be used simultaneously, which is why we decided to use three different methods.

\subsubsection{Best method for $v\sin i$}
Even though the different methods used for deriving $v\sin i$ are well established, it is always important to keep in mind that stellar properties and instrumental effects play a huge role, see Fig.~\ref{fig:vsini_comparison}. \textit{Method\,1} and \textit{method\,3} can often fail if multiple stellar components are present and clearly visible in the spectra, while \textit{method\,2} can still be used in a limited way. It is difficult to decide which method is preferable and which one should be used for the final value. Our methodology
can lead to up to 9 values in some cases, as multiple stars were observed with up to three different spectrographs. All measurements per spectrograph and per method are summarised in Table~2 only available at the Strasbourg Astronomical Data Centre (CDS). We also estimate one value of $v\sin i$ per star, as the mean of values from all available instruments and all three methods. The uncertainty of the final value $v\sin i$ is taken as the standard deviation of the available values. This final value of $v\sin i$ can also be found in the last column of Table~2. This mean value is later used in our combined analysis of the TESS photometry. 

\begin{figure}[h!]
\resizebox{\hsize}{!}{\includegraphics[angle=0]{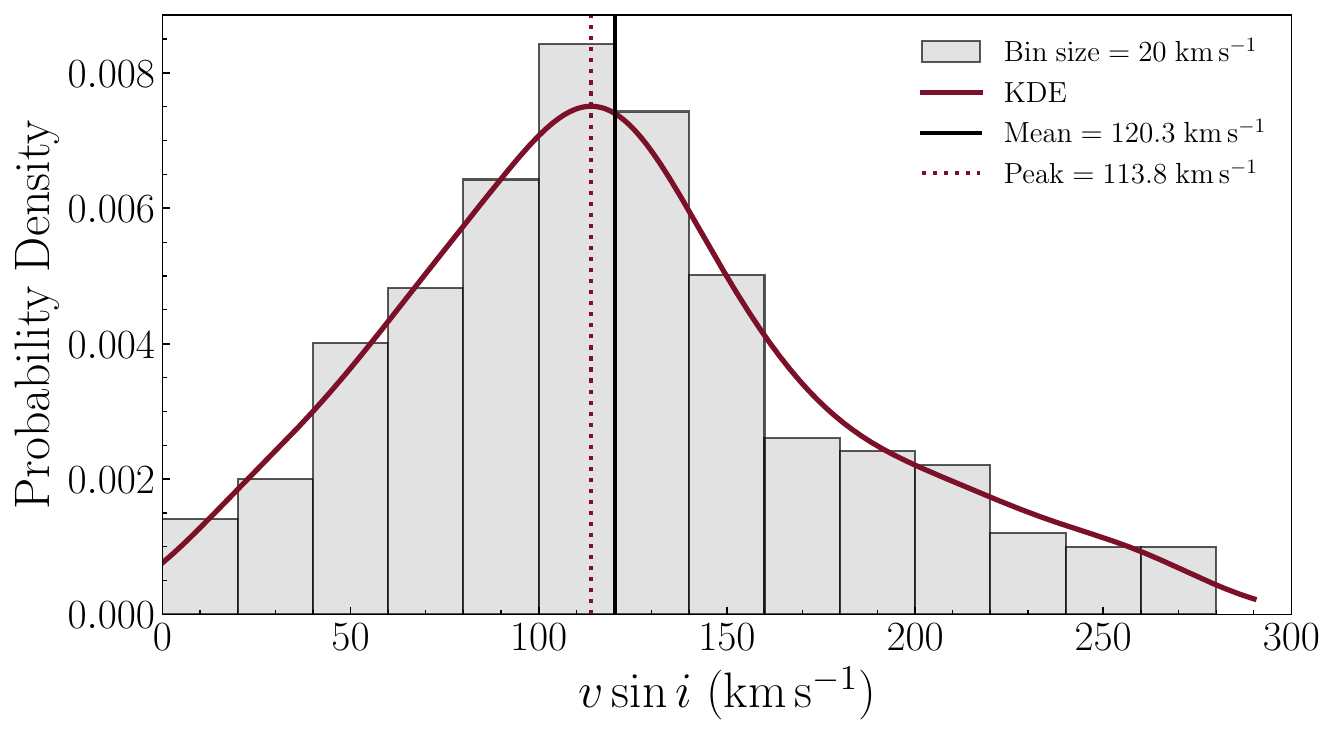}}
\caption{Distribution of measured $v\sin i$ for $\delta$~Scuti stars with bin size $\SI{20}{\kilo\meter\per\second}$, mean value $\SI{120.3}{\kilo\meter\per\second}$ and peak probability density value $\SI{113.8}{\kilo\meter\per\second}$.}
\label{fig:vsini_distribution}
\end{figure}

A different option was a weighted mean of different methods; however, this comes with varying reliabilities of errorbars, as they are almost always underestimated compared to the actual error. Using different methods allows us to quantify the spread in the values for each star. The final measured distribution of $v\sin i$ is shown in Fig.~\ref{fig:vsini_distribution}, with a mean value of \SI{120.3}{\kilo\meter\per\second}. The probability density distribution is unimodal and positively skewed \citep{Zorec2012, Wang2025}.

We note that micro and macroturbulence play an important role in determining the true $v\sin i$, mainly in slow rotators, e.g. $<\SI{20}{\kilo\meter\per\second}$. The rotational profile in \textit{method\,1} does not distinguish $v\sin i$ from macroturbulence, and we obtain an upper broadening limit. When using \textit{method\,2} with \texttt{iSpec}, all 3 parameters can be fitted, which in the $\chi^2$ minimisation may cause degeneracies, mainly for the slowest rotators. It is possible that some derived uncertainties for slower rotators are underestimated for different methods; because of that, we prefer to use the mean value from all measurements. As expected, different values at higher $v\sin i$ can be attributed to gravity darkening, which creates a different line profile than that of a simple non-rotating star. While it is possible to disentangle all three: $v\sin i$, microturbulence, and macroturbulence \citep{SimonDiaz2014}, we do not expect statistically different results. We just note that treating stars with low and very high $v\sin i$ should be approached with caution.

\subsubsection{Comparison to previous studies}
\begin{figure*}[ht!]
    \centering
     \includegraphics[width=\linewidth]{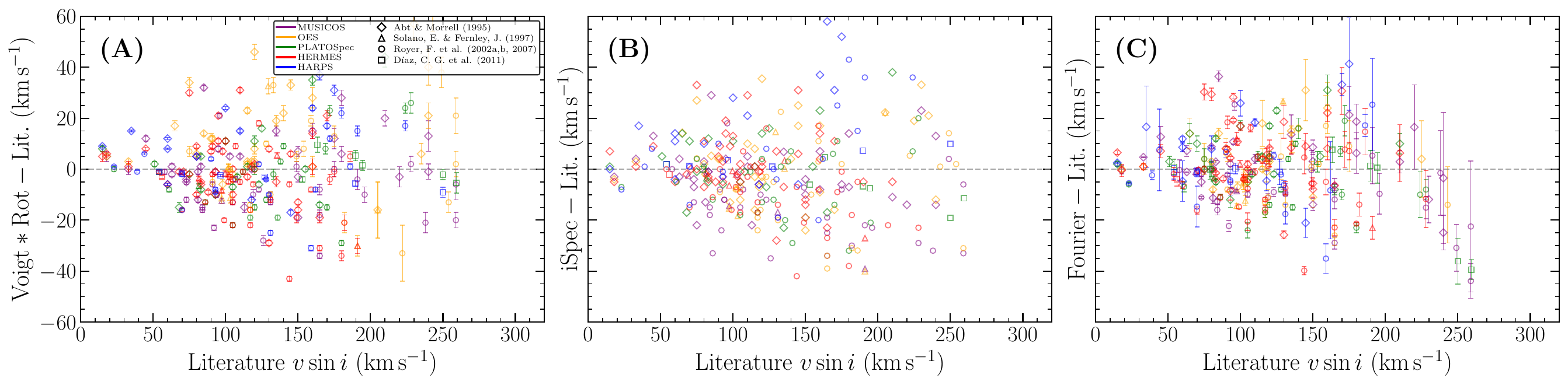}     
     \caption{Left: Comparison of $v\sin i$ from this work compared to previous studies. Left to right: \textit{Method\,1, 2} and \textit{3}. Dashed gray is the 1:1 line. Error bars in the $y$ axes are for our derived values.}
      \label{fig:vsini_comparison_previous_studies}
\end{figure*}

A fraction of the stars from our observed sample already had published values of $v\sin i$, \citep{Helmut1995,Solano1997,Royer2002_1, Royer2002_2, Royer2007, Diaz2011}, which we compare to our derived values in Fig.~\ref{fig:vsini_comparison_previous_studies}, subplots (A), (B) and (C). We do not show \citet{Kahraman2017} in the comparison, as only two of the stars, TIC\,17992601 with $v\sin i=\SI{145\pm6}{\kilo\meter\per\second}$ (our value $\SI{142\pm6}{\kilo\meter\per\second}$)
and TIC\,66693422 with $v\sin i=\SI{75\pm3}{\kilo\meter\per\second}$ (our value $\SI{75\pm1}{\kilo\meter\per\second}$), were observed in both our sample and theirs. None of the stars from \citet{Wang2025} is in our sample; therefore, we cannot make a comparison to our derived values.

Sometimes, multiple studies yield values that are off by more than \SI{30}{\kilo\meter\per\second}; therefore, we plot available values from all different sources. In the comparison of our values to previous studies, there is a much greater scatter than in the comparison of different methods (See Figs.~\ref{fig:vsini_comparison}~and  ~\ref{fig:vsini_comparison_previous_studies}), which should be expected. The only visible trend is in subplot (C), which is probably the range where gravity darkening becomes prominent. Interestingly, the Fourier method (C) also deviates from \citep{Diaz2011}, who used the same Fourier method, suggesting that other factors, such as the continuum normalisation of the spectra or the changes to the spectral line profiles due to non-radial pulsations, can play a role.

\subsubsection{Comparison of instruments}

\begin{figure*}[ht!]
\resizebox{\hsize}{!}{\includegraphics[angle=0]{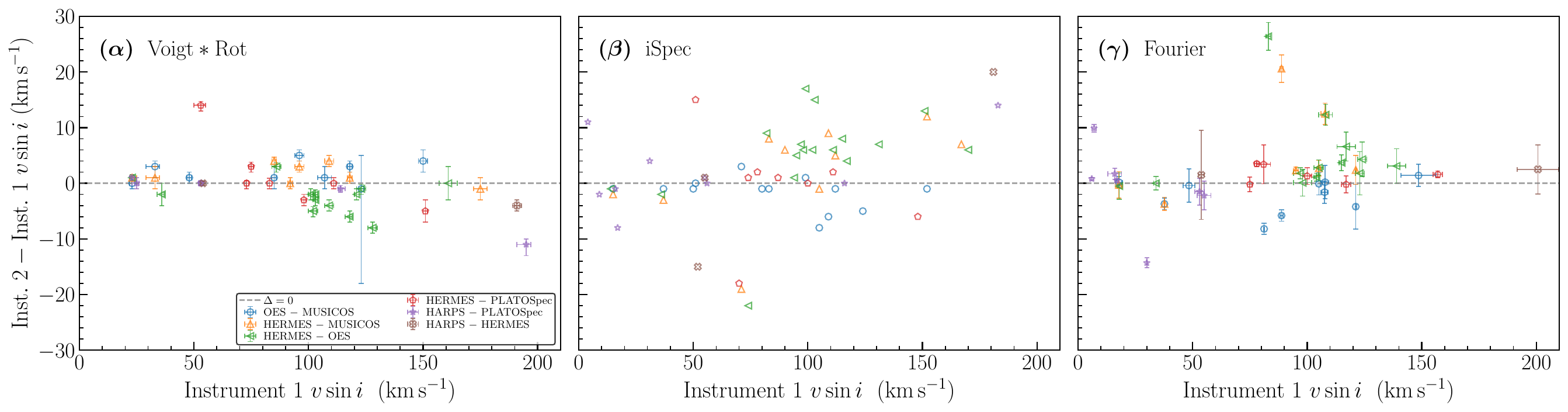}}
\caption{Comparison of $v\sin i$ measured by various spectrographs with \textit{method\,1}, \textit{2} and \textit{3}. In total 48, 63 and 45 star pairs were simultaneously observed by more instruments, sometimes not all methods for all spectra could be used. Dashed gray is the 1:1 line. Error bars in the $y$ axes are for the instrument 2.}
\label{fig:comparison_spectrographs}
\end{figure*}

A subsample of $\delta$~Sct stars was observed using up to three different instruments to verify their compatibility. The comparison of $v\sin i$ measured by different methods can be seen in Fig.~\ref{fig:comparison_spectrographs}, subplots ($\alpha$), ($\beta$) and ($\gamma$). The largest spread is for \texttt{iSpec} (\textit{method\, 2}) with $\approx\SI{12}{\kilo\meter\per\second}$, while Fourier and Voigt$*$Rot have $\approx\SI{9}{\kilo\meter\per\second}$ and $\approx\SI{8}{\kilo\meter\per\second}$ respectively. A larger spread in subplot ($\beta$) is likely caused by the dependence on the spectrograph's resolving power, as lower resolution MUSICOS and OES have a smaller spread than HERMES-MUSICOS or HERMES-OES. The biggest outliers in the subplot ($\gamma$) are the HERMES-OES and HERMES-MUSICOS pairs. Similarly to the bottom part of Fig.~\ref{fig:iSpec_Teff_vs_TIC_Teff}, we can see that the difference between spectrographs is often far greater than the error of separate measurements. 

\subsubsection{Gaia \texttt{vbroad} broadening parameter}

\begin{figure}[h!]
\resizebox{\hsize}{!}{\includegraphics[angle=0]{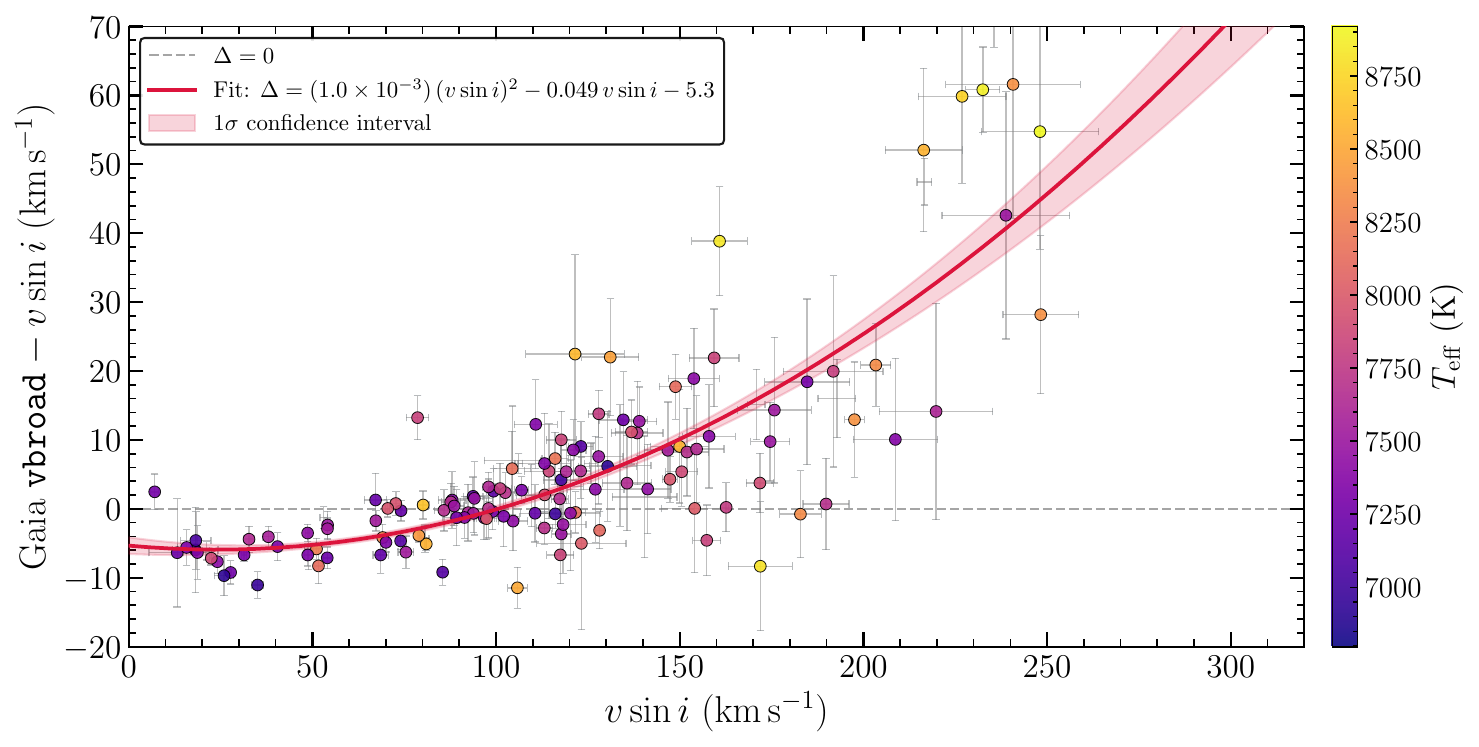}}
\caption{Comparison of measured $v\sin i$ and Gaia~DR3 \texttt{vbroad} parameter. Dashed gray is the 1:1 line. Red line is a second-order polynomial fit. Effective temperature $T_\textrm{eff}$ is obtained by fitting spectra in \texttt{iSpec}.}
\label{fig:vsini_vs_Gaia_vbroad}
\end{figure}

We have also compared our measurement of $v\sin i$ to the Gaia DR3 \texttt{vbroad} spectral line broadening parameter \citep{Fremat2023}, which combines both the effects of rotation and macroturbulence. It is well established that \texttt{vbroad} is under/overestimated compared to the projected rotational velocity at low/high $v\sin i$, and it also deviates for stars with $T_\textrm{eff}>\SI{7500}{\kelvin}$ \citep{Fremat2023,Gootkin2024,Murphy2024,Murphy2026_09}. The comparison of our measurement of $v\sin i$ to available \texttt{vbroad} is shown in Fig.~\ref{fig:vsini_vs_Gaia_vbroad} and confirms that \texttt{vbroad} in the low/high $v\sin i$ range is under/overestimated. If divided into \SI{20}{\kilo\meter\per\second} bins, the mean difference smaller than $\SI{4}{\kilo\meter\per\second}$ is only in the range \SI{60}{\kilo\meter\per\second} to \SI{120}{\kilo\meter\per\second}, while above \SI{200}{\kilo\meter\per\second}, the mean difference is $\approx\SI{45}{\kilo\meter\per\second}$. We also see that the biggest outliers are stars with $T_\textrm{eff}>\SI{8000}{\kelvin}$ and $v\sin i$ above \SI{200}{\kilo\meter\per\second}. Using the second-order polynomial $\Delta =\texttt{vbroad}-v\sin i=a\,(v\sin i)^2 + b\,(v\sin i) + c$ (red line in Fig.~\ref{fig:vsini_vs_Gaia_vbroad}), we obtain the best-fit parameters $a = (1.0 \pm 0.2) \times 10^{-3}\,\mathrm{km^{-1}\,s}$, $b = (-4.9 \pm 2.9) \times 10^{-2}$, and $c = -5.3 \pm 1.2\,\mathrm{km\,s^{-1}}$ for calibrating between $v\sin i$ and \texttt{vbroad} for the $\delta$~Sct stellar population.

\subsection{Effect of binaries}
\label{Sec:Binarity}
Binary stars, especially eclipsing binaries, are one of the best laboratories for testing stellar structure and evolution \citep{Southworth2026}. The binary fraction in A- and F-type stars can be as high as 50~\% \citep{Duchene2013, Moe2017}, and separating them from single stars can be challenging. While our sample selection avoids close eclipsing binaries and ellipsoidal variables, many non-eclipsing or even long-period binaries are present in our sample. We have checked for binarity using the SIMBAD Astronomical Database\footnote{\url{https://simbad.u-strasbg.fr/simbad/}} \citep{Wenger2000SIMBAD} and the Gaia DR3 for multiplicity \citep{GaiaDR3}. We used the \citet{Liakos2017} catalogue of $\delta$~Sct stars in binaries and crossmatched it with our sample. We have also cross-matched our sample with the catalogue of binaries detected from phase modulation of the pulsations in TESS photometry \citep{Dholakia2025} with only one star, TIC\,235709086, being in both samples. 

Out of a total of 81 binaries, we have found 5 new binary candidates. For every observed spectrum, we performed a cross correlation with the template spectrum in \texttt{iSpec}, and if more than one dominant peak is present, we assume possible binarity. We list new binary candidates in Appendix~\ref{Apendix:B}. One star, TIC\,331893210, has 3 visible components in the spectra (see Fig.~\ref{fig:Apendix:Binaries}). Some of the binaries in our sample probably remain hidden and might cause low amplitude pulsations to come from the secondary, while our measured $v\sin i$ is from the primary.

\subsection{Critically rotating stars}
\label{Sec:CriticallyRotatingStars}
\begin{figure*}[ht!]
    \centering
     {\includegraphics[width=.49\linewidth]{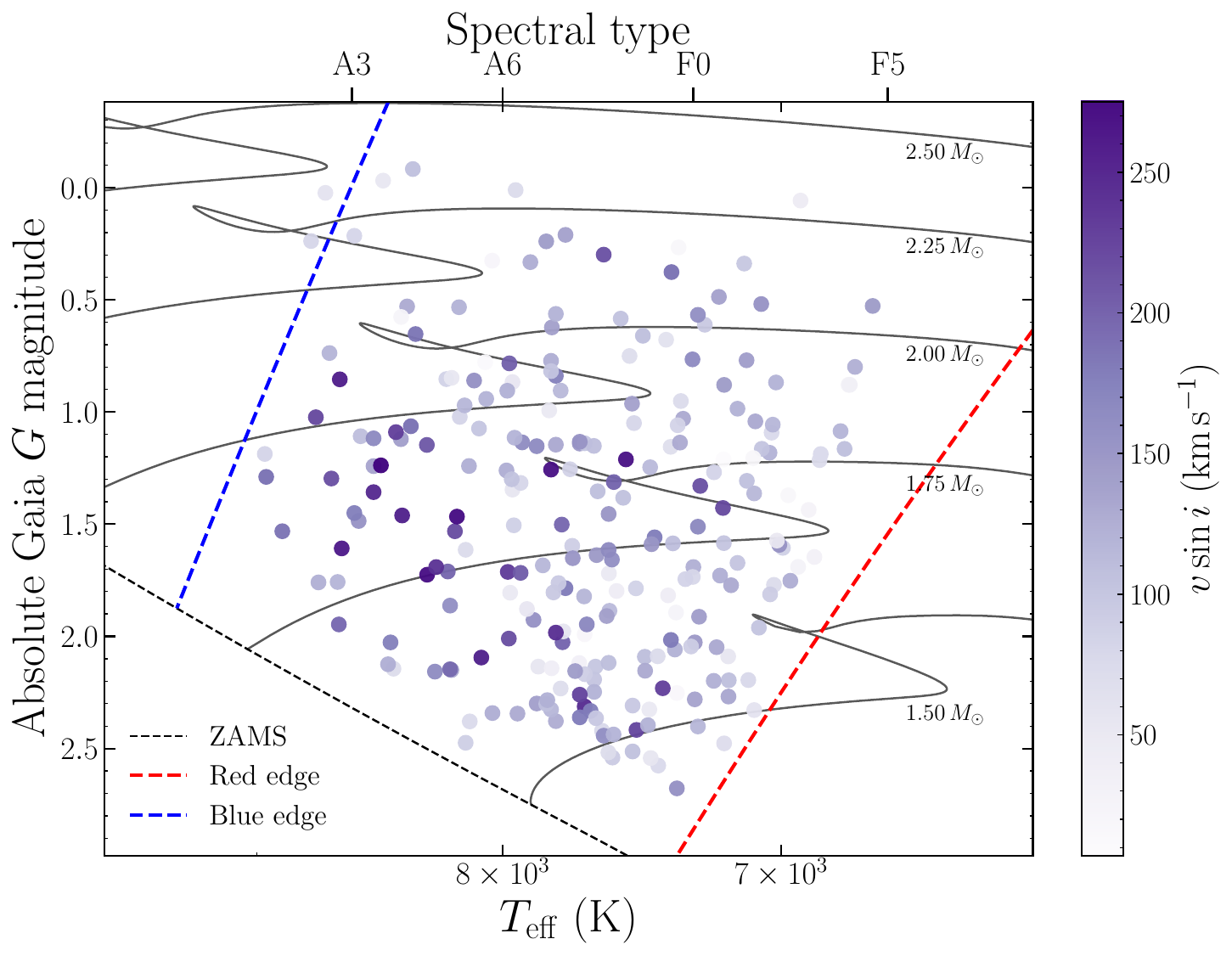}
     \includegraphics[width=.49\linewidth]{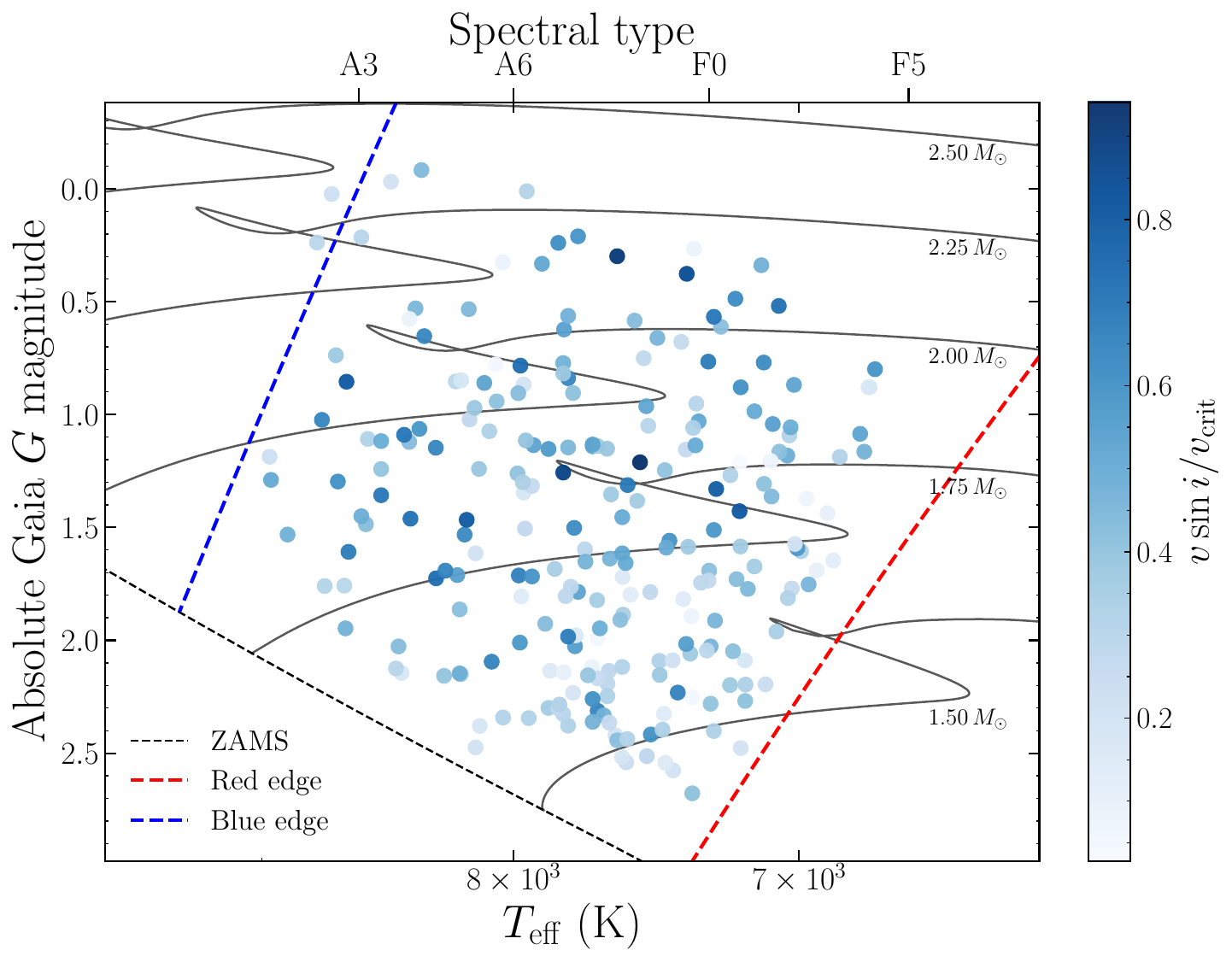}}   
     \caption{Same as Fig.~\ref{fig:HRdiagram} but with $v\sin i$ (left) and $v\sin i/v_\textrm{crit}$ ratio (right) for observed $\delta$~Scuti sample.}
      \label{fig:HR_diagram_vsini}
\end{figure*}

For stars with mass concentrated in the centre, the Roche model relates the critical rotational velocity to mass and radius as 
\begin{equation}
    v_\textrm{crit}=\sqrt{\frac{GM_*}{R_\textrm{e}}}=\sqrt{\frac{2}{3}\frac{GM_*}{R_\textrm{p}}}\,,
\end{equation}
where $G$ is the universal gravitational constant, $M_*$ is the mass of the star, and $R_\textrm{e}$ and $R_\textrm{p}$ are the equatorial and polar radii, respectively \citep{Roche1854, Maeder2000, Georgy2013}. We have used the mass $M_*=M_\textrm{TIC}$ and the polar radius value $R_\textrm{p} =  R_\textrm{TIC}$ \citep{Collins1963, Collins1965, Georgy2013, Murphy2026_08} from the TIC catalogue \citep{Stassun2019TIC} to calculate the $v_\textrm{crit}$. In the next section, it will be useful to work with the ratio $v\sin i/v_\textrm{crit}$ and not purely $v\sin i$ as the critical rotation can range from \SI{200}{\kilo\meter\per\second} to \SI{400}{\kilo\meter\per\second}. In Fig.~\ref{fig:HR_diagram_vsini}, we plot the HR diagram, colour-coded by $v\sin i$ (left) and $v\sin i/v_\textrm{crit}$ (right).

\subsection{Inclination and equatorial velocity}
It is almost impossible to obtain an inclination $i$ and therefore true equatorial rotational velocity $v_\textrm{eq}$ without major assumptions and very specific cases.

\paragraph{Mode splitting} - Asteroseismology provides an independent constraint on the inclination $i$ through the rotational splitting, which lifts the degeneracy in the azimuthal order $m$, splitting each non-radial mode ($\ell \neq 0$) into a multiplet whose spacing is proportional to the rotation frequency \citep{Ledoux1951}. Combined with the stellar radius, the measured frequency splitting gives equatorial velocity $v_{\rm eq}$ and hence $\sin i = v\sin i / v_{\rm eq}$ \citep{Aerts2010_Astroseismology}. However, this perturbative approach is valid only for slow rotators and breaks down at the equatorial velocities of $\SI{50}{}-\SI{200}{\kilo\meter\per\second}$ typical of $\delta$~Sct stars, where the multiplets become asymmetric and overlap \citep{Lignieres2006, Reese2009}.

\paragraph{Gravity darkening} - Several studies in the past have demonstrated that $v\sin i$ is underestimated due to gravity darkening \citep{Townsend2004, Fremat2005, Zorec2016, Zorec2017}, and \citet{Forteza2018} have even proposed utilising gravity darkening in the context of $\delta$~Sct stars. 

Because the morphology of the spectral line profile of a fast pole-on star $(i \to \ang{0})$ and a slowly rotating star differs, \citet{Reiners2003} proposed that by assuming rigid stellar rotation, we can find its true rotational velocity for fast rotators ($v_\textrm{eq}>\SI{200}{\kilo\meter\per\second}$) utilising the zeros of the Fourier transform of the spectra. We have tried to apply the method by \citet{Reiners2003} to our best available spectra from PLATOSpec, HERMES, and HARPS, but the change in spectral lines due to gravity darkening can also be caused by pulsations \citep{Aerts2014}, which makes this method impractical without a substantial number of high-resolution observations at different pulsation phases.

\citet{Forteza2018} proposed a method for using the scaling relation between the frequency at maximum power and temperature, where one could use the scatter from $dT$, which was supposed to be caused by gravity darkening, to obtain inclination. However, as newer studies have pointed out, \citep{Forteza2020} frequency at maximum power carries enough ambiguity, and it is hard to break the degeneracy in inclination. \citet{Murphy2024} pointed out that frequency at maximum power does not have a strong correlation with effective temperature across a homogeneous $\delta$~Sct population. We can conclude that all methods involving gravity darkening cannot be readily applied to $\delta$~Sct, as only a small fraction of stars are actually very fast rotators, in which the effects of gravity darkening would be observable in the spectra \citep{Royer2007, Zorec2012}.

\paragraph{Surface modulation} - Due to the photometric modulations caused by rotating starspots, it is possible to extract stellar rotation. If this rotational period $P_\star$ is identified properly, we can then calculate stellar rotation as $v_\textrm{eq}=\frac{2\pi R_\star}{P_\textrm{rot}}$. We have manually inspected all lightcurves for rotational modulation; however, this is complicated by the g mode frequencies excited in the $\gamma$~Dor range (0.3 to 3 d$^{-1}$) \citep{Balona2011, Garcia2022,  Skarka2022} together with instrumental artefacts from the spacecraft. We also require the presence of the rotational frequency $\nu_\textrm{rot}$ and its harmonic $2\nu_\textrm{rot}$ where $\nu_2 = 2\nu_1$ is within formal analytical error. We have not found any clear candidates, which again points to the fact that none of these methods can be used for a statistical sample, but only for very specific stars.

\section{TESS photometry}
\label{Sec:TESS_photometry}
At the beginning, we downloaded all available sectors from TESS using the \texttt{lightkurve Python} package \citep{Lightkurve2018}. We used Pre-search Data Conditioned Simple Aperture Photometry flux (PDCSAP) in the 2-minute cadence mode.  If SPOC lightcurves were not available, we used Quick-Look Pipeline (QLP) data. The data were normalised to their median value per sector, and no further detrending was done. Available sectors were stitched into contiguous light curves without removing the gaps \citep{Bedding2022}. The sectors can be even years apart, and this can lead to a spectral window contaminated with aliases \citep{Forteza2024}; however, in the statistical sense of this study, this does not induce substantial error. The longest lightcurve spans 12 sectors, while the median spans 3 sectors. An example light curve can be seen in Fig.~\ref{fig:TIC144219989_lightcurve}. For every TIC number in our sample, we have checked the \texttt{contratio}, which identifies neighbouring sources and estimates their contribution to the target flux within a nominal TESS aperture using a PSF-based contamination model. For targets with \texttt{contratio} $> 0.01$ we have manually inspected TESS target pixel files using \texttt{tpfplotter} \citep{Aller2020tpfplotter} and discarded strongly contaminated targets.

\subsection{Pulsation amplitudes}
\label{sec:PulsationAmplitudes}
\begin{figure*}[ht!]
    \centering
     \includegraphics[width=\linewidth]{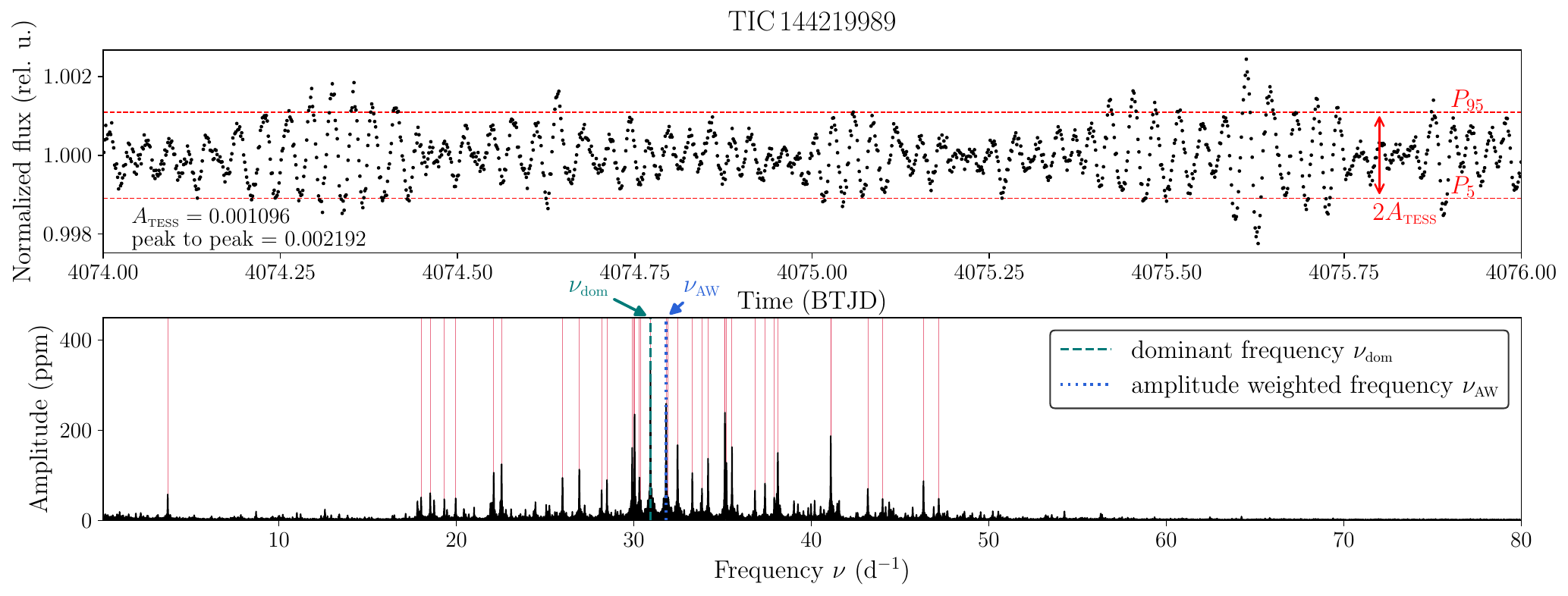}
     \caption{Top: Example of relative semi-amplitude determination. 5th ($P_5$) and 95th ($P_{95}$) percentile are shown using red dashed lines, semi-amplitude $A_\textrm{TESS}$ using red arrow. Bottom: periodogram with extracted frequencies with $S/N>4$ (red), dominant frequency $\nu_\textrm{dom}$ (green) and amplitude weighted frequency $\nu_\textrm{AW}$ (blue).}
      \label{fig:TIC144219989_lightcurve}
\end{figure*}

Because $\delta$~Sct stars can pulsate at hundreds of frequencies, with different amplitudes evolving over time, no single photometric amplitude is defined. Because of this, we have decided to use a 5th ($P_5$) and 95th ($P_{95}$) percentile approach of the normalised TESS flux to estimate the normalised semi-amplitude $A_\textrm{TESS}$. The semi-amplitude $A_\textrm{TESS}$ was calculated as half of the relative flux distance from the $P_5$ and $P_{95}$ percentile lines, see Fig.~\ref{fig:TIC144219989_lightcurve}. This simple approach allows us to define amplitude even for light curves with multiple pulsation modes and complicated behaviour.

The relation between the TESS semi-amplitude $A_\textrm{TESS}$ and the projected rotational velocity $v\sin i$ can be seen in Fig.~\ref{fig:vsini_vcrit_vs_amplitudes} subplot (VII). Although this relation between photometric amplitude and $v\sin i$ is well established, measured by \citet{Danziger1972} and \citet{Solano1997} in $V$ magnitude, their sample did not adequately cover all values of the projected rotational velocity $v\sin i$. In addition, we find that the upper envelope drops and the lower envelope rises; therefore, the range of photometric amplitudes is very limited for higher $v\sin i$; see the underpopulated regions in Subsect.~\ref{Sec:Statistical_analysis}.

\paragraph{High-amplitude $\delta$~Sct (HADS)}  \citet{McNamara2000} defined HADS as a class of radial pulsators having peak-to-peak photometric amplitudes $\Delta V>0.3\ \textrm{mag}$. They pulsate in fundamental and first-overtone radial modes and exhibit small rotational velocities $(v\sin i<\SI{40}{\kilo\meter\per\second})$; however, not all slow rotators have high amplitudes. Some HADS may even pulsate in low-amplitude non-radial pulsations \citep{Poretti2003, Poterri2005}. The nature of HADS remains poorly understood, and it is still an open question how they differ from other $\delta$~Sct stars \citep{Balona2016}. 

\begin{figure}[ht!]
    \centering
     \includegraphics[width=\linewidth]{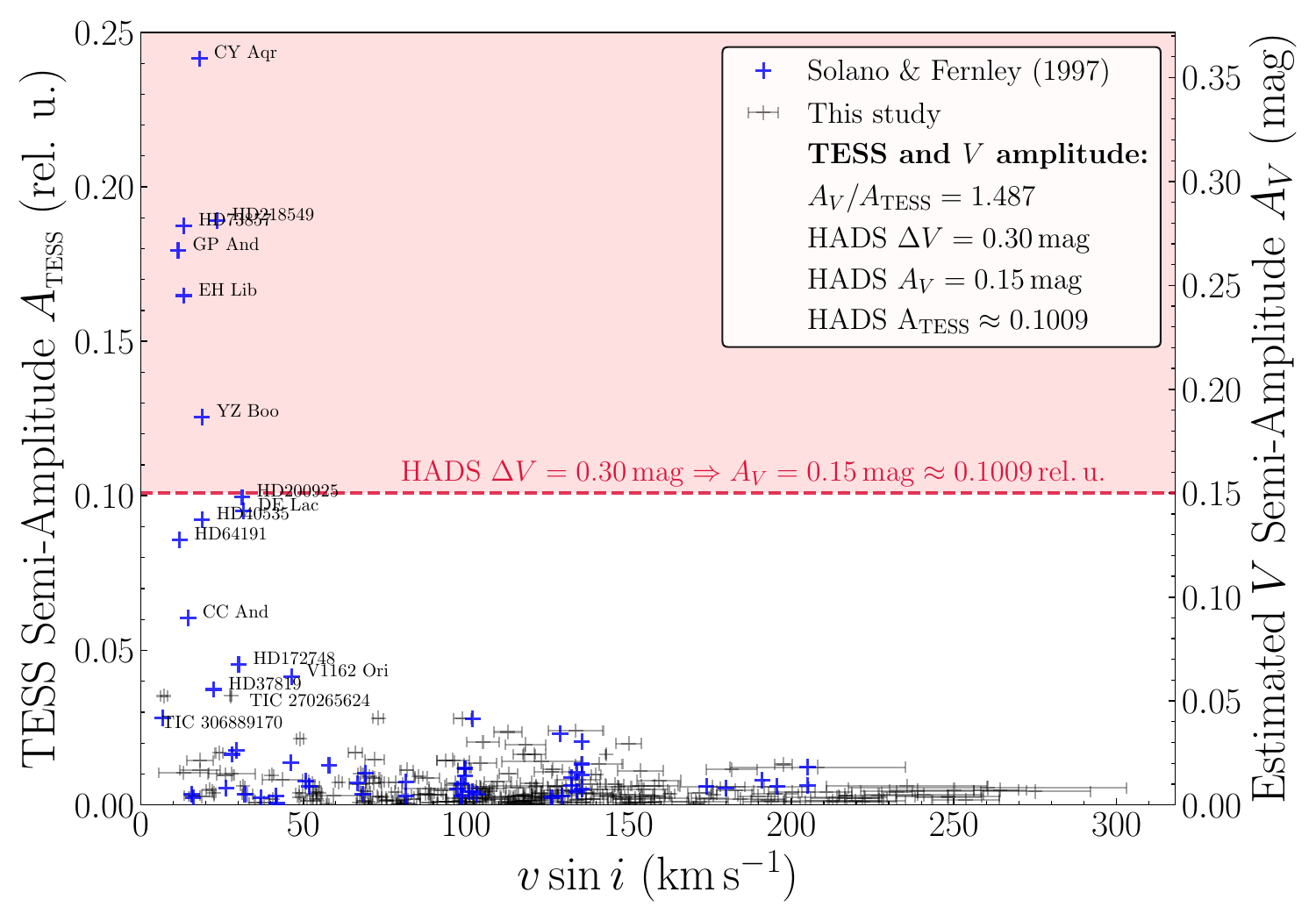}
     \caption{Photometric TESS semi-amplitude $A_\textrm{TESS}$ vs $v\sin i$ for our sample complemented with stars from \citet{Solano1997}. Determined ratio of TESS and $V$ amplitudes is in the top-right corner. Red boundary is the HADS region defined by \citet{McNamara2000}.}
      \label{fig:HADS_region}
\end{figure}

Pulsation amplitudes in the TESS passband ($\SI{600}{}$–$\SI{1000}{\nano\meter}$, $\lambda_\textrm{eff}\approx\SI{786}{\nano\meter}$) are suppressed compared to Johnson $V$. Because of this, we define HADS in the TESS passband by calibrating 59 stars from \citet{Solano1997} using our semi-amplitude determination $A_\textrm{TESS}$ to their peak-to-peak amplitude $\Delta V= 2A_V$, which is twice the semi-amplitude we use (see Fig.~\ref{fig:HADS_region}). We empirically derive the semi-amplitude ratio $A_V/A_\textrm{TESS}=1.49\pm0.05$, 
which is consistent with theoretical non-adiabatic pulsation models 
\citep{DaszynskaDaszkiewicz2003, Dupret2005} and with the empirical ratio 
$A_\textrm{TESS}/A_{Kepler}=0.74\pm0.01$ \citep{Antoci2019}. This would put HADS at $A_\textrm{TESS}> 0.1$ in our relative TESS semi-amplitude. HADS make up only a small fraction of the general $\delta$~Sct population \citep{McNamara2000, Breger2000}, and due to our sample definition, we have not observed any HADS. However, one star in our sample, TIC\,306889170, with $v\sin i=$ \SI{7\pm1}{\kilo\meter\per\second}, has HADS characteristics with much lower amplitude $A_V=0.035$. TIC\,306889170 has two dominant modes with a period ratio $P_1/P_0=0.79$, typical of the radial first overtone and the fundamental mode 
\citep{Petersen1996}. This again confirms that not all slowly rotating, double-mode radial pulsators are necessarily HADS \citep{Zima2007, Lenz2008} and that the distinction between HADS and classical $\delta$~Sct stars is still not understood.

\subsection{Frequency and amplitude analysis}
A common characteristic of $\delta$~Sct stars is that they pulsate in a variety of frequencies, combining both p and g modes \citep{Breger2000, Breger2005, Breger2011, Aerts2010_Astroseismology, Gatuam}. Given the size of our sample, individual mode identification for each star is beyond the scope of this study; therefore, we adopt a
simplified approach for the frequency analysis. For each object, we calculated Lomb-Scargle periodograms normalised to Fourier semi-amplitude \citep{Lomb1976, Scargle1982} using the \texttt{Astropy} \citep{astropy:2013, astropy:2018, astropy:2022} \texttt{Python} package and extracted all frequencies with $S/N > 4$ \citep{Breger1993} through iterative prewhitening, with all frequencies fitted and optimised using the multi-sinusoidal least-squares model \citep{LenzBreger2005}. We searched the frequency range from 0 to 150 d$^{-1}$, allowing a maximum of 50 extracted frequencies, however, this limit was never reached. We also require that each extracted frequency must lie at least one Rayleigh resolution ($1/T$, where $T$ is the length of the data set) away from every frequency already found \citep{Antoci2019}, and its amplitude must be between 0.8 and 1.25 times the amplitude in the original periodogram at the same frequency \citep{Antoci2019, VanReeth2015b}. An example periodogram with extracted frequencies is shown in Fig.~\ref{fig:TIC144219989_lightcurve}. 

\begin{figure*}[ht!]
    \centering
     \includegraphics[width=\linewidth]{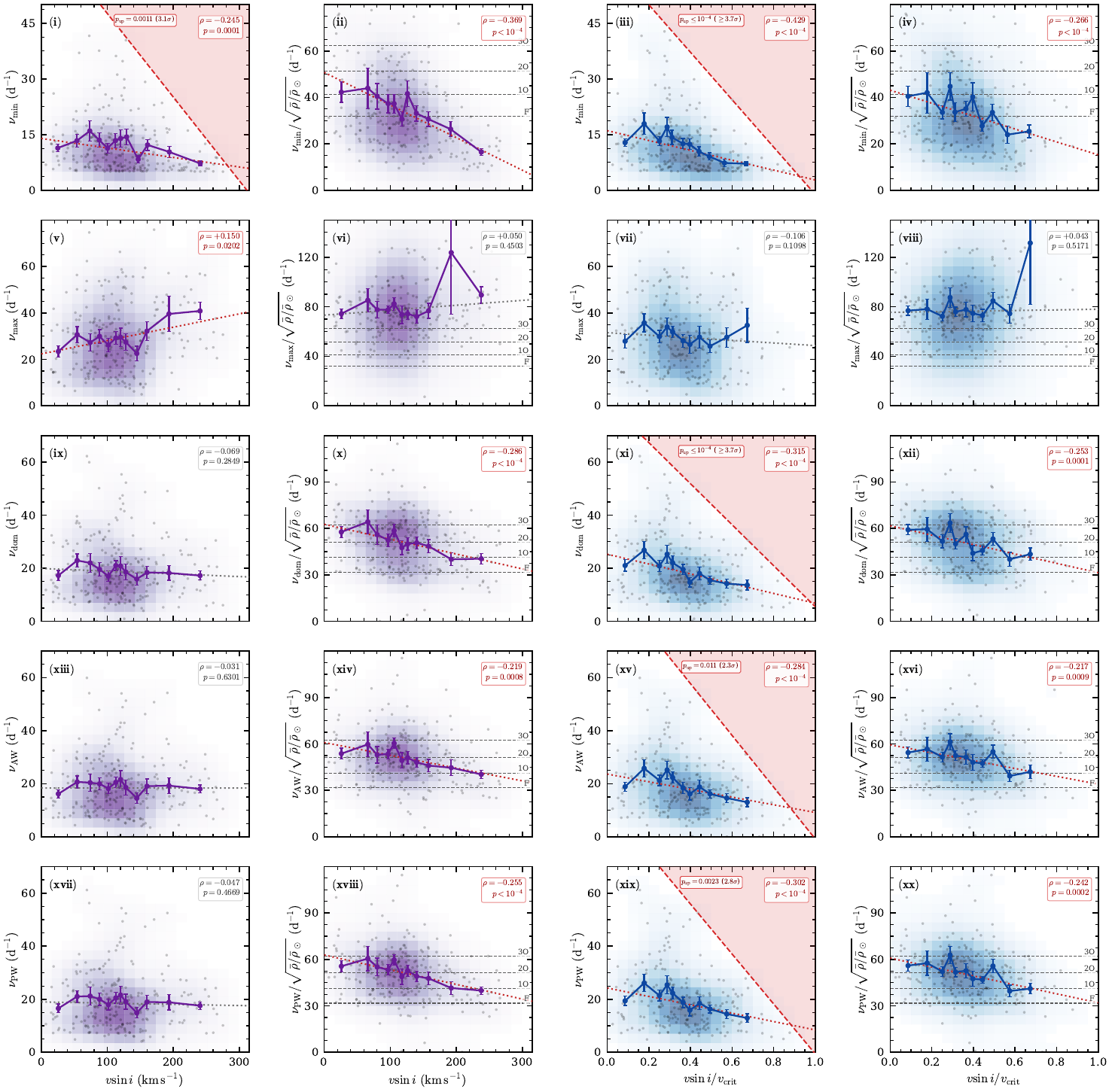}
     \caption{Column 1: Measured $v\sin i$ vs lowest extracted frequency $\nu_\textrm{min}$ in the range $>5$ d$^{-1}$, largest extracted frequency $\nu_\textrm{max}$, dominant frequency $\nu_\textrm{dom}$, amplitude weighted frequency $\nu_\textrm{AW}$, and power weighted frequency $\nu_\textrm{PW}$.
     Column 2: Same as column 1 but for mean density normalised frequencies. Column 3: Same as column 1 but for $v\sin i/v_\textrm{crit}$. Column 4: Same as column 2 but for $v\sin i/v_\textrm{crit}$. In columns 1 and 3 red regions are the upper underpopulated regions. Permutation $p_\textrm{up}$ value is printed together with $\sigma$ in the middle red squares. Uncertainty weighted straight line fits for bins with 20 stars are shown in dashed red line (statistically significant) and dashed gray line (no significance). Spearman rank correlation $\rho$ and $p$ are shown in upper right corner. Horizontal dashed lines in columns 2 and 4 is the radial fundamental mode (F), and the first, second, and third radial overtones (1O–3O). Blue and purple points are the binned data ($\textrm{bin\ size} = 20$ stars) and lines are the uncertainty weighted straight line fits.}
      \label{fig:vsini_vcrit_vs_frequencies}
\end{figure*}

Using the extracted frequencies, we define the following metrics \citep{Kallinger2010,Forteza2018, Forteza2020}: the lowest extracted frequency $\nu_\textrm{min}$ in a range $>5$ d$^{-1}$, the highest extracted frequency $\nu_\textrm{max}$, the dominant frequency $\nu_\textrm{dom}$ as the frequency with the highest amplitude, the dominant amplitude $A_\textrm{dom}$ as the highest amplitude, the amplitude weighted frequency $\nu_\textrm{AW}$, and the mean amplitude $A_\textrm{mean}$
\begin{equation}
    \nu_\textrm{AW} = \frac{\sum_i^N \nu_iA_i}{\sum_i^N A_i}\,,\quad A_\textrm{mean} = \frac{\sum_i^NA_i}{N}\,,
\end{equation}
power weighted frequency $\nu_\textrm{PW}$ and mean-square amplitude $\langle A^2\rangle$
\begin{equation}
    \nu_\textrm{PW} = \frac{\sum_i^N \nu_iA_i^2}{\sum_i^N A_i^2}\,,\quad \langle A^2\rangle = \frac{\sum_i^NA^2_i}{N}\,.
\end{equation}
Distributions of defined frequency metrics are shown in Fig.~\ref{fig:vsini_vcrit_vs_frequencies}, subplots (i), (iii), (v), (vii), (ix), (xi), (xiii), (xv), (xvii), (xix) and amplitude metrics in Fig.~\ref{fig:vsini_vcrit_vs_amplitudes}, subplots (I) to (VIII).  

\subsection{Statistical analysis}
\label{Sec:Statistical_analysis}
\begin{figure}[ht!]
    \centering
     \includegraphics[width=\linewidth]{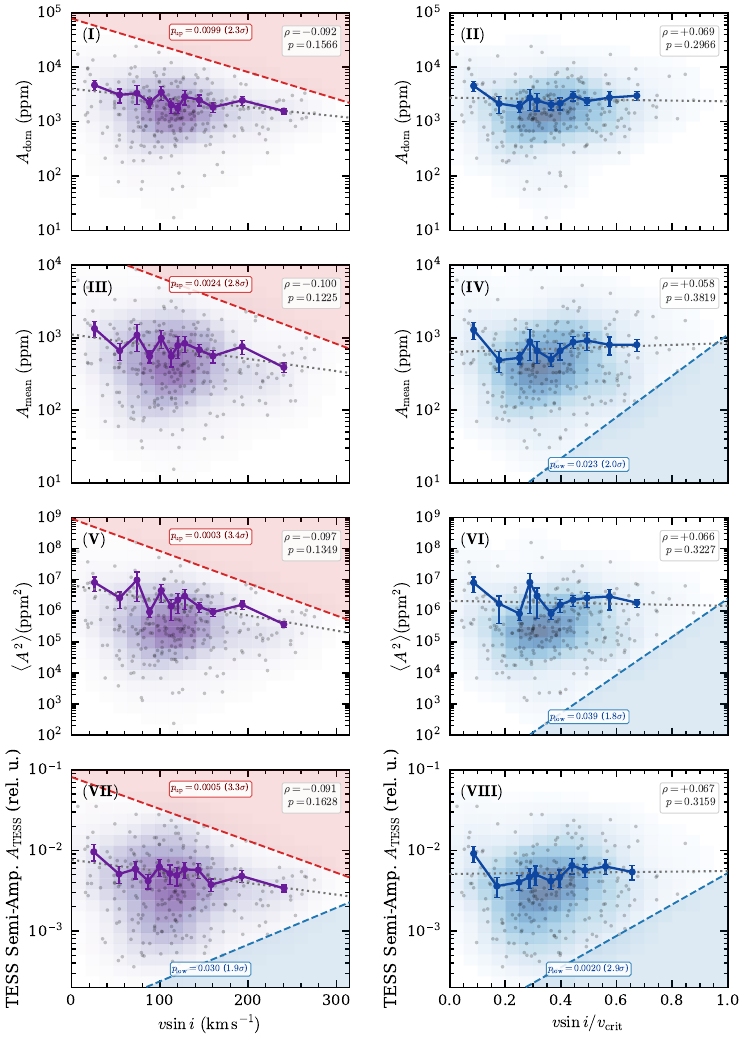}
     \caption{Same as Fig.~\ref{fig:vsini_vcrit_vs_frequencies} but for amplitudes. Left column: measured $v\sin i$ vs dominant amplitude $A_\textrm{dom}$, mean amplitude $A_\textrm{mean}$, mean-square amplitude $\langle A^2\rangle$ and TESS pulsation semi-amplitude $A_\textrm{TESS}$. Right column: same as left but for $v\sin i/v_\textrm{crit}$.}
      \label{fig:vsini_vcrit_vs_amplitudes}
\end{figure}

\paragraph{Underpopulated regions} - Visual inspection of the subplots in Fig.~\ref{fig:vsini_vcrit_vs_frequencies} and Fig.~\ref{fig:vsini_vcrit_vs_amplitudes} reveals the possibility that the frequencies decrease with increasing projected rotation $v\sin i$ and that certain regions of the plots are underpopulated. To test this hypothesis for all frequency metrics, we define the \textit{underpopulated region} using the linear boundary envelope that encloses all of the observed data ($\nu=ax+b$, Fig.~\ref{fig:vsini_vcrit_vs_frequencies}), and for amplitude metrics, the boundaries are log-linear ($\log A = ax+b$, Fig.~\ref{fig:vsini_vcrit_vs_amplitudes}). This is done for the upper and lower boundaries, represented by the red and blue dashed lines. After that, we evaluate the statistical significance of these underpopulated regions using a non-parametric Monte Carlo permutation test. We permute parameters on the $y$ axes \SI{10000}{} times, while keeping $v\sin i$ or $v\sin i/v_\textrm{crit}$ fixed. In each permutation, we refit the boundary envelope to the permuted data and record the area of the resulting empty region, so that the test accounts for the boundary itself being defined from the observed data. The $p$ value is the fraction of permutations whose empty region is at least as large as the observed one, and we express it as $\sigma$, the number of standard deviations from the mean of a normal distribution that would give the same probability (e.g. $p = 0.05$ corresponds to $2\sigma$). In Figs.~\ref{fig:vsini_vcrit_vs_frequencies} and \ref{fig:vsini_vcrit_vs_amplitudes}, we only show regions with $p<0.05$.

For $v\sin i$, the most significant underpopulated regions are found for the
minimum frequency (i) at $3.1\sigma$ and, among the amplitude metrics, for
$\langle A^2\rangle$ (V) at $3.4\sigma$ and the TESS semi-amplitude $A_\textrm{TESS}$ (VII) at $3.3\sigma$,
followed by $A_\textrm{mean}$ (III) at $2.8\sigma$ and $A_\textrm{dom}$ (I) at
$2.3\sigma$. When the frequency metrics are plotted against
$v\sin i/v_\textrm{crit}$, the trends become even more significant; in subplots
(iii) and (xi), the underpopulated regions reach $p\leq10^{-4}$ ($\geq3.7\sigma$),
followed by (xix) at $2.8\sigma$ and (xv) at $2.3\sigma$, while the maximum
extracted frequency $\nu_\textrm{max}$ (vii) remains unchanged. When pulsation
amplitudes are plotted against $v\sin i/v_\textrm{crit}$, the upper underpopulated
region disappears, and the lower becomes more prominent, most notably for the
TESS amplitude (VIII) at $2.9\sigma$. Because the effective temperature $T_\textrm{eff}$ and $v\sin i$ are intertwined, we also performed random permutations in the \SI{250}{\kelvin} temperature bins. With this temperature permutation, we can obtain even stronger statistical significance.

\paragraph{General trends} - Minimum and maximum extracted frequencies can be heavily dependent on the $S/N$ ratio in the periodograms; therefore, we also investigate the general trends using a straight line fit. We have divided the sample into bins with 20 stars in the $v\sin i$ and $v\sin i/v_\textrm{crit}$, performed an uncertainty weighted straight line fit, and calculated the non-parametric Spearman correlation test for all unbinned data. For $v\sin i$, the only statistically significant yet modest decrease is $\nu_\textrm{min}$ ($\rho= -0.245$ and $p=0.0001$), while $\nu_\textrm{max}$ ($\rho= +0.150$ and $p=0.0202$) shows an even smaller increase that is less statistically significant. All the other parameters remain invariant ($p>0.05$). When we evaluated pulsation characteristics against $v\sin i/v_\textrm{crit}$, we find that the only statistically significant results are $\nu_\textrm{min}$ ($\rho=-0.429$, $p<10^{-4}$), $\nu_\textrm{dom}$ ($\rho=-0.315$, $p<10^{-4}$), $\nu_\textrm{AW}$ ($\rho=-0.284$, $p<10^{-4}$), and $\nu_\textrm{PW}$ ($\rho=-0.302$, $p<10^{-4}$), while all other metrics remain invariant.

Frequency metrics $\nu_\textrm{min}$, $\nu_\textrm{dom}$, $\nu_\textrm{AW}$, and $\nu_\textrm{PW}$ are decreasing with $v\sin i/v_\textrm{crit}$, which can be explained by the fact that the frequencies of p modes scale with mean stellar density $\nu\propto \sqrt{\Bar{\rho}}$ \citep{Breger2000, Aerts2010_Astroseismology}, and the centrifugal force in fast rotators causes the equatorial radius to inflate while conserving mass, which lowers the mean stellar density $\Bar{\rho}$.

To investigate whether the inflation of the fast rotating star is solely responsible for this monotonic decrease in frequencies, we have normalised all frequency metrics by the square root of the mean stellar density $\sqrt{\bar{\rho}/\bar{\rho}_\odot}$, which is calculated from the $M_\textrm{TIC}$ and $R_\textrm{TIC}$ values. This can be done in the first approximation even for equator-on fast rotators (which is our case for high $v\sin i$), and the error in density would not exceed 5-10\%. As seen in Fig.~\ref{fig:vsini_vcrit_vs_frequencies} subplots (ii), (iv), (x), (xii), (xiv), (xvi), (xviii), (xx) the decreasing trends persist, which suggests that density is not the only driver of lower frequencies in fast rotators. Horizontal dashed lines in Fig.~\ref{fig:vsini_vcrit_vs_frequencies} columns 2 and 4 mark the radial fundamental mode (F), derived from the empirical pulsation constant $Q_\mathrm{F} = 0.0315$ d \citep{Barac2022, Mani2025}, and the first, second, and third radial overtones (1O–3O), obtained from period ratios $P_n/P_0$ predicted by pulsation models \citep{Stellingwerf1979}.

\section{Conclusions}
\label{Sec:Conclusions}
In this paper, we present the relationships between the rotation of $\delta$~Sct stars and their frequencies and amplitudes. Our findings show the general trends among this stellar population. We have derived projected rotational velocities $v\sin i$ for 262 bright $\delta$~Sct stars from an all-sky survey using 5 high-resolution spectrographs. Together, we have obtained 307 high-resolution spectra, with some of the stars being observed with multiple instruments. We have derived values of $v\sin i$ using three different methods and compared them to each other. We conclude that the different methods provide consistent results at low to moderate rotational velocities but begin to diverge at higher velocities, which we attribute, at least partly, to gravity darkening in rapidly rotating stars. We show that the statistical error from deriving $v\sin i$ and effective temperature $T_\textrm{eff}$ is smaller than the spread of the values derived using different spectrographs. We conclude that deriving per-star values of $v\sin i$ and effective temperature $T_\textrm{eff}$ is not only sensitive to the method but also to the instrument itself and the quality of its spectra, mainly the normalisation of the continuum and $S/N$. We demonstrate that instrumental effects and drifts can appear between different spectrographs (see Fig.~\ref{fig:vsini_comparison}, subplot (b)), highlighting the need for cross-checking to ensure homogeneous parameter estimates. We discussed the different possibilities of finding true $v_\textrm{eq}$ and showed that the methods are either highly impractical or almost impossible for $\delta$~Sct stars.

We have also compared our derived $v\sin i$ to the literature values (Fig.~\ref{fig:vsini_comparison_previous_studies}), and different spectrographs with each other (Fig.~\ref{fig:comparison_spectrographs}), which shows that lower resolution and higher resolution spectra yield different values depending on the methods used. This behaviour can reflect the fact that some methods are more effective in decoupling the intrinsic rotational broadening from the spectrograph's instrumental profile. The measured $v\sin i$ is compared to the Gaia \texttt{vbroad} spectral line broadening parameter (Fig.~\ref{fig:vsini_vs_Gaia_vbroad}), which confirms that for low/high $v\sin i$, the \texttt{vbroad} parameter is under/overestimated. Using the second-order polynomial, we obtain the calibration relation $
    \Delta [\SI{}{\kilo\meter\per\second}] = \texttt{vbroad} - v\sin i = (1.0 \pm 0.2)\times 10^{-3}\,(v\sin i)^2 - (4.9 \pm 2.9)\times 10^{-2}\,(v\sin i) - (5.3 \pm 1.2).
$ This indicates that additional spectroscopic information is needed to reliably determine rotational velocities across the full range.

The distribution of $v\sin i$ is \textit{unimodal}, with a mean value of \SI{120.3}{\kilo\meter\per\second} and a peak probability at \SI{113.8}{\kilo\meter\per\second} (Fig.~\ref{fig:vsini_distribution}).

\citet{Bedding2020} showed that fast rotators viewed pole-on $(i \to \ang{0})$ have a simplified pulsation spectrum, as the disc-integrated visibility arises primarily from zonal $(m=0)$ and pure radial $(\ell=0)$ modes. Gravity darkening in fast rotators causes shifts in temperature and apparent luminosity, and moves them in the HR diagram. Naturally, we will have some pole-on pulsators located in the low $v\sin i$ range, but  the probability distribution of randomly orientated inclinations is $P(i)\,di=\sin i\, d i$, so intermediate-to-high inclinations are statistically much more likely than pole-on configurations. This allows us to use $v\sin i$ not only as a firm lower limit but also as a robust proxy for the intrinsic equatorial rotation velocity $v_{\mathrm{eq}}$.

Using all available TESS photometry, we have derived photometric amplitudes, extracted pulsation frequencies and amplitudes, and compared them to our defined pseudo metrics (Figs.~\ref{fig:vsini_vcrit_vs_frequencies} and \ref{fig:vsini_vcrit_vs_amplitudes}). We confirm that the photometric amplitude decreases with $v\sin i$ and shows a modest decrease even for stars that are not high-amplitude $\delta$~Sct. We find that the value of photometric amplitude is constrained from the lower region with high $v\sin i$. We find that stars with high $v\sin i$ have a much more limited range of allowed amplitudes than slow rotators. We calibrate the $V$ magnitude with TESS flux, obtaining the semi-amplitude ratio $A_V/A_\textrm{TESS}=1.49\pm0.05$. Using the definition by \citet{McNamara2000}, we can define HADS in the TESS passband as stars with a relative semi-amplitude $A_\textrm{TESS}>0.10$.

We find that, apart from the maximum frequency $\nu_\textrm{max}$, all frequency metrics \textit{decrease} with rotation. The minimum extracted frequency $\nu_\textrm{min}$ decreases with $v\sin i$, and the empty region of fast rotators with high frequencies has a significance of 3.1$\sigma$, while the largest frequency $\nu_\textrm{max}$ does not decrease with any statistical significance. The dominant frequency $\nu_\textrm{dom}$, amplitude weighted frequency $\nu_\textrm{AW}$, and power weighted frequency $\nu_\textrm{PW}$ decrease with $v\sin i/v_\textrm{crit}$. We show that the amplitudes of the extracted frequencies and their following metrics ($A_\textrm{dom}$, $A_\textrm{mean}$, $\langle A^2\rangle$) are also constrained in the region of fast rotators, and that pulsation amplitudes have a more limited range with higher rotation. At high $v\sin i$, the TESS photometric amplitude is bounded from above and, more weakly, from below.

Frequencies of p modes scale with the average density $\nu\propto\sqrt{\Bar{\rho}}$. Centrifugal force in fast rotators causes the equatorial radius to inflate while conserving mass, which results in lower densities and, therefore, lower frequencies. We show that the change in stellar density cannot be solely responsible for the decrease in observed frequencies. Dividing frequencies by the $\sqrt{\bar{\rho}/\bar{\rho}_\odot}$, we remove the first-order effect of rotational inflation. This property then measures the pulsation constant $\nu/\sqrt{\bar{\rho}/\bar{\rho}_\odot}=1/Q$, which is nearly the same for all $\delta$~Sct stars for a given radial mode. 

We suggest that rapid rotation lowers the characteristic radial order of excited p modes, which, together with a decrease in mean stellar density, causes lower observed frequencies. This agrees with \citet{Murphy2024}, who found that rapid rotators in the Cep–Her complex populate the low-order ridge of the $\nu_\mathrm{max}$–$T_\mathrm{eff}$ diagram, which is dominated by modes near the radial fundamental.

Even though the frequency normalisation relies on TIC masses and radii, the individual stars can be uncertain by half a radial order, whereas the shift to lower frequencies is significant for faster rotators. The highest extracted frequency lies well above the third overtone and is independent of $v\sin i$. These cannot be low-order radial p modes, but they may be higher-order radial overtones, non-radial p modes, or combination frequencies; distinguishing between these possibilities requires mode identification, which is beyond the scope of this study. The lowest extracted frequencies have values below the expected  fundamental radial mode ($\approx 32 \textrm{d}^{-1}$), so they cannot be low-order radial modes. We attribute these peaks to non-radial p modes, mainly low-degree retrograde components shifted to lower frequencies in the inertial frame by rotational splitting, and suggest that fast rotators excite a larger fraction of non-radial p modes at low radial orders. The excited frequencies of fast rotators tend to be more complex than those of slow rotators, often caused by rotational splitting and the coupling of modes of different degrees \citep{Aerts2010_Astroseismology, Bedding2020,Murphy2024}.
 
In this study, we have demonstrated that the pulsation frequencies of $\delta$~Sct stars depend on stellar rotation. Our findings demonstrate the need for asteroseismic modelling that takes rotation into account. We show the need for large spectroscopic samples in the era of high-precision space photometry. Relations between $v\sin i$ and pulsation characteristics established in this study provide a spectroscopic reference which can be used and compared for a much larger population of $\delta$~Sct stars observed by the upcoming PLATO mission \citep{rauer2025plato}.

\section*{Data availability}
Table 2 is available only in electronic form at the CDS via anonymous ftp to cdsarc.u-strasbg.fr (130.79.128.5) or via http://cdsweb.u-strasbg.fr/cgi-bin/qcat?J/A+A/.

\begin{acknowledgements}
      We are grateful to all observers for their work. MZ acknowledges the support from EU MSCA EXOWORLD project ID:101086149. MS and JZ acknowledge support by PRODEX agreement no. 400012791. MV and TP acknowledge support from the Slovak Research and Development Agency supported this work under contract No. APVV-24-0160 and the VEGA grant of the Slovak Academy of Sciences No. 2/0033/26. We acknowledge a bilateral mobility project between Slovak Academy of Sciences and Czech Academy of Sciences SAV-25-10 (CAS-SAS-2024-01). The research of PG was supported by the Slovak Research and Development Agency under contract No. APVV-24-0160. This study is based on new spectroscopic observations with the MUSICOS, OES, and PLATOSpec échelle spectrographs and the ESO and MELCHIORS spectral library archives.  
      The study uses space photometry from the Transiting Exoplanet Survey Satellite (TESS), which is publicly available from the Mikulski Archive for Space Telescopes (MAST). Funding for the TESS mission is provided by NASA’s Science Mission Directorate. This work made use of \texttt{tpfplotter} by J. Lillo-Box (publicly available in www.github.com/jlillo/tpfplotter), which also made use of the \texttt{Python} packages \texttt{astropy}, \texttt{lightkurve}, \texttt{matplotlib} and \texttt{numpy}. Finally, we acknowledge the use of the electronic database from the CDS, Strasbourg, and the electronic bibliography maintained by the NASA/ADS system.
\end{acknowledgements}

\bibliographystyle{bibtex/aa} 
\bibliography{bibtex/bibliography}

@ARTICLE{Brahm2017,
       author = {{Brahm}, Rafael and {Jord{\'a}n}, Andr{\'e}s and {Espinoza}, N{\'e}stor},
        title = "{CERES: A Set of Automated Routines for Echelle Spectra}",
      journal = {\pasp},
         year = 2017,
        month = mar,
       volume = {129},
       number = {973},
        pages = {034002},
          doi = {10.1088/1538-3873/aa5455},
archivePrefix = {arXiv},
       eprint = {1609.02279},
 primaryClass = {astro-ph.IM},
       adsurl = {https://ui.adsabs.harvard.edu/abs/2017PASP..129c4002B}
}

@BOOK{Aerts2010_Astroseismology,
       author = {{Aerts}, Conny and {Christensen-Dalsgaard}, J{\o}rgen and {Kurtz}, Donald W.},
        title = "{Asteroseismology}",
         year = 2010,
          doi = {10.1007/978-1-4020-5803-5},
       adsurl = {https://ui.adsabs.harvard.edu/abs/2010aste.book.....A}
}

@article{Aerts2014,
	author = {Aerts, C. and Simón-Díaz, S. and Groot, P. J. and Degroote, P.},
	title = {On the use of the Fourier transform to determine the projected rotational velocity of line-profile variable B stars⋆},
	DOI= "10.1051/0004-6361/201424012",
	url= "https://doi.org/10.1051/0004-6361/201424012",
	journal = {\aap},
	year = 2014,
	volume = 569,
	pages = "A118",
	month = "",
}

@article{Aller2020tpfplotter,
	author = {Aller, A. and Lillo-Box, J. and Jones, D. and Miranda, L. F. and Barceló Forteza, S.},
	title = {Planetary nebulae seen with TESS: Discovery of new binary central star candidates from Cycle 1},
	DOI= "10.1051/0004-6361/201937118",
	url= "https://doi.org/10.1051/0004-6361/201937118",
	journal = {\aap},
	year = 2020,
	volume = 635,
	pages = "A128",
}

@article{Antoci2014,
doi = {10.1088/0004-637X/796/2/118},
url = {https://doi.org/10.1088/0004-637X/796/2/118},
year = {2014},
month = {nov},
publisher = {The American Astronomical Society},
volume = {796},
number = {2},
pages = {118},
author = {Antoci, V. and Cunha, M. and Houdek, G. and Kjeldsen, H. and Trampedach, R. and Handler, G. and Lüftinger, T. and Arentoft, T. and Murphy, S.},
title = {THE ROLE OF TURBULENT PRESSURE AS A COHERENT PULSATIONAL DRIVING MECHANISM: THE CASE OF THE δ SCUTI STAR HD 187547},
journal = {\apj}
}

@ARTICLE{Antoci2019,
       author = {{Antoci}, V. and {Cunha}, M.~S. and {Bowman}, D.~M. and {Murphy}, S.~J. and {Kurtz}, D.~W. and {Bedding}, T.~R. and {Borre}, C.~C. and {Christophe}, S. and {Daszy{\'n}ska-Daszkiewicz}, J. and {Fox-Machado}, L. and {Garc{\'\i}a Hern{\'a}ndez}, A. and {Ghasemi}, H. and {Handberg}, R. and {Hansen}, H. and {Hasanzadeh}, A. and {Houdek}, G. and {Johnston}, C. and {Justesen}, A.~B. and {Kahraman Alicavus}, F. and {Kotysz}, K. and {Latham}, D. and {Matthews}, J.~M. and {M{\o}nster}, J. and {Niemczura}, E. and {Paunzen}, E. and {S{\'a}nchez Arias}, J.~P. and {Pigulski}, A. and {Pepper}, J. and {Richey-Yowell}, T. and {Safari}, H. and {Seager}, S. and {Smalley}, B. and {Shutt}, T. and {S{\'o}dor}, A. and {Su{\'a}rez}, J.-C. and {Tkachenko}, A. and {Wu}, T. and {Zwintz}, K. and {Barcel{\'o} Forteza}, S. and {Brunsden}, E. and {Bogn{\'a}r}, Z. and {Buzasi}, D.~L. and {Chowdhury}, S. and {De Cat}, P. and {Evans}, J.~A. and {Guo}, Z. and {Guzik}, J.~A. and {Jevtic}, N. and {Lampens}, P. and {Lares Martiz}, M. and {Lovekin}, C. and {Li}, G. and {Mirouh}, G.~M. and {Mkrtichian}, D. and {Monteiro}, M.~J.~P.~F.~G. and {Nemec}, J.~M. and {Ouazzani}, R.-M. and {Pascual-Granado}, J. and {Reese}, D.~R. and {Rieutord}, M. and {Rodon}, J.~R. and {Skarka}, M. and {Sowicka}, P. and {Stateva}, I. and {Szab{\'o}}, R. and {Weiss}, W.~W.},
        title = "{The first view of {\ensuremath{\delta}} Scuti and {\ensuremath{\gamma}} Doradus stars with the TESS mission}",
      journal = {\mnras},
         year = 2019,
        month = dec,
       volume = {490},
       number = {3},
        pages = {4040-4059},
          doi = {10.1093/mnras/stz2787},
archivePrefix = {arXiv},
       eprint = {1909.12018},
 primaryClass = {astro-ph.SR},
       adsurl = {https://ui.adsabs.harvard.edu/abs/2019MNRAS.490.4040A}
}

@article{Antoci2026,
	author = {Antoci, V. and Labadie-Bartz, J. and Święch, M. and Dürfeldt-Pedros, O. and Murphy, S. J. and Kurtz, D. W. and Bedding, T. R. and Handler, G. and Fuller, J. and Ouazzani, R.-M. and Kjeldsen, H. and Fellay, L. and Gade Pedersen, M. and Niemczura, E. and Bowman, D. M. and Deal, M. and Mani, P.},
	title = {Wildly oscillating stars - Unexplained dense ridge-like frequency agglomerations in A- and F-type pulsators},
	DOI= "10.1051/0004-6361/202660809",
	url= "https://doi.org/10.1051/0004-6361/202660809",
	journal = {\aap},
	year = 2026,
	volume = 712,
	pages = "A220",
}

@article{astropy:2013,
        Adsurl = {https://adsabs.harvard.edu/abs/2013A%26A...558A..33A},
        Archiveprefix = {arXiv},
        Author = {{Astropy Collaboration} and {Robitaille}, T.~P. and {Tollerud}, E.~J. and {Greenfield}, P. and {Droettboom}, M. and {Bray}, E. and {Aldcroft}, T. and {Davis}, M. and {Ginsburg}, A. and {Price-Whelan}, A.~M. and {Kerzendorf}, W.~E. and {Conley}, A. and {Crighton}, N. and {Barbary}, K. and {Muna}, D. and {Ferguson}, H. and {Grollier}, F. and {Parikh}, M.~M. and {Nair}, P.~H. and {Unther}, H.~M. and {Deil}, C. and {Woillez}, J. and {Conseil}, S. and {Kramer}, R. and {Turner}, J.~E.~H. and {Singer}, L. and {Fox}, R. and {Weaver}, B.~A. and {Zabalza}, V. and {Edwards}, Z.~I. and {Azalee Bostroem}, K. and {Burke}, D.~J. and {Casey}, A.~R. and {Crawford}, S.~M. and {Dencheva}, N. and {Ely}, J. and {Jenness}, T. and {Labrie}, K. and {Lim}, P.~L. and {Pierfederici}, F. and {Pontzen}, A. and {Ptak}, A. and {Refsdal}, B. and {Servillat}, M. and {Streicher}, O.},
        Doi = {10.1051/0004-6361/201322068},
        Eid = {A33},
        Eprint = {1307.6212},
        Journal = {\aap},
        Month = oct,
        Pages = {A33},
        Primaryclass = {astro-ph.IM},
        Title = {{Astropy: A community Python package for astronomy}},
        Volume = 558,
        Year = 2013}

@ARTICLE{astropy:2018,
               author = {{Astropy Collaboration} and {Price-Whelan}, A.~M. and
                 {Sip{\H{o}}cz}, B.~M. and {G{\"u}nther}, H.~M. and {Lim}, P.~L. and
                 {Crawford}, S.~M. and {Conseil}, S. and {Shupe}, D.~L. and
                 {Craig}, M.~W. and {Dencheva}, N. and {Ginsburg}, A. and {Vand
                erPlas}, J.~T. and {Bradley}, L.~D. and {P{\'e}rez-Su{\'a}rez}, D. and
                 {de Val-Borro}, M. and {Aldcroft}, T.~L. and {Cruz}, K.~L. and
                 {Robitaille}, T.~P. and {Tollerud}, E.~J. and {Ardelean}, C. and
                 {Babej}, T. and {Bach}, Y.~P. and {Bachetti}, M. and {Bakanov}, A.~V. and
                 {Bamford}, S.~P. and {Barentsen}, G. and {Barmby}, P. and
                 {Baumbach}, A. and {Berry}, K.~L. and {Biscani}, F. and {Boquien}, M. and
                 {Bostroem}, K.~A. and {Bouma}, L.~G. and {Brammer}, G.~B. and
                 {Bray}, E.~M. and {Breytenbach}, H. and {Buddelmeijer}, H. and
                 {Burke}, D.~J. and {Calderone}, G. and {Cano Rodr{\'\i}guez}, J.~L. and
                 {Cara}, M. and {Cardoso}, J.~V.~M. and {Cheedella}, S. and {Copin}, Y. and
                 {Corrales}, L. and {Crichton}, D. and {D'Avella}, D. and {Deil}, C. and
                 {Depagne}, {\'E}. and {Dietrich}, J.~P. and {Donath}, A. and
                 {Droettboom}, M. and {Earl}, N. and {Erben}, T. and {Fabbro}, S. and
                 {Ferreira}, L.~A. and {Finethy}, T. and {Fox}, R.~T. and
                 {Garrison}, L.~H. and {Gibbons}, S.~L.~J. and {Goldstein}, D.~A. and
                 {Gommers}, R. and {Greco}, J.~P. and {Greenfield}, P. and
                 {Groener}, A.~M. and {Grollier}, F. and {Hagen}, A. and {Hirst}, P. and
                 {Homeier}, D. and {Horton}, A.~J. and {Hosseinzadeh}, G. and {Hu}, L. and
                 {Hunkeler}, J.~S. and {Ivezi{\'c}}, {\v{Z}}. and {Jain}, A. and
                 {Jenness}, T. and {Kanarek}, G. and {Kendrew}, S. and {Kern}, N.~S. and
                 {Kerzendorf}, W.~E. and {Khvalko}, A. and {King}, J. and {Kirkby}, D. and
                 {Kulkarni}, A.~M. and {Kumar}, A. and {Lee}, A. and {Lenz}, D. and
                 {Littlefair}, S.~P. and {Ma}, Z. and {Macleod}, D.~M. and
                 {Mastropietro}, M. and {McCully}, C. and {Montagnac}, S. and
                 {Morris}, B.~M. and {Mueller}, M. and {Mumford}, S.~J. and {Muna}, D. and
                 {Murphy}, N.~A. and {Nelson}, S. and {Nguyen}, G.~H. and
                 {Ninan}, J.~P. and {N{\"o}the}, M. and {Ogaz}, S. and {Oh}, S. and
                 {Parejko}, J.~K. and {Parley}, N. and {Pascual}, S. and {Patil}, R. and
                 {Patil}, A.~A. and {Plunkett}, A.~L. and {Prochaska}, J.~X. and
                 {Rastogi}, T. and {Reddy Janga}, V. and {Sabater}, J. and
                 {Sakurikar}, P. and {Seifert}, M. and {Sherbert}, L.~E. and
                 {Sherwood-Taylor}, H. and {Shih}, A.~Y. and {Sick}, J. and
                 {Silbiger}, M.~T. and {Singanamalla}, S. and {Singer}, L.~P. and
                 {Sladen}, P.~H. and {Sooley}, K.~A. and {Sornarajah}, S. and
                 {Streicher}, O. and {Teuben}, P. and {Thomas}, S.~W. and
                 {Tremblay}, G.~R. and {Turner}, J.~E.~H. and {Terr{\'o}n}, V. and
                 {van Kerkwijk}, M.~H. and {de la Vega}, A. and {Watkins}, L.~L. and
                 {Weaver}, B.~A. and {Whitmore}, J.~B. and {Woillez}, J. and
                 {Zabalza}, V. and {Astropy Contributors}},
                title = "{The Astropy Project: Building an Open-science Project and Status of the v2.0 Core Package}",
              journal = {\aj},
                 year = 2018,
                month = sep,
               volume = {156},
               number = {3},
                  eid = {123},
                pages = {123},
                  doi = {10.3847/1538-3881/aabc4f},
        archivePrefix = {arXiv},
               eprint = {1801.02634},
         primaryClass = {astro-ph.IM},
               adsurl = {https://ui.adsabs.harvard.edu/abs/2018AJ....156..123A}
        }

@ARTICLE{astropy:2022,
               author = {{Astropy Collaboration} and {Price-Whelan}, Adrian M. and {Lim}, Pey Lian and {Earl}, Nicholas and {Starkman}, Nathaniel and {Bradley}, Larry and {Shupe}, David L. and {Patil}, Aarya A. and {Corrales}, Lia and {Brasseur}, C.~E. and {N{"o}the}, Maximilian and {Donath}, Axel and {Tollerud}, Erik and {Morris}, Brett M. and {Ginsburg}, Adam and {Vaher}, Eero and {Weaver}, Benjamin A. and {Tocknell}, James and {Jamieson}, William and {van Kerkwijk}, Marten H. and {Robitaille}, Thomas P. and {Merry}, Bruce and {Bachetti}, Matteo and {G{"u}nther}, H. Moritz and {Aldcroft}, Thomas L. and {Alvarado-Montes}, Jaime A. and {Archibald}, Anne M. and {B{'o}di}, Attila and {Bapat}, Shreyas and {Barentsen}, Geert and {Baz{'a}n}, Juanjo and {Biswas}, Manish and {Boquien}, M{'e}d{'e}ric and {Burke}, D.~J. and {Cara}, Daria and {Cara}, Mihai and {Conroy}, Kyle E. and {Conseil}, Simon and {Craig}, Matthew W. and {Cross}, Robert M. and {Cruz}, Kelle L. and {D'Eugenio}, Francesco and {Dencheva}, Nadia and {Devillepoix}, Hadrien A.~R. and {Dietrich}, J{"o}rg P. and {Eigenbrot}, Arthur Davis and {Erben}, Thomas and {Ferreira}, Leonardo and {Foreman-Mackey}, Daniel and {Fox}, Ryan and {Freij}, Nabil and {Garg}, Suyog and {Geda}, Robel and {Glattly}, Lauren and {Gondhalekar}, Yash and {Gordon}, Karl D. and {Grant}, David and {Greenfield}, Perry and {Groener}, Austen M. and {Guest}, Steve and {Gurovich}, Sebastian and {Handberg}, Rasmus and {Hart}, Akeem and {Hatfield-Dodds}, Zac and {Homeier}, Derek and {Hosseinzadeh}, Griffin and {Jenness}, Tim and {Jones}, Craig K. and {Joseph}, Prajwel and {Kalmbach}, J. Bryce and {Karamehmetoglu}, Emir and {Ka{l}uszy{'n}ski}, Miko{l}aj and {Kelley}, Michael S.~P. and {Kern}, Nicholas and {Kerzendorf}, Wolfgang E. and {Koch}, Eric W. and {Kulumani}, Shankar and {Lee}, Antony and {Ly}, Chun and {Ma}, Zhiyuan and {MacBride}, Conor and {Maljaars}, Jakob M. and {Muna}, Demitri and {Murphy}, N.~A. and {Norman}, Henrik and {O'Steen}, Richard and {Oman}, Kyle A. and {Pacifici}, Camilla and {Pascual}, Sergio and {Pascual-Granado}, J. and {Patil}, Rohit R. and {Perren}, Gabriel I. and {Pickering}, Timothy E. and {Rastogi}, Tanuj and {Roulston}, Benjamin R. and {Ryan}, Daniel F. and {Rykoff}, Eli S. and {Sabater}, Jose and {Sakurikar}, Parikshit and {Salgado}, Jes{'u}s and {Sanghi}, Aniket and {Saunders}, Nicholas and {Savchenko}, Volodymyr and {Schwardt}, Ludwig and {Seifert-Eckert}, Michael and {Shih}, Albert Y. and {Jain}, Anany Shrey and {Shukla}, Gyanendra and {Sick}, Jonathan and {Simpson}, Chris and {Singanamalla}, Sudheesh and {Singer}, Leo P. and {Singhal}, Jaladh and {Sinha}, Manodeep and {Sip{H{o}}cz}, Brigitta M. and {Spitler}, Lee R. and {Stansby}, David and {Streicher}, Ole and {{{S}}umak}, Jani and {Swinbank}, John D. and {Taranu}, Dan S. and {Tewary}, Nikita and {Tremblay}, Grant R. and {Val-Borro}, Miguel de and {Van Kooten}, Samuel J. and {Vasovi{'c}}, Zlatan and {Verma}, Shresth and {de Miranda Cardoso}, Jos{'e} Vin{'i}cius and {Williams}, Peter K.~G. and {Wilson}, Tom J. and {Winkel}, Benjamin and {Wood-Vasey}, W.~M. and {Xue}, Rui and {Yoachim}, Peter and {Zhang}, Chen and {Zonca}, Andrea and {Astropy Project Contributors}},
                title = "{The Astropy Project: Sustaining and Growing a Community-oriented Open-source Project and the Latest Major Release (v5.0) of the Core Package}",
              journal = {\apj},
                 year = 2022,
                month = aug,
               volume = {935},
               number = {2},
                  eid = {167},
                pages = {167},
                  doi = {10.3847/1538-4357/ac7c74},
        archivePrefix = {arXiv},
               eprint = {2206.14220},
         primaryClass = {astro-ph.IM},
               adsurl = {https://ui.adsabs.harvard.edu/abs/2022ApJ...935..167A}
        }

@article{Auvergne2009corot,
  author  = {Auvergne, M. and Bodin, P. and Boisnard, L. and Buey, J.-T. and Chaintreuil, S. and Epstein, G. and Jouret, M. and Lam-Trong, T. and Levacher, P. and Magnan, A. and Perez, R. and Plasson, P. and Plesseria, J. and Peter, G. and Steller, M. and Tiph{\`e}ne, D. and Baglin, A. and others},
  title   = {{The CoRoT satellite in flight: description and performance}},
  journal = {A\&A},
  volume  = {506},
  number  = {1},
  pages   = {411--424},
  year    = {2009},
  doi     = {10.1051/0004-6361/200810860}
}

@article{Balona2011,
    author = {Balona, L. A. and Guzik, J. A. and Uytterhoeven, K. and Smith, J. C. and Tenenbaum, P. and Twicken, J. D.},
    title = {The Kepler view of γ Doradus stars},
    journal = {\mnras},
    volume = {415},
    number = {4},
    pages = {3531-3538},
    year = {2011},
    month = {08},
    issn = {0035-8711},
    doi = {10.1111/j.1365-2966.2011.18973.x},
    url = {https://doi.org/10.1111/j.1365-2966.2011.18973.x},
    eprint = {https://academic.oup.com/mnras/article-pdf/415/4/3531/4914137/mnras0415-3531.pdf},
}

@article{Balona2014a,
    author = {Balona, L. A.},
    title = {Low frequencies in Kepler δ Scuti stars},
    journal = {\mnras},
    volume = {437},
    number = {2},
    pages = {1476-1484},
    year = {2014},
    month = {01},
    issn = {0035-8711},
    doi = {10.1093/mnras/stt1981},
    url = {https://doi.org/10.1093/mnras/stt1981},
    eprint = {https://academic.oup.com/mnras/article-pdf/437/2/1476/3854982/stt1981.pdf},
}

@article{Balona2015,
    author = {Balona, L. A. and Daszyńska-Daszkiewicz, J. and Pamyatnykh, A. A.},
    title = {Pulsation frequency distribution in δ Scuti stars},
    journal = {\mnras},
    volume = {452},
    number = {3},
    pages = {3073-3084},
    year = {2015},
    month = {09},
    issn = {0035-8711},
    doi = {10.1093/mnras/stv1513},
    url = {https://doi.org/10.1093/mnras/stv1513},
    eprint = {https://academic.oup.com/mnras/article-pdf/452/3/3073/4932209/stv1513.pdf},
}

@article{Balona2016,
    author = {Balona, L. A.},
    title = {Combination frequencies in high-amplitude δ Scuti stars},
    journal = {\mnras},
    volume = {459},
    number = {1},
    pages = {1097-1103},
    year = {2016},
    month = {06},
    issn = {0035-8711},
    doi = {10.1093/mnras/stw671},
    url = {https://doi.org/10.1093/mnras/stw671},
    eprint = {https://academic.oup.com/mnras/article-pdf/459/1/1097/8115783/stw671.pdf},
}

@article{Balona2020,
    author = {Balona, L A and Ozuyar, D},
    title = {Pulsation among TESS A and B stars and the Maia variables},
    journal = {\mnras},
    volume = {493},
    number = {4},
    pages = {5871-5879},
    year = {2020},
    month = {04},
    issn = {0035-8711},
    doi = {10.1093/mnras/staa670},
    url = {https://doi.org/10.1093/mnras/staa670},
    eprint = {https://academic.oup.com/mnras/article-pdf/493/4/5871/32978303/staa670.pdf},
}

@article{Barac2022,
    author = {Barac, Natascha and Bedding, Timothy R and Murphy, Simon J and Hey, Daniel R},
    title = {Revisiting bright δ Scuti stars and their period–luminosity relation with TESS and Gaia DR3},
    journal = {\mnras},
    volume = {516},
    number = {2},
    pages = {2080-2094},
    year = {2022},
    month = {10},
    issn = {0035-8711},
    doi = {10.1093/mnras/stac2132},
    url = {https://doi.org/10.1093/mnras/stac2132},
    eprint = {https://academic.oup.com/mnras/article-pdf/516/2/2080/45787657/stac2132.pdf},
}

@article{Bedding2022,
doi = {10.3847/2515-5172/ac8f94},
url = {https://doi.org/10.3847/2515-5172/ac8f94},
year = {2022},
month = {sep},
publisher = {The American Astronomical Society},
volume = {6},
number = {9},
pages = {202},
author = {Bedding, Timothy R. and Kjeldsen, Hans},
title = {Dealing with Large Gaps in Asteroseismic Time Series},
journal = {Research Notes of the AAS}
}

@article{Bedding2020,
  author  = {Bedding, Timothy R. and Murphy, Simon J. and Hey, Daniel R. and Huber, Daniel and Li, Tanda and Smalley, Barry and Stello, Dennis and White, Timothy R. and Ball, Warrick H. and Chaplin, William J. and Colman, Isabel L. and Fuller, Jim and Gaidos, Eric and Harbeck, Daniel R. and Hermes, J. J. and Holdsworth, Daniel L. and Li, Gang and Li, Yaguang and Mann, Andrew W. and Reese, Daniel R. and Sekaran, Sanjay and Yu, Jie and Antoci, Victoria and Bergmann, Christoph and Brown, Timothy M.},
  title   = {Very regular high-frequency pulsation modes in young intermediate-mass stars},
  journal = {Nature},
  year    = {2020},
  volume  = {581},
  pages   = {147--151},
  doi     = {10.1038/s41586-020-2226-8}
}

@article{Berry2026,
doi = {10.3847/1538-4357/ae9764},
url = {https://doi.org/10.3847/1538-4357/ae9764},
year = {2026},
month = {sep},
publisher = {The American Astronomical Society},
volume = {1009},
number = {2},
pages = {243},
author = {Berry, Ian and Huber, Daniel and Li, Yaguang and Hey, Daniel and Bedding, Timothy R. and Murphy, Simon J.},
title = {The δ Scuti Pulsator Occurrence as a Function of Age, Teff, Rotation, and Metallicity},
journal = {\apjl}
}

@article{Blanco2014iSpec,
	author = {Blanco-Cuaresma, S. and Soubiran, C. and Heiter, U. and Jofr\'e, P.},
	title = {Determining stellar atmospheric parameters and chemical abundances of FGK stars with iSpec},
	DOI= "10.1051/0004-6361/201423945",
	url= "https://doi.org/10.1051/0004-6361/201423945",
	journal = {A\&A},
	year = 2014,
	volume = 569,
	pages = "A111",
	month = "",
}

@article{Blanco2019iSpec,
    author = {Blanco-Cuaresma, Sergi},
    title = {Modern stellar spectroscopy caveats},
    journal = {\mnras},
    volume = {486},
    number = {2},
    pages = {2075-2101},
    year = {2019},
    month = {04},
    issn = {0035-8711},
    doi = {10.1093/mnras/stz549},
    url = {https://doi.org/10.1093/mnras/stz549},
    eprint = {https://academic.oup.com/mnras/article-pdf/486/2/2075/28488167/stz549.pdf},
}

@ARTICLE{Breger1993,
       author = {{Breger}, M. and {Stich}, J. and {Garrido}, R. and {Martin}, B. and {Jiang}, S.-Y. and {Li}, Z.-P. and {Hube}, D.~P. and {Ostermann}, W. and {Paparo}, M. and {Scheck}, M.},
        title = "{Nonradial pulsation of the delta Scuti star BU CANCRI in the Praesepe cluster.}",
      journal = {\aap},
         year = 1993,
        month = apr,
       volume = {271},
        pages = {482-486},
       adsurl = {https://ui.adsabs.harvard.edu/abs/1993A&A...271..482B}
}

@INPROCEEDINGS{Breger2000,
       author = {{Breger}, M.},
        title = "{{\ensuremath{\delta}} Scuti stars (Review)}",
    booktitle = {Delta Scuti and Related Stars},
         year = 2000,
       editor = {{Breger}, Michel and {Montgomery}, Michael},
       series = {Astronomical Society of the Pacific Conference Series},
       volume = {210},
        month = jan,
        pages = {3},
       adsurl = {https://ui.adsabs.harvard.edu/abs/2000ASPC..210....3B}
}

@article{Breger2005,
	author = {Breger, M. and Lenz, P. and Antoci, V. and Guggenberger, E. and Shobbrook, R. R. and Handler, G. and Ngwato, B. and Rodler, F. and Rodriguez, E. and {P. L\'opez de Coca} and Rolland, A. and Costa, V.},
	title = {Detection of 75+ pulsation frequencies  in the ti star FG~Virginis},
	DOI= "10.1051/0004-6361:20042480",
	url= "https://doi.org/10.1051/0004-6361:20042480",
	journal = {A\&A},
	year = 2005,
	volume = 435,
	number = 3,
	pages = "955-965",
}

@article{Breger2011,
    author = {Breger, M. and Balona, L. and Lenz, P. and Hollek, J. K. and Kurtz, D. W. and Catanzaro, G. and Marconi, M. and Pamyatnykh, A. A. and Smalley, B. and Suárez, J. C. and Szabo, R. and Uytterhoeven, K. and Ripepi, V. and Christensen-Dalsgaard, J. and Kjeldsen, H. and Fanelli, M. N. and Ibrahim, K. A. and Uddin, K.},
    title = {Regularities in frequency spacings of δ Scuti stars: the Kepler star KIC 9700322},
    journal = {\mnras},
    volume = {414},
    number = {2},
    pages = {1721-1731},
    year = {2011},
    month = {06},
    issn = {0035-8711},
    doi = {10.1111/j.1365-2966.2011.18508.x},
    url = {https://doi.org/10.1111/j.1365-2966.2011.18508.x},
    eprint = {https://academic.oup.com/mnras/article-pdf/414/2/1721/3016439/mnras0414-1721.pdf},
}

@INPROCEEDINGS{Castelli2003,
       author = {{Castelli}, F. and {Kurucz}, R.~L.},
        title = "{New Grids of ATLAS9 Model Atmospheres}",
    booktitle = {Modelling of Stellar Atmospheres},
         year = 2003,
       editor = {{Piskunov}, N. and {Weiss}, W.~W. and {Gray}, D.~F.},
       series = {IAU Symposium},
       volume = {210},
        month = jan,
        pages = {A20},
          doi = {10.48550/arXiv.astro-ph/0405087},
archivePrefix = {arXiv},
       eprint = {astro-ph/0405087},
 primaryClass = {astro-ph},
       adsurl = {https://ui.adsabs.harvard.edu/abs/2003IAUS..210P.A20C}
}

@ARTICLE{Chavalier1971,
       author = {{Chevalier}, C.},
        title = "{Short-period variables. VIII. Evolution and pulsation of delta Scuti stars.}",
      journal = {\aap},
         year = 1971,
        month = sep,
       volume = {14},
        pages = {24},
       adsurl = {https://ui.adsabs.harvard.edu/abs/1971A\&A....14...24C}
}

@ARTICLE{Collins1963,
       author = {Collins, George W., II},
        title = "{Continuum Emission from a Rapidly Rotating Stellar Atmosphere.}",
      journal = {\apj},
         year = 1963,
        month = nov,
       volume = {138},
        pages = {1134},
          doi = {10.1086/147711}
}

@ARTICLE{Collins1965,
       author = {Collins, George W., II},
        title = "{Continuum Emission from a Rotating Non-Gray Stellar Atmosphere.}",
      journal = {\apj},
         year = 1965,
        month = jul,
       volume = {142},
        pages = {265},
          doi = {10.1086/148282}
}

@ARTICLE{Cox1963,
       author = {{Cox}, John P.},
        title = "{On Second Helium Ionization as a Cause of Pulsational Instability in Stars.}",
      journal = {\apj},
         year = 1963,
        month = aug,
       volume = {138},
        pages = {487},
          doi = {10.1086/147661},
       adsurl = {https://ui.adsabs.harvard.edu/abs/1963ApJ...138..487C}
}

@article{Cunha2019,
    author = {Cunha, M S and Antoci, V and Holdsworth, D L and Kurtz, D W and Balona, L A and Bognár, Zs and Bowman, D M and Guo, Z and Kołaczek-Szymański, P A and Lares-Martiz, M and Paunzen, E and Skarka, M and Smalley, B and Sódor, Á and Kochukhov, O and Pepper, J and Richey-Yowell, T and Ricker, G R and Seager, S and Buzasi, D L and Fox-Machado, L and Hasanzadeh, A and Niemczura, E and Quitral-Manosalva, P and Monteiro, M J P F G and Stateva, I and De Cat, P and García Hernández, A and Ghasemi, H and Handler, G and Hey, D and Matthews, J M and Nemec, J M and Pascual-Granado, J and Safari, H and Suárez, J C and Szabó, R and Tkachenko, A and Weiss, W W},
    title = {Rotation and pulsation in Ap stars: first light results from TESS sectors 1 and 2},
    journal = {\mnras},
    volume = {487},
    number = {3},
    pages = {3523-3549},
    year = {2019},
    month = {06},
    issn = {0035-8711},
    doi = {10.1093/mnras/stz1332},
    url = {https://doi.org/10.1093/mnras/stz1332},
    eprint = {https://academic.oup.com/mnras/article-pdf/487/3/3523/28850735/stz1332.pdf},
}

@article{DaszynskaDaszkiewicz2003,
	author = {Daszyńska-Daszkiewicz, J. and Dziembowski, W. A. and Pamyatnykh, A. A.},
	title = {Constraints on stellar convection from multi-colour photometry of δ Scuti stars},
	DOI= "10.1051/0004-6361:20030947",
	url= "https://doi.org/10.1051/0004-6361:20030947",
	journal = {\aap},
	year = 2003,
	volume = 407,
	number = 3,
	pages = "999-1006",
}

@ARTICLE{Danziger1972,
       author = {{Danziger}, I.~J. and {Faber}, S.~M.},
        title = "{Rotation of evolving A and F stars.}",
      journal = {\aap},
         year = 1972,
        month = may,
       volume = {18},
        pages = {428},
       adsurl = {https://ui.adsabs.harvard.edu/abs/1972A&A....18..428D}
}

@article{Dholakia2025,
    author = {Dholakia, Shishir and Murphy, Simon J and Huang, Chelsea X and Venner, Alexander and Wright, Duncan},
    title = {A catalogue of binary stars from phase modulation in the first four years of TESS mission photometry},
    journal = {\mnras},
    volume = {536},
    number = {3},
    pages = {2313-2323},
    year = {2025},
    month = {01},
    issn = {0035-8711},
    doi = {10.1093/mnras/stae2551},
    url = {https://doi.org/10.1093/mnras/stae2551},
    eprint = {https://academic.oup.com/mnras/article-pdf/536/3/2313/60651867/stae2551.pdf},
}

@article{Diaz2011,
	author = {D\'{\i}az, C. G. and Gonz\'alez, J. F. and Levato, H. and Grosso, M.},
	title = {Accurate stellar rotational velocities using the Fourier  transform of the cross correlation maximum},
	DOI= "10.1051/0004-6361/201016386",
	url= "https://doi.org/10.1051/0004-6361/201016386",
	journal = {\aap},
	year = 2011,
	volume = 531,
	pages = "A143",
	month = "",
}

@article{Dupret2005,
	author = {Dupret, M.-A. and Grigahcène, A. and Garrido, R. and Gabriel, M. and Scuflaire, R.},
	title = {Convection-pulsation coupling - II. Excitation and stabilization mechanisms in 
δ Sct and γ Dor stars},
	DOI= "10.1051/0004-6361:20041817",
	url= "https://doi.org/10.1051/0004-6361:20041817",
	journal = {\aap},
	year = 2005,
	volume = 435,
	number = 3,
	pages = "927-939",
}

@article{SimonDiaz2014,
	author = {Simón-Díaz, S. and Herrero, A.},
	title = {The IACOB project⋆⋆⋆ - I. Rotational velocities in northern Galactic O- and early B-type stars revisited. The impact of other sources of line-broadening},
	DOI= "10.1051/0004-6361/201322758",
	url= "https://doi.org/10.1051/0004-6361/201322758",
	journal = {\aap},
	year = 2014,
	volume = 562,
	pages = "A135",
	month = "",
}

@ARTICLE{Duchene2013,
       author = {{Duch{\^e}ne}, Gaspard and {Kraus}, Adam},
        title = "{Stellar Multiplicity}",
      journal = {\araa},
         year = 2013,
        month = aug,
       volume = {51},
       number = {1},
        pages = {269-310},
          doi = {10.1146/annurev-astro-081710-102602},
archivePrefix = {arXiv},
       eprint = {1303.3028},
 primaryClass = {astro-ph.SR},
       adsurl = {https://ui.adsabs.harvard.edu/abs/2013ARA&A..51..269D}
}

@article{Forteza2018,
	author = {Barceló Forteza, S. and Roca Cortés, T. and García, R. A.},
	title = {The envelope of the power spectra of over a thousand δ Scuti stars - The T̅eff − νmax
 scaling relation★},
	DOI= "10.1051/0004-6361/201731803",
	url= "https://doi.org/10.1051/0004-6361/201731803",
	journal = {\aap},
	year = 2018,
	volume = 614,
	pages = "A46",
}

@article{Forteza2020,
	author = {Barceló Forteza, S. and Moya, A. and Barrado, D. and Solano, E. and Martín-Ruiz, S. and Suárez, J. C. and García Hernández, A.},
	title = {Unveiling the power spectra of δ Scuti stars with TESS⋆ - The temperature, gravity, and frequency scaling relation},
	DOI= "10.1051/0004-6361/201937262",
	url= "https://doi.org/10.1051/0004-6361/201937262",
	journal = {\aap},
	year = 2020,
	volume = 638,
	pages = "A59",
}

@article{Forteza2024,
    author = {Barceló Forteza, S and Pascual-Granado, J and Suárez, J C and García Hernández, A and Mirouh, G M and Lares-Martiz, M},
    title = {Characterizing the features of the low-amplitude peaks in delta Scuti stars with TESS},
    journal = {\mnras},
    volume = {535},
    number = {3},
    pages = {2189-2209},
    year = {2024},
    month = {12},
    issn = {0035-8711},
    doi = {10.1093/mnras/stae2439},
    url = {https://doi.org/10.1093/mnras/stae2439},
    eprint = {https://academic.oup.com/mnras/article-pdf/535/3/2189/60685907/stae2439.pdf},
}

@article{Fremat2005,
	author = {Frémat, Y. and Zorec, J. and Hubert, A.-M. and Floquet, M.},
	title = {Effects of gravitational darkening on the determination   of fundamental
parameters in fast-rotating B-type stars
},
	DOI= "10.1051/0004-6361:20042229",
	url= "https://doi.org/10.1051/0004-6361:20042229",
	journal = {\aap},
	year = 2005,
	volume = 440,
	number = 1,
	pages = "305-320",
}

@article{Fremat2023,
	author = {Fr{\'e}mat, Y. and Royer, F. and Marchal, O. and Blomme, R. and Sartoretti, P. and Guerrier, A. and Panuzzo, P. and Katz, D. and Seabroke, G. M. and Th{\'e}venin, F. and Cropper, M. and Benson, K. and Damerdji, Y. and Haigron, R. and Lobel, A. and Smith, M. and Baker, S. G. and Chemin, L. and David, M. and Dolding, C. and Gosset, E. and Jan{\ss}en, K. and Jasniewicz, G. and Plum, G. and Samaras, N. and Snaith, O. and Soubiran, C. and Vanel, O. and Zorec, J. and Zwitter, T. and Brouillet, N. and Caffau, E. and Crifo, F. and Fabre, C. and Fragkoudi, F. and Huckle, H. E. and Lasne, Y. and Leclerc, N. and Mastrobuono-Battisti, A. and {Jean-Antoine Piccolo}, A. and Viala, Y.},
	title = {{Gaia Data Release 3 - Properties of the line-broadening parameter derived with the Radial Velocity Spectrometer (RVS)}},
	DOI = "10.1051/0004-6361/202243809",
	url = "https://doi.org/10.1051/0004-6361/202243809",
	journal = {\aap},
	year = 2023,
	volume = 674,
	pages = "A8",
}

@article{Gatuam,
    author = {Gautam, Anuj and Murphy, Simon J and Bedding, Timothy R},
    title = {Modelling δ Scuti pulsations: a new grid of p, g, and f modes across pre-main-sequence to post-main-sequence evolution},
    journal = {\mnras},
    volume = {545},
    number = {4},
    pages = {staf2001},
    year = {2026},
    month = {02},
    issn = {0035-8711},
    doi = {10.1093/mnras/staf2001},
    url = {https://doi.org/10.1093/mnras/staf2001},
    eprint = {https://academic.oup.com/mnras/article-pdf/545/4/staf2001/66376218/staf2001.pdf},
}

@article{GaiaDR2,
	author = {{Gaia Collaboration} and Brown, A. G. A. and Vallenari, A. and Prusti, T. and de Bruijne, J. H. J. and Babusiaux, C. and Bailer-Jones, C. A. L. and Biermann, M. and Evans, D. W. and Eyer, L. and Jansen, F. and Jordi, C. and Klioner, S. A. and Lammers, U. and Lindegren, L. and Luri, X. and Mignard, F. and Panem, C. and Pourbaix, D. and Randich, S. and Sartoretti, P. and Siddiqui, H. I. and Soubiran, C. and van Leeuwen, F. and Walton, N. A. and Arenou, F. and Bastian, U. and Cropper, M. and Drimmel, R. and Katz, D. and Lattanzi, M. G. and Bakker, J. and Cacciari, C. and Casta\~neda, J. and Chaoul, L. and Cheek, N. and De Angeli, F. and Fabricius, C. and Guerra, R. and Holl, B. and Masana, E. and Messineo, R. and Mowlavi, N. and Nienartowicz, K. and Panuzzo, P. and Portell, J. and Riello, M. and Seabroke, G. M. and Tanga, P. and Th\'evenin, F. and Gracia-Abril, G. and Comoretto, G. and Garcia-Reinaldos, M. and Teyssier, D. and Altmann, M. and Andrae, R. and Audard, M. and Bellas-Velidis, I. and Benson, K. and Berthier, J. and Blomme, R. and Burgess, P. and Busso, G. and Carry, B. and Cellino, A. and Clementini, G. and Clotet, M. and Creevey, O. and Davidson, M. and De Ridder, J. and Delchambre, L. and Dell\'{}Oro, A. and Ducourant, C. and Fern\'andez-Hern\'andez, J. and Fouesneau, M. and Fr\'emat, Y. and Galluccio, L. and Garc\'{\i}a-Torres, M. and Gonz\'alez-N\'u\~nez, J. and Gonz\'alez-Vidal, J. J. and Gosset, E. and Guy, L. P. and Halbwachs, J.-L. and Hambly, N. C. and Harrison, D. L. and Hern\'andez, J. and Hestroffer, D. and Hodgkin, S. T. and Hutton, A. and Jasniewicz, G. and Jean-Antoine-Piccolo, A. and Jordan, S. and Korn, A. J. and Krone-Martins, A. and Lanzafame, A. C. and Lebzelter, T. and L\"offler, W. and Manteiga, M. and Marrese, P. M. and Mart\'{\i}n-Fleitas, J. M. and Moitinho, A. and Mora, A. and Muinonen, K. and Osinde, J. and Pancino, E. and Pauwels, T. and Petit, J.-M. and Recio-Blanco, A. and Richards, P. J. and Rimoldini, L. and Robin, A. C. and Sarro, L. M. and Siopis, C. and Smith, M. and Sozzetti, A. and S\"uveges, M. and Torra, J. and van Reeven, W. and Abbas, U. and Abreu Aramburu, A. and Accart, S. and Aerts, C. and Altavilla, G. and \'Alvarez, M. A. and Alvarez, R. and Alves, J. and Anderson, R. I. and Andrei, A. H. and Anglada Varela, E. and Antiche, E. and Antoja, T. and Arcay, B. and Astraatmadja, T. L. and Bach, N. and Baker, S. G. and Balaguer-N\'u\~nez, L. and Balm, P. and Barache, C. and Barata, C. and Barbato, D. and Barblan, F. and Barklem, P. S. and Barrado, D. and Barros, M. and Barstow, M. A. and Bartholom\'e Mu\~noz, S. and Bassilana, J.-L. and Becciani, U. and Bellazzini, M. and Berihuete, A. and Bertone, S. and Bianchi, L. and Bienaym\'e, O. and Blanco-Cuaresma, S. and Boch, T. and Boeche, C. and Bombrun, A. and Borrachero, R. and Bossini, D. and Bouquillon, S. and Bourda, G. and Bragaglia, A. and Bramante, L. and Breddels, M. A. and Bressan, A. and Brouillet, N. and Br\"usemeister, T. and Brugaletta, E. and Bucciarelli, B. and Burlacu, A. and Busonero, D. and Butkevich, A. G. and Buzzi, R. and Caffau, E. and Cancelliere, R. and Cannizzaro, G. and Cantat-Gaudin, T. and Carballo, R. and Carlucci, T. and Carrasco, J. M. and Casamiquela, L. and Castellani, M. and Castro-Ginard, A. and Charlot, P. and Chemin, L. and Chiavassa, A. and Cocozza, G. and Costigan, G. and Cowell, S. and Crifo, F. and Crosta, M. and Crowley, C. and Cuypers+, J. and Dafonte, C. and Damerdji, Y. and Dapergolas, A. and David, P. and David, M. and de Laverny, P. and De Luise, F. and De March, R. and de Martino, D. and de Souza, R. and de Torres, A. and Debosscher, J. and del Pozo, E. and Delbo, M. and Delgado, A. and Delgado, H. E. and Di Matteo, P. and Diakite, S. and Diener, C. and Distefano, E. and Dolding, C. and Drazinos, P. and Dur\'an, J. and Edvardsson, B. and Enke, H. and Eriksson, K. and Esquej, P. and Eynard Bontemps, G. and Fabre, C. and Fabrizio, M. and Faigler, S. and Falc\~ao, A. J. and Farr\`as Casas, M. and Federici, L. and Fedorets, G. and Fernique, P. and Figueras, F. and Filippi, F. and Findeisen, K. and Fonti, A. and Fraile, E. and Fraser, M. and Fr\'ezouls, B. and Gai, M. and Galleti, S. and Garabato, D. and Garc\'{\i}a-Sedano, F. and Garofalo, A. and Garralda, N. and Gavel, A. and Gavras, P. and Gerssen, J. and Geyer, R. and Giacobbe, P. and Gilmore, G. and Girona, S. and Giuffrida, G. and Glass, F. and Gomes, M. and Granvik, M. and Gueguen, A. and Guerrier, A. and Guiraud, J. and Guti\'errez-S\'anchez, R. and Haigron, R. and Hatzidimitriou, D. and Hauser, M. and Haywood, M. and Heiter, U. and Helmi, A. and Heu, J. and Hilger, T. and Hobbs, D. and Hofmann, W. and Holland, G. and Huckle, H. E. and Hypki, A. and Icardi, V. and Jan\ss{}en, K. and Jevardat de Fombelle, G. and Jonker, P. G. and Juh\'asz, \'A. L. and Julbe, F. and Karampelas, A. and Kewley, A. and Klar, J. and Kochoska, A. and Kohley, R. and Kolenberg, K. and Kontizas, M. and Kontizas, E. and Koposov, S. E. and Kordopatis, G. and Kostrzewa-Rutkowska, Z. and Koubsky, P. and Lambert, S. and Lanza, A. F. and Lasne, Y. and Lavigne, J.-B. and Le Fustec, Y. and Le Poncin-Lafitte, C. and Lebreton, Y. and Leccia, S. and Leclerc, N. and Lecoeur-Taibi, I. and Lenhardt, H. and Leroux, F. and Liao, S. and Licata, E. and Lindstr\o{}m, H. E. P. and Lister, T. A. and Livanou, E. and Lobel, A. and L\'opez, M. and Managau, S. and Mann, R. G. and Mantelet, G. and Marchal, O. and Marchant, J. M. and Marconi, M. and Marinoni, S. and Marschalk\'o, G. and Marshall, D. J. and Martino, M. and Marton, G. and Mary, N. and Massari, D. and Matijevic, G. and Mazeh, T. and McMillan, P. J. and Messina, S. and Michalik, D. and Millar, N. R. and Molina, D. and Molinaro, R. and Moln\'ar, L. and Montegriffo, P. and Mor, R. and Morbidelli, R. and Morel, T. and Morris, D. and Mulone, A. F. and Muraveva, T. and Musella, I. and Nelemans, G. and Nicastro, L. and Noval, L. and O\'{}Mullane, W. and Ord\'enovic, C. and Ord\'o\~nez-Blanco, D. and Osborne, P. and Pagani, C. and Pagano, I. and Pailler, F. and Palacin, H. and Palaversa, L. and Panahi, A. and Pawlak, M. and Piersimoni, A. M. and Pineau, F.-X. and Plachy, E. and Plum, G. and Poggio, E. and Poujoulet, E. and Prsa, A. and Pulone, L. and Racero, E. and Ragaini, S. and Rambaux, N. and Ramos-Lerate, M. and Regibo, S. and Reyl\'e, C. and Riclet, F. and Ripepi, V. and Riva, A. and Rivard, A. and Rixon, G. and Roegiers, T. and Roelens, M. and Romero-G\'omez, M. and Rowell, N. and Royer, F. and Ruiz-Dern, L. and Sadowski, G. and Sagrist\`a Sell\'es, T. and Sahlmann, J. and Salgado, J. and Salguero, E. and Sanna, N. and Santana-Ros, T. and Sarasso, M. and Savietto, H. and Schultheis, M. and Sciacca, E. and Segol, M. and Segovia, J. C. and S\'egransan, D. and Shih, I-C. and Siltala, L. and Silva, A. F. and Smart, R. L. and Smith, K. W. and Solano, E. and Solitro, F. and Sordo, R. and Soria Nieto, S. and Souchay, J. and Spagna, A. and Spoto, F. and Stampa, U. and Steele, I. A. and Steidelm\"uller, H. and Stephenson, C. A. and Stoev, H. and Suess, F. F. and Surdej, J. and Szabados, L. and Szegedi-Elek, E. and Tapiador, D. and Taris, F. and Tauran, G. and Taylor, M. B. and Teixeira, R. and Terrett, D. and Teyssandier, P. and Thuillot, W. and Titarenko, A. and Torra Clotet, F. and Turon, C. and Ulla, A. and Utrilla, E. and Uzzi, S. and Vaillant, M. and Valentini, G. and Valette, V. and van Elteren, A. and Van Hemelryck, E. and van Leeuwen, M. and Vaschetto, M. and Vecchiato, A. and Veljanoski, J. and Viala, Y. and Vicente, D. and Vogt, S. and von Essen, C. and Voss, H. and Votruba, V. and Voutsinas, S. and Walmsley, G. and Weiler, M. and Wertz, O. and Wevers, T. and Wyrzykowski, L. and Yoldas, A. and Zerjal, M. and Ziaeepour, H. and Zorec, J. and Zschocke, S. and Zucker, S. and Zurbach, C. and Zwitter, T.},
	title = {Gaia Data Release 2 - Summary of the contents and survey properties},
	DOI= "10.1051/0004-6361/201833051",
	url= "https://doi.org/10.1051/0004-6361/201833051",
	journal = {\aap},
	year = 2018,
	volume = 616,
	pages = "A1",
}

@article{GaiaDR3,
	author = {{Gaia Collaboration} and Vallenari, A. and Brown, A. G. A. and Prusti, T. and de Bruijne, J. H. J. and Arenou, F. and Babusiaux, C. and Biermann, M. and Creevey, O. L. and Ducourant, C. and Evans, D. W. and Eyer, L. and Guerra, R. and Hutton, A. and Jordi, C. and Klioner, S. A. and Lammers, U. L. and Lindegren, L. and Luri, X. and Mignard, F. and Panem, C. and Pourbaix, D. and Randich, S. and Sartoretti, P. and Soubiran, C. and Tanga, P. and Walton, N. A. and Bailer-Jones, C. A. L. and Bastian, U. and Drimmel, R. and Jansen, F. and Katz, D. and Lattanzi, M. G. and van Leeuwen, F. and Bakker, J. and Cacciari, C. and Casta\~neda, J. and De Angeli, F. and Fabricius, C. and Fouesneau, M. and Fr\'emat, Y. and Galluccio, L. and Guerrier, A. and Heiter, U. and Masana, E. and Messineo, R. and Mowlavi, N. and Nicolas, C. and Nienartowicz, K. and Pailler, F. and Panuzzo, P. and Riclet, F. and Roux, W. and Seabroke, G. M. and Sordo, R. and Th\'evenin, F. and Gracia-Abril, G. and Portell, J. and Teyssier, D. and Altmann, M. and Andrae, R. and Audard, M. and Bellas-Velidis, I. and Benson, K. and Berthier, J. and Blomme, R. and Burgess, P. W. and Busonero, D. and Busso, G. and C\'anovas, H. and Carry, B. and Cellino, A. and Cheek, N. and Clementini, G. and Damerdji, Y. and Davidson, M. and de Teodoro, P. and Nu\~nez Campos, M. and Delchambre, L. and Dell\'{}Oro, A. and Esquej, P. and Fern\'andez-Hern\'andez, J. and Fraile, E. and Garabato, D. and Garc\'{\i}a-Lario, P. and Gosset, E. and Haigron, R. and Halbwachs, J.-L. and Hambly, N. C. and Harrison, D. L. and Hern\'andez, J. and Hestroffer, D. and Hodgkin, S. T. and Holl, B. and Jan\ss{}en, K. and Jevardat de Fombelle, G. and Jordan, S. and Krone-Martins, A. and Lanzafame, A. C. and L\"offler, W. and Marchal, O. and Marrese, P. M. and Moitinho, A. and Muinonen, K. and Osborne, P. and Pancino, E. and Pauwels, T. and Recio-Blanco, A. and Reyl\'e, C. and Riello, M. and Rimoldini, L. and Roegiers, T. and Rybizki, J. and Sarro, L. M. and Siopis, C. and Smith, M. and Sozzetti, A. and Utrilla, E. and van Leeuwen, M. and Abbas, U. and \'Abrah\'am, P. and Abreu Aramburu, A. and Aerts, C. and Aguado, J. J. and Ajaj, M. and Aldea-Montero, F. and Altavilla, G. and \'Alvarez, M. A. and Alves, J. and Anders, F. and Anderson, R. I. and Anglada Varela, E. and Antoja, T. and Baines, D. and Baker, S. G. and Balaguer-N\'u\~nez, L. and Balbinot, E. and Balog, Z. and Barache, C. and Barbato, D. and Barros, M. and Barstow, M. A. and Bartolom\'e, S. and Bassilana, J.-L. and Bauchet, N. and Becciani, U. and Bellazzini, M. and Berihuete, A. and Bernet, M. and Bertone, S. and Bianchi, L. and Binnenfeld, A. and Blanco-Cuaresma, S. and Blazere, A. and Boch, T. and Bombrun, A. and Bossini, D. and Bouquillon, S. and Bragaglia, A. and Bramante, L. and Breedt, E. and Bressan, A. and Brouillet, N. and Brugaletta, E. and Bucciarelli, B. and Burlacu, A. and Butkevich, A. G. and Buzzi, R. and Caffau, E. and Cancelliere, R. and Cantat-Gaudin, T. and Carballo, R. and Carlucci, T. and Carnerero, M. I. and Carrasco, J. M. and Casamiquela, L. and Castellani, M. and Castro-Ginard, A. and Chaoul, L. and Charlot, P. and Chemin, L. and Chiaramida, V. and Chiavassa, A. and Chornay, N. and Comoretto, G. and Contursi, G. and Cooper, W. J. and Cornez, T. and Cowell, S. and Crifo, F. and Cropper, M. and Crosta, M. and Crowley, C. and Dafonte, C. and Dapergolas, A. and David, M. and David, P. and de Laverny, P. and De Luise, F. and De March, R. and De Ridder, J. and de Souza, R. and de Torres, A. and del Peloso, E. F. and del Pozo, E. and Delbo, M. and Delgado, A. and Delisle, J.-B. and Demouchy, C. and Dharmawardena, T. E. and Di Matteo, P. and Diakite, S. and Diener, C. and Distefano, E. and Dolding, C. and Edvardsson, B. and Enke, H. and Fabre, C. and Fabrizio, M. and Faigler, S. and Fedorets, G. and Fernique, P. and Fienga, A. and Figueras, F. and Fournier, Y. and Fouron, C. and Fragkoudi, F. and Gai, M. and Garcia-Gutierrez, A. and Garcia-Reinaldos, M. and Garc\'{\i}a-Torres, M. and Garofalo, A. and Gavel, A. and Gavras, P. and Gerlach, E. and Geyer, R. and Giacobbe, P. and Gilmore, G. and Girona, S. and Giuffrida, G. and Gomel, R. and Gomez, A. and Gonz\'alez-N\'u\~nez, J. and Gonz\'alez-Santamar\'{\i}a, I. and Gonz\'alez-Vidal, J. J. and Granvik, M. and Guillout, P. and Guiraud, J. and Guti\'errez-S\'anchez, R. and Guy, L. P. and Hatzidimitriou, D. and Hauser, M. and Haywood, M. and Helmer, A. and Helmi, A. and Sarmiento, M. H. and Hidalgo, S. L. and Hilger, T. and Hladczuk, N. and Hobbs, D. and Holland, G. and Huckle, H. E. and Jardine, K. and Jasniewicz, G. and Jean-Antoine Piccolo, A. and Jim\'enez-Arranz, \'O. and Jorissen, A. and Juaristi Campillo, J. and Julbe, F. and Karbevska, L. and Kervella, P. and Khanna, S. and Kontizas, M. and Kordopatis, G. and Korn, A. J. and K\'osp\'al, \'A and Kostrzewa-Rutkowska, Z. and Kruszy\'{}nska, K. and Kun, M. and Laizeau, P. and Lambert, S. and Lanza, A. F. and Lasne, Y. and Le Campion, J.-F. and Lebreton, Y. and Lebzelter, T. and Leccia, S. and Leclerc, N. and Lecoeur-Taibi, I. and Liao, S. and Licata, E. L. and Lindstr\o{}m, H. E. P. and Lister, T. A. and Livanou, E. and Lobel, A. and Lorca, A. and Loup, C. and Madrero Pardo, P. and Magdaleno Romeo, A. and Managau, S. and Mann, R. G. and Manteiga, M. and Marchant, J. M. and Marconi, M. and Marcos, J. and Marcos Santos, M. M. S. and Mar\'{\i}n Pina, D. and Marinoni, S. and Marocco, F. and Marshall, D. J. and Martin Polo, L. and Mart\'{\i}n-Fleitas, J. M. and Marton, G. and Mary, N. and Masip, A. and Massari, D. and Mastrobuono-Battisti, A. and Mazeh, T. and McMillan, P. J. and Messina, S. and Michalik, D. and Millar, N. R. and Mints, A. and Molina, D. and Molinaro, R. and Moln\'ar, L. and Monari, G. and Mongui\'o, M. and Montegriffo, P. and Montero, A. and Mor, R. and Mora, A. and Morbidelli, R. and Morel, T. and Morris, D. and Muraveva, T. and Murphy, C. P. and Musella, I. and Nagy, Z. and Noval, L. and Oca\~na, F. and Ogden, A. and Ordenovic, C. and Osinde, J. O. and Pagani, C. and Pagano, I. and Palaversa, L. and Palicio, P. A. and Pallas-Quintela, L. and Panahi, A. and Payne-Wardenaar, S. and Pe\~nalosa Esteller, X. and Penttil\"a, A. and Pichon, B. and Piersimoni, A. M. and Pineau, F.-X. and Plachy, E. and Plum, G. and Poggio, E. and Prsa, A. and Pulone, L. and Racero, E. and Ragaini, S. and Rainer, M. and Raiteri, C. M. and Rambaux, N. and Ramos, P. and Ramos-Lerate, M. and Re Fiorentin, P. and Regibo, S. and Richards, P. J. and Rios Diaz, C. and Ripepi, V. and Riva, A. and Rix, H.-W. and Rixon, G. and Robichon, N. and Robin, A. C. and Robin, C. and Roelens, M. and Rogues, H. R. O. and Rohrbasser, L. and Romero-G\'omez, M. and Rowell, N. and Royer, F. and Ruz Mieres, D. and Rybicki, K. A. and Sadowski, G. and S\'aez N\'u\~nez, A. and Sagrist\`a Sell\'es, A. and Sahlmann, J. and Salguero, E. and Samaras, N. and Sanchez Gimenez, V. and Sanna, N. and Santove\~na, R. and Sarasso, M. and Schultheis, M. and Sciacca, E. and Segol, M. and Segovia, J. C. and S\'egransan, D. and Semeux, D. and Shahaf, S. and Siddiqui, H. I. and Siebert, A. and Siltala, L. and Silvelo, A. and Slezak, E. and Slezak, I. and Smart, R. L. and Snaith, O. N. and Solano, E. and Solitro, F. and Souami, D. and Souchay, J. and Spagna, A. and Spina, L. and Spoto, F. and Steele, I. A. and Steidelm\"uller, H. and Stephenson, C. A. and S\"uveges, M. and Surdej, J. and Szabados, L. and Szegedi-Elek, E. and Taris, F. and Taylor, M. B. and Teixeira, R. and Tolomei, L. and Tonello, N. and Torra, F. and Torra, J. and Torralba Elipe, G. and Trabucchi, M. and Tsounis, A. T. and Turon, C. and Ulla, A. and Unger, N. and Vaillant, M. V. and van Dillen, E. and van Reeven, W. and Vanel, O. and Vecchiato, A. and Viala, Y. and Vicente, D. and Voutsinas, S. and Weiler, M. and Wevers, T. and Wyrzykowski, L. and Yoldas, A. and Yvard, P. and Zhao, H. and Zorec, J. and Zucker, S. and Zwitter, T.},
	title = {Gaia Data Release 3 - Summary of the content and survey properties},
	DOI= "10.1051/0004-6361/202243940",
	url= "https://doi.org/10.1051/0004-6361/202243940",
	journal = {\aap},
	year = 2023,
	volume = 674,
	pages = "A1",
}

@article{Garcia2022,
	author = {Garcia, S. and Van Reeth, T. and De Ridder, J. and Tkachenko, A. and IJspeert, L. and Aerts, C.},
	title = {Detection of period-spacing patterns due to the gravity modes of rotating dwarfs in the TESS southern continuous viewing zone★},
	DOI= "10.1051/0004-6361/202141926",
	url= "https://doi.org/10.1051/0004-6361/202141926",
	journal = {\aap},
	year = 2022,
	volume = 662,
	pages = "A82",
}

@article{Georgy2013,
	author = {Georgy, C. and Ekström, S. and Granada, A. and Meynet, G. and Mowlavi, N. and Eggenberger, P. and Maeder, A.},
	title = {Populations of rotating stars - I. Models from 1.7 to 15 M⊙ at Z
          = 0.014, 0.006, and 0.002 with Ω/Ωcrit between 0
          and 1⋆⋆⋆},
	DOI= "10.1051/0004-6361/201220558",
	url= "https://doi.org/10.1051/0004-6361/201220558",
	journal = {\aap},
	year = 2013,
	volume = 553,
	pages = "A24",
	month = "",
}

@article{Grevesse1998,
  author  = {Grevesse, N. and Sauval, A. J.},
  title   = {Standard Solar Composition},
  journal = {Space Science Reviews},
  year    = {1998},
  volume  = {85},
  pages   = {161--174},
  doi     = {10.1023/A:1005161325181}
}

@article{Gootkin2024,
doi = {10.3847/1538-4357/ad5282},
url = {https://doi.org/10.3847/1538-4357/ad5282},
year = {2024},
month = {sep},
publisher = {The American Astronomical Society},
volume = {972},
number = {2},
pages = {137},
author = {Gootkin, Keyan and Hon, Marc and Huber, Daniel and Hey, Daniel R. and Bedding, Timothy R. and Murphy, Simon J.},
title = {A New Catalog of 100,000 Variable TESS A-F Stars Reveals a Correlation between δ Scuti Pulsator Fraction and Stellar Rotation},
journal = {\apj}
}

@BOOK{Gray1992,
       author = {{Gray}, D.~F.},
        title = "{The observation and analysis of stellar photospheres.}",
         year = 1992,
       volume = {20},
       adsurl = {https://ui.adsabs.harvard.edu/abs/1992oasp.book.....G}
}

@article{Grigahcene2010,
doi = {10.1088/2041-8205/713/2/L192},
url = {https://doi.org/10.1088/2041-8205/713/2/L192},
year = {2010},
month = {mar},
publisher = {The American Astronomical Society},
volume = {713},
number = {2},
pages = {L192},
author = {Grigahcène, A. and Antoci, V. and Balona, L. and Catanzaro, G. and Daszyńska-Daszkiewicz, J. and Guzik, J. A. and Handler, G. and Houdek, G. and Kurtz, D. W. and Marconi, M. and Monteiro, M. J. P. F. G. and Moya, A. and Ripepi, V. and Suárez, J.-C. and Uytterhoeven, K. and Borucki, W. J. and Brown, T. M. and Christensen-Dalsgaard, J. and Gilliland, R. L. and Jenkins, J. M. and Kjeldsen, H. and Koch, D. and Bernabei, S. and Bradley, P. and Breger, M. and Criscienzo, M. Di and Dupret, M.-A. and García, R. A. and Hernández, A. García and Jackiewicz, J. and Kaiser, A. and Lehmann, H. and Martín-Ruiz, S. and Mathias, P. and Molenda-Żakowicz, J. and Nemec, J. M. and Nuspl, J. and Paparó, M. and Roth, M. and Szabó, R. and Suran, M. D. and Ventura, R.},
title = {HYBRID γ DORADUS–δ SCUTI PULSATORS: NEW INSIGHTS INTO THE PHYSICS OF THE OSCILLATIONS FROM KEPLER OBSERVATIONS},
journal = {\apjl}
}

@article{Handler2002,
    author = {Handler, G. and Shobbrook, R. R.},
    title = {On the relationship between the δ Scuti and γ Doradus pulsators},
    journal = {\mnras},
    volume = {333},
    number = {2},
    pages = {251-261},
    year = {2002},
    month = {06},
    issn = {0035-8711},
    doi = {10.1046/j.1365-8711.2002.05401.x},
    url = {https://doi.org/10.1046/j.1365-8711.2002.05401.x},
    eprint = {https://academic.oup.com/mnras/article-pdf/333/2/251/18411307/333-2-251.pdf},
}

@INPROCEEDINGS{HARPS,
       author = {{Pepe}, Francesco and {Mayor}, Michel and {Delabre}, Bernard and {Kohler}, Dominique and {Lacroix}, Daniel and {Queloz}, Didier and {Udry}, Stephane and {Benz}, Willy and {Bertaux}, Jean-Loup and {Sivan}, Jean-Pierre},
        title = "{HARPS: a new high-resolution spectrograph for the search of extrasolar planets}",
    booktitle = {Optical and IR Telescope Instrumentation and Detectors},
         year = 2000,
       editor = {{Iye}, Masanori and {Moorwood}, Alan F.},
       series = {Society of Photo-Optical Instrumentation Engineers (SPIE) Conference Series},
       volume = {4008},
        month = aug,
        pages = {582-592},
          doi = {10.1117/12.395516},
       adsurl = {https://ui.adsabs.harvard.edu/abs/2000SPIE.4008..582P}
}

@ARTICLE{Helmut1995,
       author = {{Abt}, Helmut A. and {Morrell}, Nidia I.},
        title = "{The Relation between Rotational Velocities and Spectral Peculiarities among A-Type Stars}",
      journal = {\apjs},
         year = 1995,
        month = jul,
       volume = {99},
        pages = {135},
          doi = {10.1086/192182},
       adsurl = {https://ui.adsabs.harvard.edu/abs/1995ApJS...99..135A}
}

@article{Henriksen2023,
    author = {Henriksen, Andreea I and Antoci, Victoria and Saio, Hideyuki and Grundahl, Frank and Kjeldsen, Hans and Van Reeth, Timothy and Bowman, Dominic M and Pápics, Péter I and De Cat, Peter and Krüger, Joachim and Andersen, M Fredslund and Pallé, P L},
    title = {Unresolved Rossby and gravity modes in 214 A and F stars showing rotational modulation},
    journal = {\mnras},
    volume = {524},
    number = {3},
    pages = {4196-4211},
    year = {2023},
    month = {09},
    issn = {0035-8711},
    doi = {10.1093/mnras/stad1971},
    url = {https://doi.org/10.1093/mnras/stad1971},
    eprint = {https://academic.oup.com/mnras/article-pdf/524/3/4196/50977016/stad1971.pdf},
}

@ARTICLE{Kabath2020,
       author = {{Kab{\'a}th}, P. and {Skarka}, M. and {Sabotta}, S. and {Guenther}, E. and {Jones}, D. and {Klocov{\'a}}, T. and {{\v{S}}ubjak}, J. and {{\v{Z}}{\'a}k}, Ji{\v{r}}{\'\i} and {{\v{S}}pokov{\'a}}, M. and {Bla{\v{z}}ek}, M. and {Dvo{\v{r}}{\'a}kov{\'a}}, J. and {Dupkala}, D. and {Fuchs}, J. and {Hatzes}, A. and {Kortusov{\'a}}, E. and {Novotn{\'y}}, R. and {Pl{\'a}valov{\'a}}, E. and {{\v{R}}ezba}, L. and {Sloup}, J. and {{\v{S}}koda}, P. and {{\v{S}}lechta}, M.},
        title = "{Ond{\v{r}}ejov Echelle Spectrograph, Ground Based Support Facility for Exoplanet Missions}",
      journal = {\pasp},
         year = 2020,
        month = mar,
       volume = {132},
       number = {1009},
          eid = {035002},
        pages = {035002},
          doi = {10.1088/1538-3873/ab6752},
archivePrefix = {arXiv},
       eprint = {2001.01001},
 primaryClass = {astro-ph.IM},
       adsurl = {https://ui.adsabs.harvard.edu/abs/2020PASP..132c5002K}
}

@article{Kallinger2010,
	author = {Kallinger, T. and Weiss, W. W. and Barban, C. and Baudin, F. and Cameron, C. and Carrier, F. and De Ridder, J. and Goupil, M.-J. and Gruberbauer, M. and Hatzes, A. and Hekker, S. and Samadi, R. and Deleuil, M.},
	title = {Oscillating red giants in the CoRoT exofield: asteroseismic mass and radius determination*},
	DOI= "10.1051/0004-6361/200811437",
	url= "https://doi.org/10.1051/0004-6361/200811437",
	journal = {\aap},
	year = 2010,
	volume = 509,
	pages = "A77",
	month = "",
}

@ARTICLE{Ledoux1951,
       author = {Ledoux, P.},
        title = "{The Nonradial Oscillations of Gaseous Stars and the Problem of Beta Canis Majoris.}",
      journal = {\apj},
         year = 1951,
        month = nov,
       volume = {114},
        pages = {373},
          doi = {10.1086/145477},
       adsurl = {https://ui.adsabs.harvard.edu/abs/1951ApJ...114..373L}
}

@article{Lenz2008,
	author = {Lenz, P. and Pamyatnykh, A. A. and Breger, M. and Antoci, V.},
	title = {An asteroseismic study of the δ Scuti star 44 Tauri},
	DOI= "10.1051/0004-6361:20078376",
	url= "https://doi.org/10.1051/0004-6361:20078376",
	journal = {\aap},
	year = 2008,
	volume = 478,
	number = 3,
	pages = "855-863",
}

@article{Lignieres2006,
	author = {Lignières, F. and Rieutord, M. and Reese, D.},
	title = {Acoustic oscillations of rapidly rotating polytropic stars - I. Effects of the centrifugal distortion},
	DOI= "10.1051/0004-6361:20065015",
	url= "https://doi.org/10.1051/0004-6361:20065015",
	journal = {\aap},
	year = 2006,
	volume = 455,
	number = 2,
	pages = "607-620",
}

@ARTICLE{Kabath2026,
       author = {{Kab{\'a}th}, P. and {Skarka}, M. and {Hatzes}, A. and {Guenther}, E. and {Vanzi}, L. and {Brahm}, R. and {Jan{\'\i}k}, J. and {Pintr}, P. and {Gajdo{\v{s}}}, P. and {Lipt{\'a}k}, J. and {{\v{Z}}{\'a}k}, J. and {Boffin}, H.~M.~J. and {Antonucci}, L. and {Avila}, G. and {Balk{\'o}ov{\'a}}, Z. and {Ball}, M.~E. and {Flores}, M. and {Fuentes}, A. and {Fuchs}, J. and {Greimel}, R. and {Gajardo}, A. and {Ivanov}, V.~D. and {K{\"o}hler}, J. and {Leitzinger}, M. and {Ludwig}, M.~R. and {Morav{\v{c}}{\'\i}k}, T. and {Ne{\v{c}}{\'a}sek}, J. and {Neubert}, R.~J.~U. and {Odert}, P. and {Olgu{\'\i}n}, G. and {Pinto}, M. Tala and {Roth}, M. and {{\v{R}}ezba}, L. and {Schaffenroth}, V. and {Sigwarth}, M. and {Srba}, J. and {Su{\'a}rez}, A. and {{\v{S}}koda}, P. and {{\v{S}}ubjak}, J. and {V{\'a}clav{\'\i}k}, J. and {Vesel{\'y}}, M. and {Vesel{\'y}}, R. and {V{\'\i}tkov{\'a}}, M. and {Winkler}, J.~U. and {Zummer}, M. and {{\v{Z}}{\v{d}}{\'a}rsk{\'a}}, E.},
        title = "{PLATOSpec: a precise spectrograph in support of space missions}",
      journal = {\mnras},
         year = 2026,
        month = jan,
       volume = {545},
       number = {1},
          eid = {staf1972},
        pages = {staf1972},
          doi = {10.1093/mnras/staf1972},
archivePrefix = {arXiv},
       eprint = {2510.11961},
 primaryClass = {astro-ph.IM},
       adsurl = {https://ui.adsabs.harvard.edu/abs/2026MNRAS.545f1972K}
}

@article{Kahraman2017,
    author = {Kahraman Aliçavuş, F. and Niemczura, E. and Polińska, M. and Hełminiak, K. G. and Lampens, P. and Molenda-Żakowicz, J. and Ukita, N. and Kambe, E.},
    title = {High-resolution spectroscopy and abundance analysis of δ Scuti stars near the γ Doradus instability strip},
    journal = {\mnras},
    volume = {470},
    number = {4},
    pages = {4408-4420},
    year = {2017},
    month = {10},
    issn = {0035-8711},
    doi = {10.1093/mnras/stx1498},
    url = {https://doi.org/10.1093/mnras/stx1498},
    eprint = {https://academic.oup.com/mnras/article-pdf/470/4/4408/19179358/stx1498.pdf},
}

@ARTICLE{Koubsky2004,
       author = {{Koubsk{\'y}}, P. and {Mayer}, P. and {{\v{C}}{\'a}p}, J. and {{\v{Z}}{\v{d}}{\'a}rsk{\'y}}, F. and {Zeman}, J. and {P{\'\i}na}, L. and {Melich}, Z.},
        title = "{Ond{\v{r}}ejov Echelle Spectrograph - OES}",
      journal = {Publ. Astron. Inst. Czech. Acad. Sci.},
         year = 2004,
        month = jan,
       volume = {92},
        pages = {37-43},
       adsurl = {https://ui.adsabs.harvard.edu/abs/2004PAICz..92...37K}
}

@article{Kurtz2014,
    author = {Kurtz, Donald W. and Saio, Hideyuki and Takata, Masao and Shibahashi, Hiromoto and Murphy, Simon J. and Sekii, Takashi},
    title = {Asteroseismic measurement of surface-to-core rotation in a main-sequence A star, KIC 11145123},
    journal = {\mnras},
    volume = {444},
    number = {1},
    pages = {102-116},
    year = {2014},
    month = {10},
    issn = {0035-8711},
    doi = {10.1093/mnras/stu1329},
    url = {https://doi.org/10.1093/mnras/stu1329},
    eprint = {https://academic.oup.com/mnras/article-pdf/444/1/102/18501410/stu1329.pdf},
}

@article{Kurtz2022,
   author = "Kurtz, Donald W.",
   title = "Asteroseismology Across the Hertzsprung–Russell Diagram", 
   journal= "Annual Review of Astronomy and Astrophysics",
   year = "2022",
   volume = "60",
   number = "Volume 60, 2022",
   pages = "31-71",
   doi = "https://doi.org/10.1146/annurev-astro-052920-094232",
   url = "https://www.annualreviews.org/content/journals/10.1146/annurev-astro-052920-094232",
   publisher = "Annual Reviews",
   issn = "1545-4282",
   type = "Journal Article",
  }

@ARTICLE{LenzBreger2005,
       author = {{Lenz}, P. and {Breger}, M.},
        title = "{Period04 User Guide}",
      journal = {Communications in Asteroseismology},
         year = 2005,
        month = jun,
       volume = {146},
        pages = {53-136},
          doi = {10.1553/cia146s53},
       adsurl = {https://ui.adsabs.harvard.edu/abs/2005CoAst.146...53L}
}

@article{Liakos2017,
    author = {Liakos, Alexios and Niarchos, Panagiotis},
    title = {Catalogue and properties of δ Scuti stars in binaries},
    journal = {\mnras},
    volume = {465},
    number = {1},
    pages = {1181-1200},
    year = {2017},
    month = {02},
    issn = {0035-8711},
    doi = {10.1093/mnras/stw2756},
    url = {https://doi.org/10.1093/mnras/stw2756},
    eprint = {https://academic.oup.com/mnras/article-pdf/465/1/1181/8593641/stw2756.pdf},
}

@MISC{Lightkurve2018,
   author = {{Lightkurve Collaboration} and {Cardoso}, J.~V.~d.~M. and
             {Hedges}, C. and {Gully-Santiago}, M. and {Saunders}, N. and
             {Cody}, A.~M. and {Barclay}, T. and {Hall}, O. and
             {Sagear}, S. and {Turtelboom}, E. and {Zhang}, J. and
             {Tzanidakis}, A. and {Mighell}, K. and {Coughlin}, J. and
             {Bell}, K. and {Berta-Thompson}, Z. and {Williams}, P. and
             {Dotson}, J. and {Barentsen}, G.},
    title = "{Lightkurve: Kepler and TESS time series analysis in Python}",
howpublished = {Astrophysics Source Code Library},
     year = 2018,
    month = dec,
archivePrefix = "ascl",
   eprint = {1812.013},
   adsurl = {http://adsabs.harvard.edu/abs/2018ascl.soft12013L},
}

@ARTICLE{Lomb1976,
       author = {{Lomb}, N.~R.},
        title = "{Least-Squares Frequency Analysis of Unequally Spaced Data}",
      journal = {\apss},
         year = 1976,
        month = feb,
       volume = {39},
       number = {2},
        pages = {447-462},
          doi = {10.1007/BF00648343},
       adsurl = {https://ui.adsabs.harvard.edu/abs/1976Ap&SS..39..447L}
}

@ARTICLE{Maeder2000,
       author = {Maeder, A. and Meynet, G.},
        title = "{Stellar evolution with rotation. VI. The Eddington and Omega -limits, the rotational mass loss for OB and LBV stars}",
      journal = {\aap},
         year = 2000,
        month = sep,
       volume = {361},
        pages = {159-166},
          doi = {10.48550/arXiv.astro-ph/0006405},
archivePrefix = {arXiv},
       eprint = {astro-ph/0006405},
 primaryClass = {astro-ph},
       adsurl = {https://ui.adsabs.harvard.edu/abs/2000A&A...361..159M}
}

@article{Mani2025,
    author = {Mani, Prasad and Bedding, Timothy R and Bernizzoni, Mara and Murphy, Simon J and Hey, Daniel},
    title = {Characterizing bright δ Scuti pulsators using TESS light curves},
    journal = {\mnras},
    volume = {542},
    number = {4},
    pages = {2866-2876},
    year = {2025},
    month = {10},
    issn = {0035-8711},
    doi = {10.1093/mnras/staf1400},
    url = {https://doi.org/10.1093/mnras/staf1400},
    eprint = {https://academic.oup.com/mnras/article-pdf/542/4/2866/64138209/staf1400.pdf},
}

@INPROCEEDINGS{Markwardt2009,
       author = {{Markwardt}, C.~B.},
        title = "{Non-linear Least-squares Fitting in IDL with MPFIT}",
    booktitle = {Astronomical Data Analysis Software and Systems XVIII},
         year = 2009,
       editor = {{Bohlender}, D.~A. and {Durand}, D. and {Dowler}, P.},
       series = {Astronomical Society of the Pacific Conference Series},
       volume = {411},
        month = sep,
        pages = {251},
          doi = {10.48550/arXiv.0902.2850},
archivePrefix = {arXiv},
       eprint = {0902.2850},
 primaryClass = {astro-ph.IM},
       adsurl = {https://ui.adsabs.harvard.edu/abs/2009ASPC..411..251M}
}

@INPROCEEDINGS{McNamara2000,
       author = {McNamara, D.~H.},
        title = "{The High-Amplitude {\ensuremath{\delta}} Scuti Stars}",
    booktitle = {Delta Scuti and Related Stars},
         year = 2000,
       editor = {{Breger}, Michel and {Montgomery}, Michael},
       series = {Astronomical Society of the Pacific Conference Series},
       volume = {210},
        month = jan,
        pages = {373},
       adsurl = {https://ui.adsabs.harvard.edu/abs/2000ASPC..210..373M}
}

@article{Moe2017,
doi = {10.3847/1538-4365/aa6fb6},
url = {https://doi.org/10.3847/1538-4365/aa6fb6},
year = {2017},
month = {jun},
publisher = {The American Astronomical Society},
volume = {230},
number = {2},
pages = {15},
author = {Moe, Maxwell and Di Stefano, Rosanne},
title = {Mind Your Ps and Qs: The Interrelation between Period (P) and Mass-ratio (Q) Distributions of Binary Stars},
journal = {\apjs}
}

@article{Murphy2019,
    author = {Murphy, Simon J and Hey, Daniel and Van Reeth, Timothy and Bedding, Timothy R},
    title = {Gaia-derived luminosities of Kepler A/F stars and the pulsator fraction across the δ Scuti instability strip},
    journal = {\mnras},
    volume = {485},
    number = {2},
    pages = {2380-2400},
    year = {2019},
    month = {05},
    issn = {0035-8711},
    doi = {10.1093/mnras/stz590},
    url = {https://doi.org/10.1093/mnras/stz590},
    eprint = {https://academic.oup.com/mnras/article-pdf/485/2/2380/28249225/stz590.pdf},
}

@article{Murphy2024,
    author = {Murphy, Simon J and Bedding, Timothy R and Gautam, Anuj and Kerr, Ronan P and Mani, Prasad},
    title = {The δ Scuti stars of the Cep–Her Complex – I. Pulsator fraction, rotation, asteroseismic large spacings, and the νmax relation},
    journal = {\mnras},
    volume = {534},
    number = {4},
    pages = {3022-3039},
    year = {2024},
    month = {11},
    issn = {0035-8711},
    doi = {10.1093/mnras/stae2226},
    url = {https://doi.org/10.1093/mnras/stae2226},
    eprint = {https://academic.oup.com/mnras/article-pdf/534/4/3022/59830969/stae2226.pdf},
}

@article{Murphy2026_08,
    author = {Murphy, Simon J and Gautam, Anuj and Claytor, Zachary R},
    title = {Quantifying isochrone-based age uncertainties for rapidly rotating A-type stars},
    journal = {\mnras},
    volume = {550},
    number = {2},
    pages = {stag1237},
    year = {2026},
    month = {08},
    issn = {0035-8711},
    doi = {10.1093/mnras/stag1237},
    url = {https://doi.org/10.1093/mnras/stag1237},
    eprint = {https://academic.oup.com/mnras/article-pdf/550/2/stag1237/68670588/stag1237.pdf},
}

@article{Murphy2026_09,
    author = {Murphy, Simon J and Gautam, Anuj},
    title = {Population synthesis of delta Scuti stars: the instability strip, period–luminosity relation, and large-separation distribution},
    journal = {\mnras},
    pages = {stag1825},
    year = {2026},
    month = {09},
    issn = {0035-8711},
    doi = {10.1093/mnras/stag1825},
    url = {https://doi.org/10.1093/mnras/stag1825},
    eprint = {https://academic.oup.com/mnras/advance-article-pdf/doi/10.1093/mnras/stag1825/71403644/stag1825.pdf},
}

@ARTICLE{Petersen1996,
       author = {{Petersen}, J.~O. and {Christensen-Dalsgaard}, J.},
        title = "{Pulsation models of {\ensuremath{\delta}} Scuti variables. I. The high-amplitude double-mode stars.}",
      journal = {\aap},
         year = 1996,
        month = aug,
       volume = {312},
        pages = {463-474},
       adsurl = {https://ui.adsabs.harvard.edu/abs/1996A&A...312..463P}
}

@article{Poretti2003,
	author = {Poretti, E.},
	title = {Asteroseismology of HADS stars: V974 Oph, a radial pulsator
flavoured by nonradial components*},
	DOI= "10.1051/0004-6361:20031223",
	url= "https://doi.org/10.1051/0004-6361:20031223",
	journal = {\aap},
	year = 2003,
	volume = 409,
	number = 3,
	pages = "1031-1035",
}

@article{Poterri2005,
	author = {Poretti, E. and Suárez, J. C. and Niarchos, P. G. and Gazeas, K. D. and Manimanis, V. N. and Van Cauteren, P. and Lampens, P. and Wils, P. and Alonso, R. and Amado, P. J. and Belmonte, J. A. and Butterworth, N. D. and Martignoni, M. and Martín-Ruiz, S. and Moskalik, P. and Robertson, C. W.},
	title = {The double-mode nature of the HADS star GSC 00144-03031   and
the Petersen diagram of the class},
	DOI= "10.1051/0004-6361:20053463",
	url= "https://doi.org/10.1051/0004-6361:20053463",
	journal = {\aap},
	year = 2005,
	volume = 440,
	number = 3,
	pages = "1097-1104",
}

@ARTICLE{Pribulla2024,
       author = {{Pribulla}, T. and {Va{\v{n}}ko}, M. and {Kom{\v{z}}{\'\i}k}, R. and {Sivani{\v{c}}}, P.},
        title = "{High-resolution {\'e}chelle spectrograph at Skalnat{\'e} Pleso Observatory}",
      journal = {Contrib. Astron. Observ. Skalnate Pleso},
         year = 2024,
        month = feb,
       volume = {54},
       number = {2},
        pages = {43-46},
          doi = {10.31577/caosp.2024.54.2.43},
       adsurl = {https://ui.adsabs.harvard.edu/abs/2024CoSka..54b..43P}
}

@article{Raskin2011HERMES,
	author = {Raskin, G. and Van Winckel, H. and Hensberge, H. and Jorissen, A. and Lehmann, H. and Waelkens, C. and Avila, G. and De Cuyper, J.-P. and Degroote, P. and Dubosson, R. and Dumortier, L. and Fr\'emat, Y. and Laux, U. and Michaud, B. and Morren, J. and Perez Padilla, J. and Pessemier, W. and Prins, S. and Smolders, K. and Van Eck, S. and Winkler, J.},
	title = {HERMES: a high-resolution fibre-fed spectrograph   for the Mercator telescope},
	DOI= "10.1051/0004-6361/201015435",
	url= "https://doi.org/10.1051/0004-6361/201015435",
	journal = {A\&A},
	year = 2011,
	volume = 526,
	pages = "A69",
	month = "",
}

@article{Reiners2003,
	author = {Reiners, A.},
	title = {The effects of inclination, gravity darkening and differential
 rotation on absorption profiles of fast rotators},
	DOI= "10.1051/0004-6361:20030966",
	url= "https://doi.org/10.1051/0004-6361:20030966",
	journal = {\aap},
	year = 2003,
	volume = 408,
	number = 2,
	pages = "707-713",
}

@article{Reese2009,
	author = {Reese, D. R. and MacGregor, K. B. and Jackson, S. and Skumanich, A. and Metcalfe, T. S.},
	title = {Pulsation modes in rapidly rotating stellar models based   on the
 self-consistent field method },
	DOI= "10.1051/0004-6361/200811510",
	url= "https://doi.org/10.1051/0004-6361/200811510",
	journal = {\aap},
	year = 2009,
	volume = 506,
	number = 1,
	pages = "189-201",
}

@ARTICLE{Roche1854,
       author = {{Roche}, {\'E}douard},
        title = "{M{\'e}moire sur la figure des atmosph{\`e}res des corps c{\'e}lestes}",
      journal = {M{\'e}moires de l'Acad{\'e}mie des Sciences et Lettres de Montpellier, Section des Sciences},
         year = 1854,
       volume = {2},
        pages = {399--439}
}

@article{Royer2002_1,
	author = {Royer, F. and Gerbaldi, M. and Faraggiana, R. and G\'omez, A. E.},
	title = {Rotational velocities of A-type stars* - I. Measurement of \$v\backslash{}sin i\$ in the southern hemisphere},
	DOI= "10.1051/0004-6361:20011422",
	url= "https://doi.org/10.1051/0004-6361:20011422",
	journal = {A\&A},
	year = 2002,
	volume = 381,
	number = 1,
	pages = "105-121",
}

@article{Royer2002_2,
	author = {Royer, F. and Grenier, S. and Baylac, M.-O. and G\'omez, A. E. and Zorec, J.},
	title = {Rotational velocities of A-type stars *** - II. Measurement of \$\{v\backslash{}sin i\}\$ in the northern hemisphere},
	DOI= "10.1051/0004-6361:20020943",
	url= "https://doi.org/10.1051/0004-6361:20020943",
	journal = {A\&A},
	year = 2002,
	volume = 393,
	number = 3,
	pages = "897-911",
}

@article{Royer2007,
	author = {Royer, F. and Zorec, J. and G\'omez, A. E.},
	title = {Rotational velocities of A-type stars *** - III. Velocity distributions},
	DOI= "10.1051/0004-6361:20065224",
	url= "https://doi.org/10.1051/0004-6361:20065224",
	journal = {A\&A},
	year = 2007,
	volume = 463,
	number = 2,
	pages = "671-682",
}

@article{Royer2024MELCHIORS,
	author = {Royer, P. and Merle, T. and Dsilva, K. and Sekaran, S. and Van Winckel, H. and Fr\'emat, Y. and Van der Swaelmen, M. and Gebruers, S. and Tkachenko, A. and Laverick, M. and Dirickx, M. and Raskin, G. and Hensberge, H. and Abdul-Masih, M. and Acke, B. and Alonso, M. L. and Bandhu Mahato, S. and Beck, P. G. and Behara, N. and Bloemen, S. and Buysschaert, B. and Cox, N. and Debosscher, J. and De Cat, P. and Degroote, P. and De Nutte, R. and De Smedt, K. and de Vries, B. and Dumortier, L. and Escorza, A. and Exter, K. and Goriely, S. and Gorlova, N. and Hillen, M. and Homan, W. and Jorissen, A. and Kamath, D. and Karjalainen, M. and Karjalainen, R. and Lampens, P. and Lobel, A. and Lombaert, R. and Marcos-Arenal, P. and Menu, J. and Merges, F. and Moravveji, E. and Nemeth, P. and Neyskens, P. and Ostensen, R. and P\'apics, P. I. and Perez, J. and Prins, S. and Royer, S. and Samadi-Ghadim, A. and Sana, H. and Sans Fuentes, A. and Scaringi, S. and Schmid, V. and Siess, L. and Siopis, C. and Smolders, K. and S\'odor, \'A. and Thoul, A. and Triana, S. and Vandenbussche, B. and Van de Sande, M. and Van De Steene, G. and Van Eck, S. and van Hoof, P. A. M. and Van Marle, A. J. and Van Reeth, T. and Vermeylen, L. and Volpi, D. and Vos, J. and Waelkens, C.},
	title = {MELCHIORS - The Mercator Library of High Resolution Stellar Spectroscopy},
	DOI= "10.1051/0004-6361/202346847",
	url= "https://doi.org/10.1051/0004-6361/202346847",
	journal = {A\&A},
	year = 2024,
	volume = 681,
	pages = "A107",
}

@article{Rozanski2022SUPPnet,
	author = {R\'oza\'nski, T. and Niemczura, E. and Lemiesz, J. and Posilek, N. and R\'oza\'nski, P.},
	title = {SUPPNet: Neural network for stellar spectrum normalisation},
	DOI= "10.1051/0004-6361/202141480",
	url= "https://doi.org/10.1051/0004-6361/202141480",
	journal = {A\&A},
	year = 2022,
	volume = 659,
	pages = "A199",
}

@ARTICLE{Saio2018,
       author = {{Saio}, Hideyuki and {Kurtz}, Donald W. and {Murphy}, Simon J. and {Antoci}, Victoria L. and {Lee}, Umin},
        title = "{Theory and evidence of global Rossby waves in upper main-sequence stars: r-mode oscillations in many Kepler stars}",
      journal = {\mnras},
         year = 2018,
        month = feb,
       volume = {474},
       number = {2},
        pages = {2774-2786},
          doi = {10.1093/mnras/stx2962},
archivePrefix = {arXiv},
       eprint = {1711.04908},
 primaryClass = {astro-ph.SR},
       adsurl = {https://ui.adsabs.harvard.edu/abs/2018MNRAS.474.2774S}
}

@ARTICLE{Schwarzschild1906,
       author = {{Schwarzschild}, K. and {Villiger}, W.},
        title = "{On the Distribution of Brightness of the Ultra-Violet Light on the Sun's Disk}",
      journal = {\apj},
         year = 1906,
        month = may,
       volume = {23},
        pages = {284},
          doi = {10.1086/141343},
       adsurl = {https://ui.adsabs.harvard.edu/abs/1906ApJ....23..284S}
}

@ARTICLE{Scargle1982,
       author = {{Scargle}, J.~D.},
        title = "{Studies in astronomical time series analysis. II. Statistical aspects of spectral analysis of unevenly spaced data.}",
      journal = {\apj},
         year = 1982,
        month = dec,
       volume = {263},
        pages = {835-853},
          doi = {10.1086/160554},
       adsurl = {https://ui.adsabs.harvard.edu/abs/1982ApJ...263..835S}
}

@ARTICLE{Slettebak1954,
       author = {{Slettebak}, Arne},
        title = "{The Spectra and Rotational Velocities of the Bright Stars of Draper Types B8-A2.}",
      journal = {\apj},
         year = 1954,
        month = jan,
       volume = {119},
        pages = {146},
          doi = {10.1086/145804},
       adsurl = {https://ui.adsabs.harvard.edu/abs/1954ApJ...119..146S}
}

@ARTICLE{Slettebak1955,
       author = {{Slettebak}, Arne},
        title = "{The Spectra and Rotational Velocities of the Bright Stars of Draper Types A3-G0.}",
      journal = {\apj},
         year = 1955,
        month = may,
       volume = {121},
        pages = {653},
          doi = {10.1086/146031},
       adsurl = {https://ui.adsabs.harvard.edu/abs/1955ApJ...121..653S}
}

@ARTICLE{Slettebak1956,
       author = {{Slettebak}, Arne},
        title = "{Line Broadening in the Spectra of o- and Early B-Type Stars.}",
      journal = {\apj},
         year = 1956,
        month = jul,
       volume = {124},
        pages = {173},
          doi = {10.1086/146213},
       adsurl = {https://ui.adsabs.harvard.edu/abs/1956ApJ...124..173S}
}

@ARTICLE{Slettebak1975,
       author = {{Slettebak}, A. and {Collins}, II, G.~W. and {Boyce}, P.~B. and {White}, N.~M. and {Parkinson}, T.~D.},
        title = "{A system of standard stars for rotational velocity determinations.}",
      journal = {\apjs},
         year = 1975,
        month = may,
       volume = {29},
        pages = {137-159},
          doi = {10.1086/190338},
       adsurl = {https://ui.adsabs.harvard.edu/abs/1975ApJS...29..137S}
}

@article{Skarka2022,
	author = {Skarka, M. and Z\'ak, J. and Fedurco, M. and Paunzen, E. and Henzl, Z. and Masek, M. and Karjalainen, R. and Sanchez Arias, J. P. and S\'odor, \'A. and Auer, R. F. and Kab\'ath, P. and Karjalainen, M. and Liska, J. and Stegner, D.},
	title = {Periodic variable A-F spectral type stars in the northern TESS continuous viewing zone - I. Identification and classification},
	DOI= "10.1051/0004-6361/202244037",
	url= "https://doi.org/10.1051/0004-6361/202244037",
	journal = {\aap},
	year = 2022,
	volume = 666,
	pages = "A142",
}

@article{Skarka2024,
	author = {Skarka, M. and Henzl, Z.},
	title = {Periodic variable A-F spectral type stars in the southern TESS continuous viewing zone - I. Identification and classification},
	DOI= "10.1051/0004-6361/202450711",
	url= "https://doi.org/10.1051/0004-6361/202450711",
	journal = {A\&A},
	year = 2024,
	volume = 688,
	pages = "A25",
}

@phdthesis{Sneden1973,
  author = {Sneden, Christopher Alan},
  title  = {Carbon and Nitrogen Abundances in Metal-Poor Stars},
  school = {The University of Texas at Austin},
  year   = {1973}
}

@misc{Sneden2012,
  author       = {Sneden, Chris and Bean, Jacob and Ivans, Inese
                  and Lucatello, Sara and Sobeck, Jennifer},
  title        = {{MOOG}: {LTE} line analysis and spectrum synthesis},
  howpublished = {Astrophysics Source Code Library, record ascl:1202.009},
  year         = {2012}
}

@article{Solano1997,
	author = {Solano, E. and Fernley, J.},
	title = {Spectroscopic survey of ti stars* - I. Rotation velocities and effective temperatures},
	DOI= "10.1051/aas:1997329",
	url= "https://doi.org/10.1051/aas:1997329",
	journal = {\aaps},
	year = 1997,
	volume = 122,
	number = 1,
	pages = "131-147",
}

@article{Southworth2026,
   author = "Southworth, John and Bowman, Dominic M.",
   title = "Pulsations in Binary Star Systems", 
   journal= "Annual Review of Astronomy and Astrophysics",
   year = "2026",
   volume = "64",
   number = "Volume 64, 2026",
   pages = "223-260",
   doi = "https://doi.org/10.1146/annurev-astro-120425-015551",
   url = "https://www.annualreviews.org/content/journals/10.1146/annurev-astro-120425-015551",
   publisher = "Annual Reviews",
   issn = "1545-4282",
   type = "Journal Article",
  }

\begin{appendix}
\onecolumn
\section{Examples of derived $v\sin i$}
\label{Apendix:A}
In Fig.~\ref{fig:Apendix:SingleLineFit} we present the examples of fitted spectral line Fe\,\textsc{i}\,4957.613\,\AA\ for various objects (\textit{method\,1}). In Figs.~\ref{fig:Apendix:iSpec1} and \ref{fig:Apendix:iSpec2} we provide examples of spectra fitted with \texttt{iSpec} (\textit{method\,2}). In Fig.~\ref{fig:Apendix:FourierTransform} we show Fourier transforms of spectral regions \SI{500}{} to \SI{580}{\nano\meter} (left) and extracted rotational profiles (right) (\textit{method\,3}).
\begin{figure*}[ht!]
    \centering
    {\includegraphics[width=.3\linewidth]{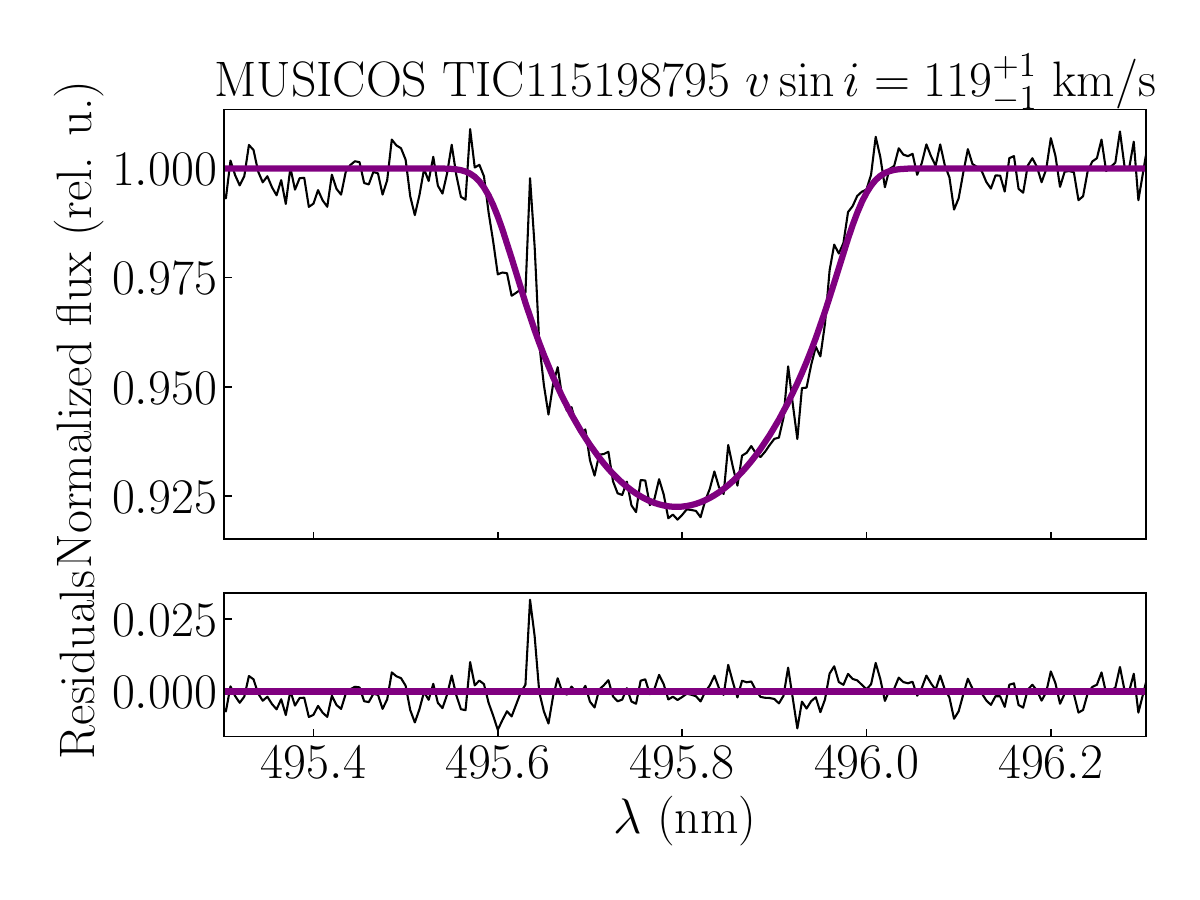}}
     {\includegraphics[width=.3\linewidth]{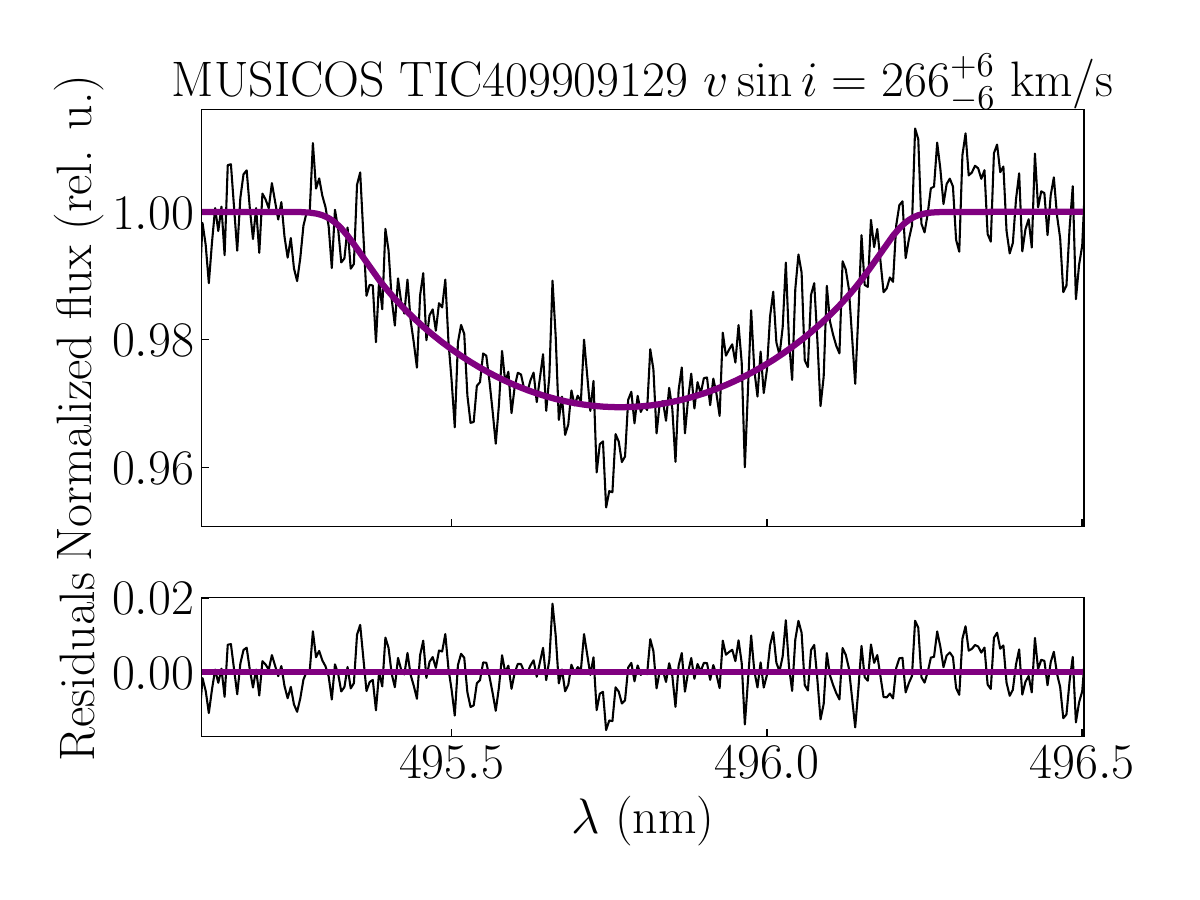}}\\
     {\includegraphics[width=.3\linewidth]{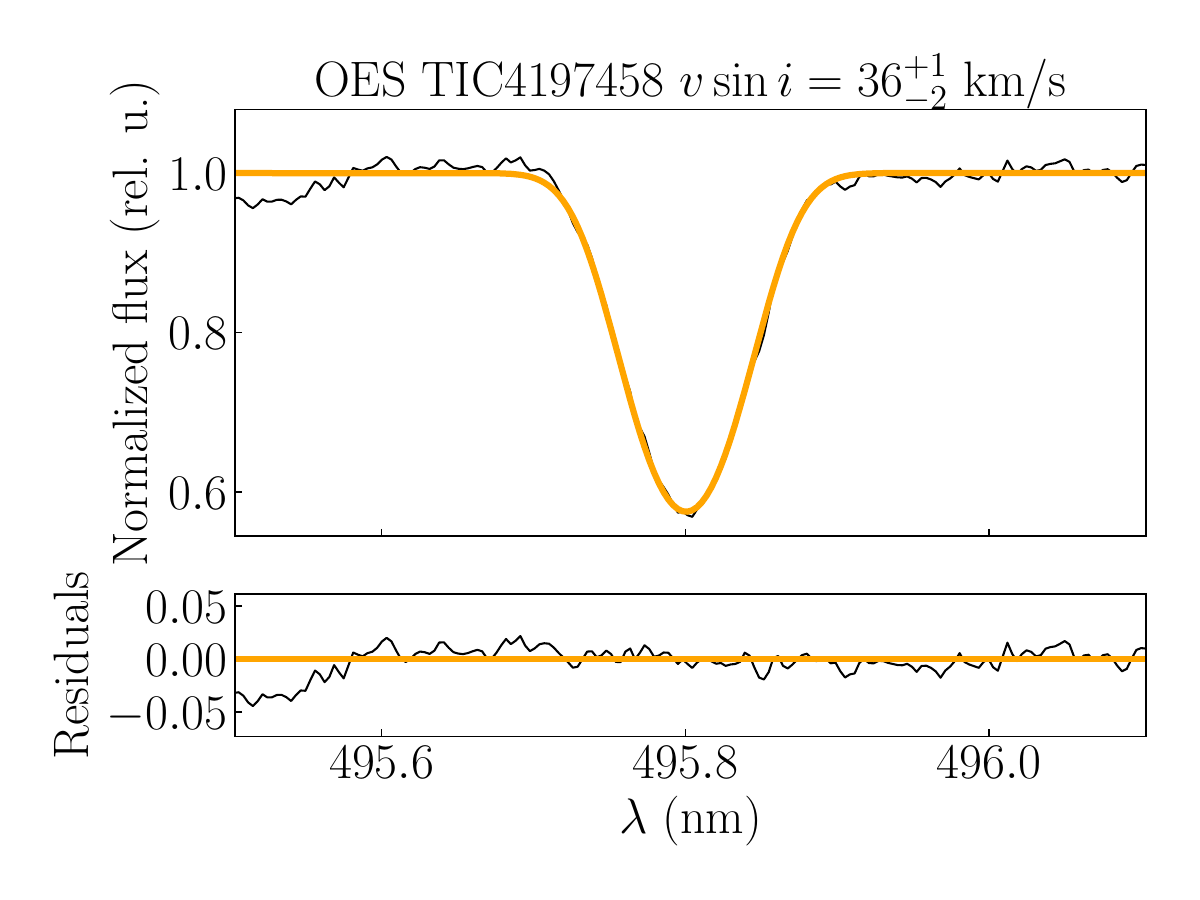}}
     {\includegraphics[width=.3\linewidth]{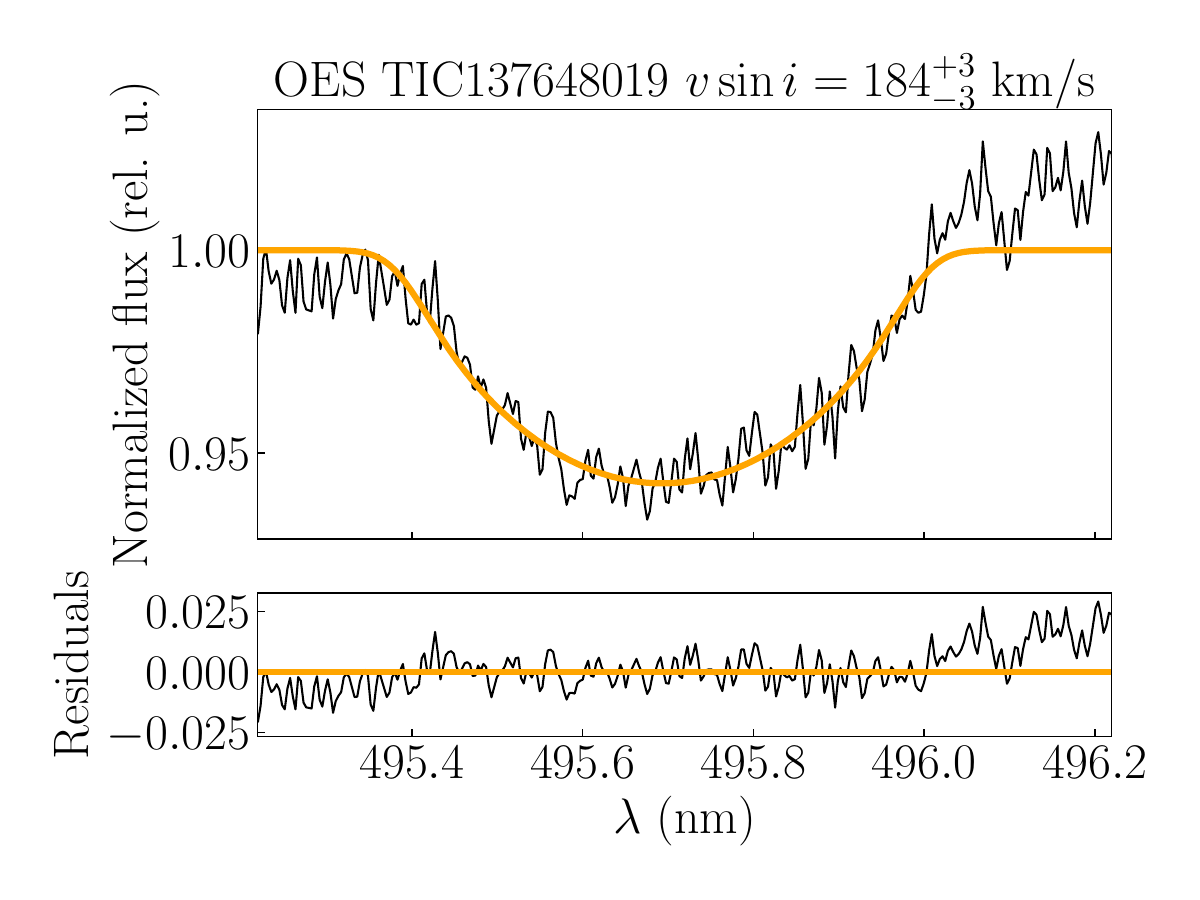}}\\
     {\includegraphics[width=.3\linewidth]{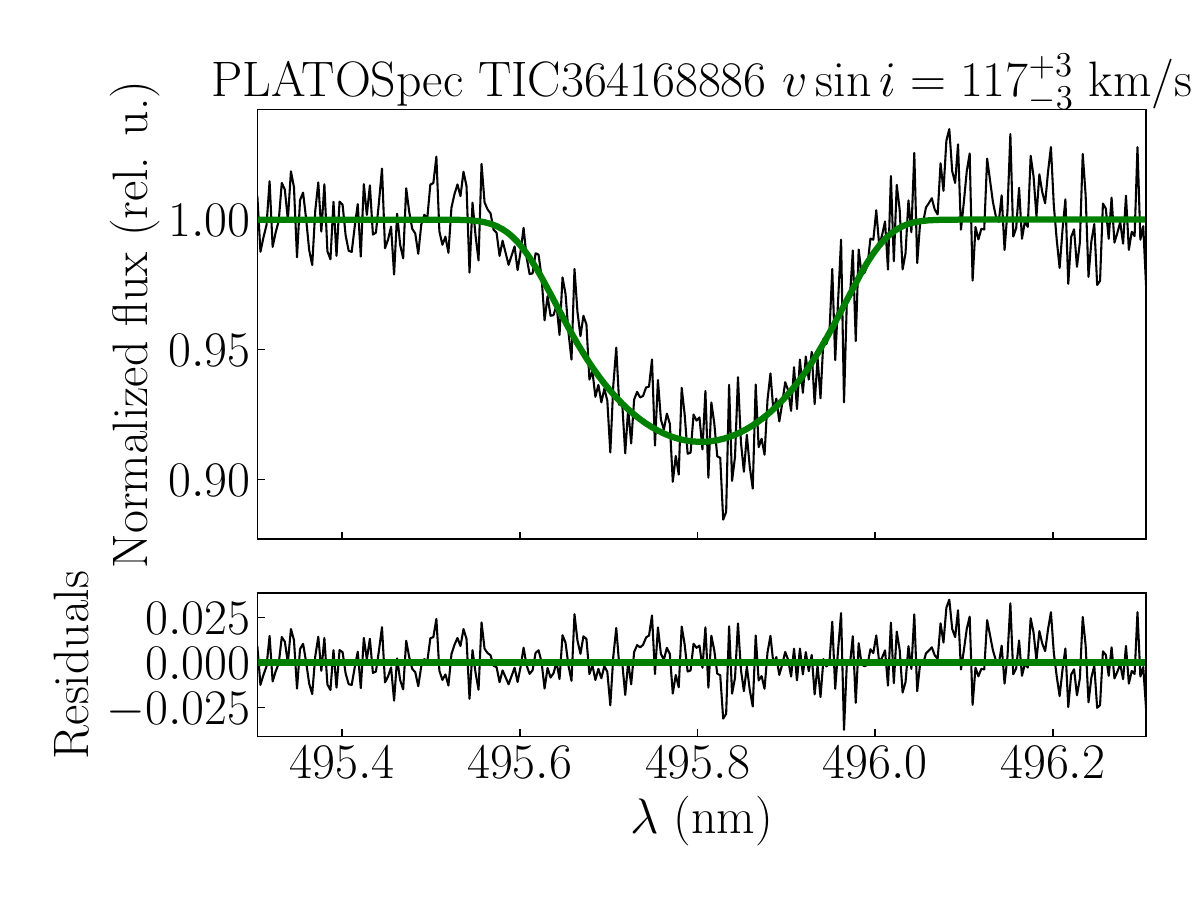}}
     {\includegraphics[width=.3\linewidth]{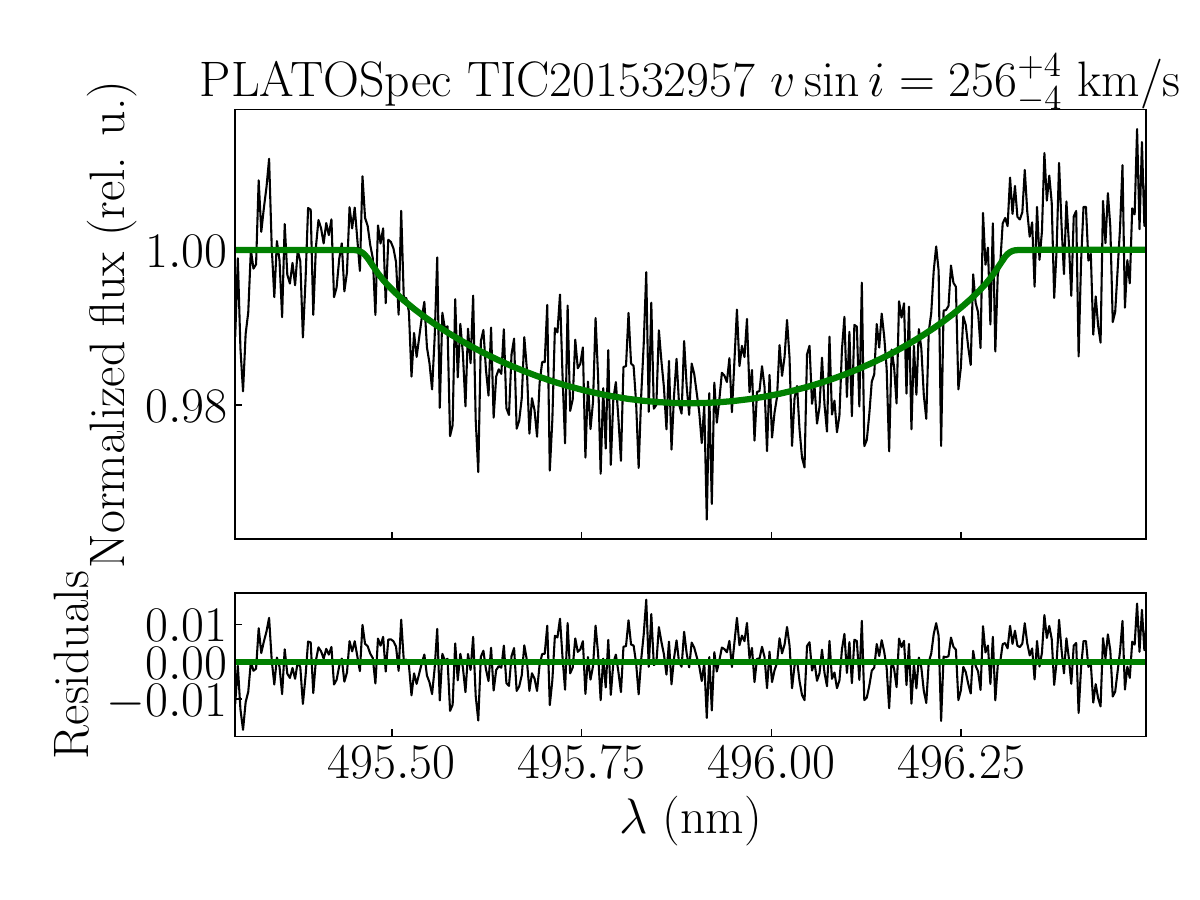}}\\
     {\includegraphics[width=.3\linewidth]{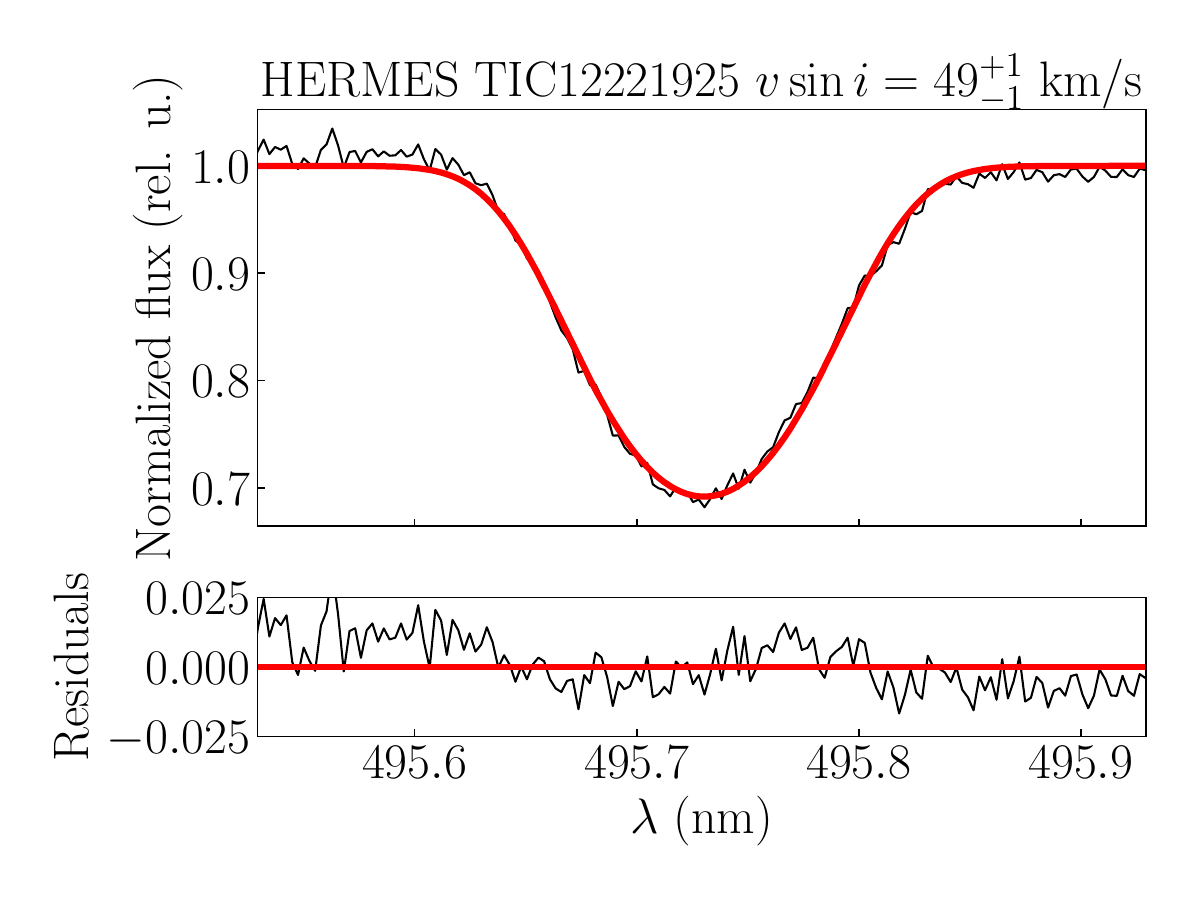}}
     {\includegraphics[width=.3\linewidth]{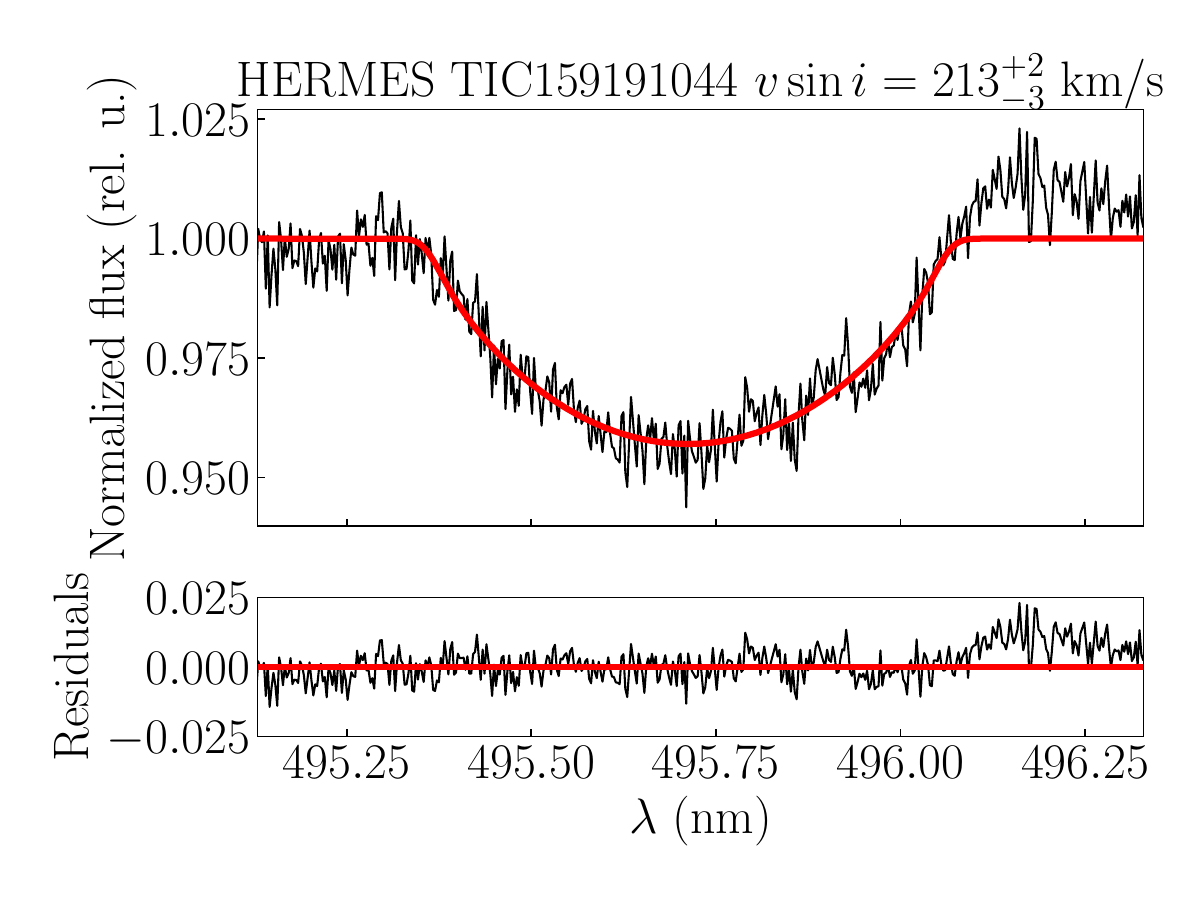}} \\
     {\includegraphics[width=.3\linewidth]{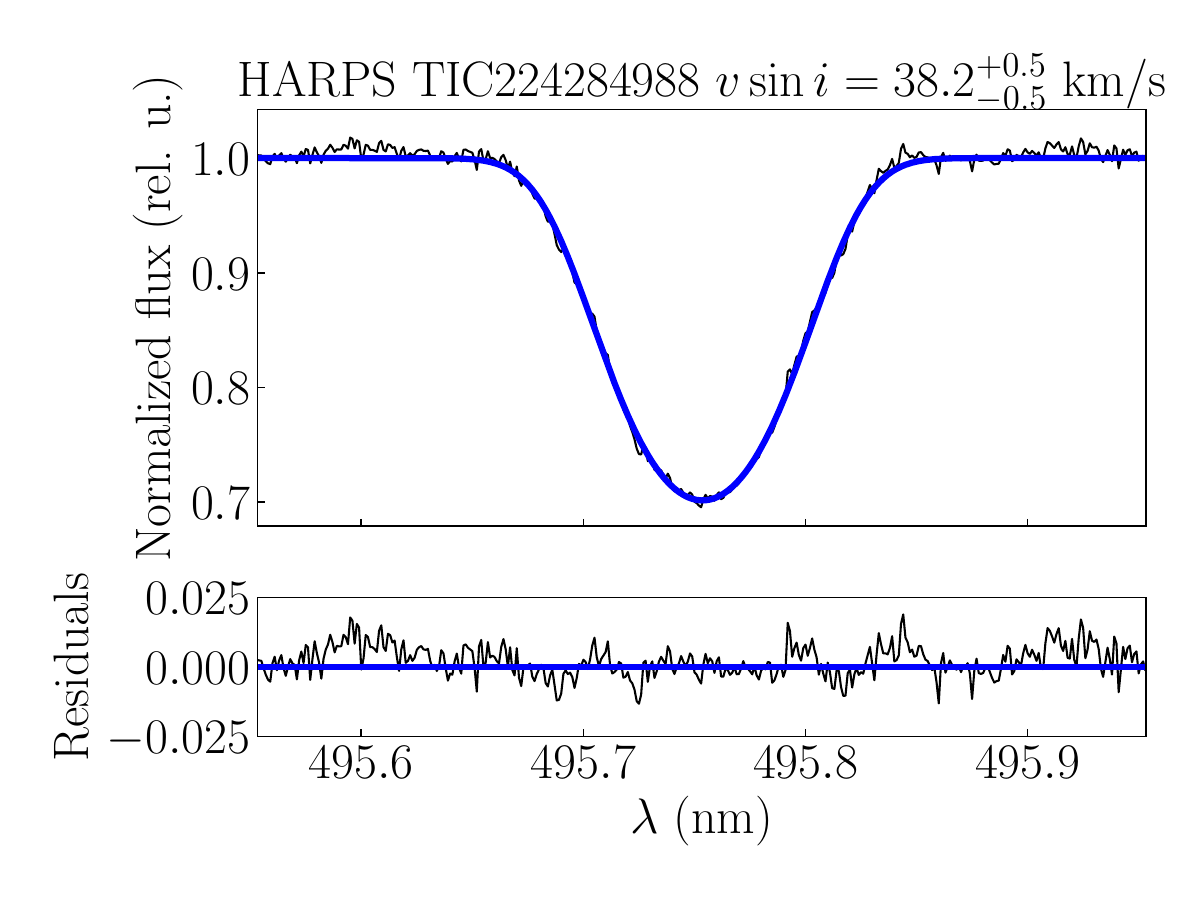}}
     {\includegraphics[width=.3\linewidth]{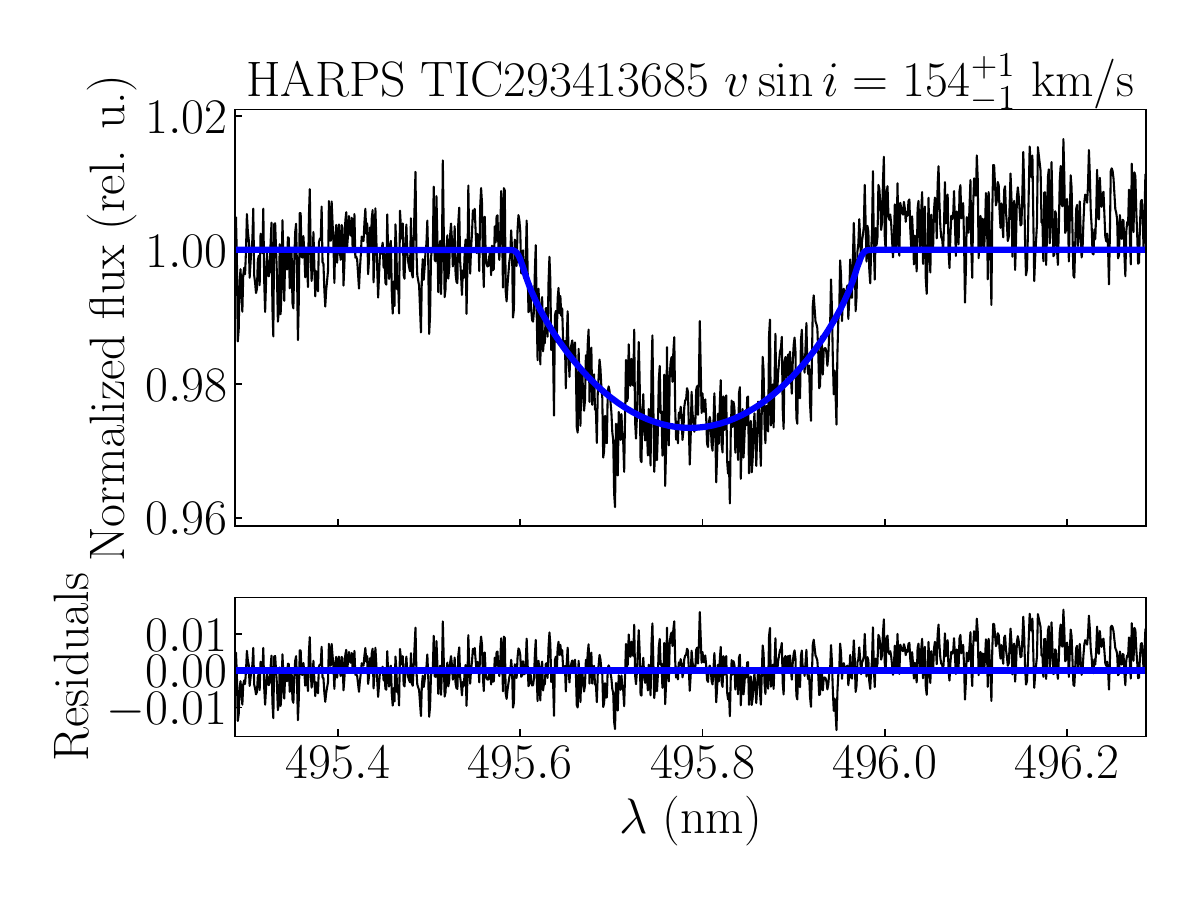}}
    \caption{Examples of \textit{method~1} on line Fe\,\textsc{i}\,4957.613\,\AA\ fitted with the convolution of the Voigt profile and rotational kernel.}
      \label{fig:Apendix:SingleLineFit} 
\end{figure*}

\begin{figure*}[ht!]
    \centering
     {\includegraphics[width=\linewidth]{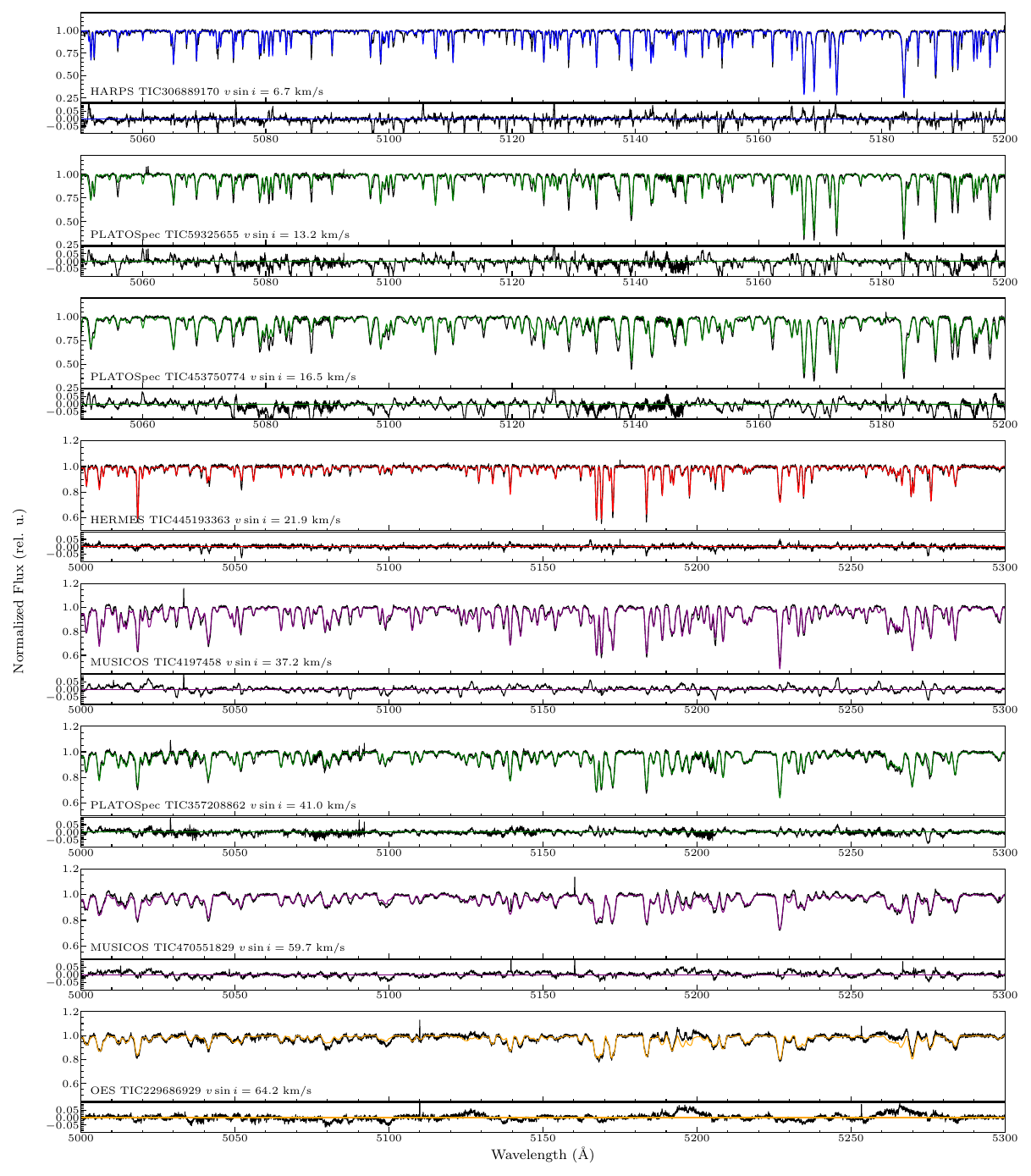}}
     \caption{Examples of \textit{method~2}. Spectra from different instruments fitted with \texttt{iSpec} are ordered in increasing $v\sin i$.}
      \label{fig:Apendix:iSpec1}
\end{figure*}

\begin{figure*}[ht!]
    \centering
     {\includegraphics[width=\linewidth]{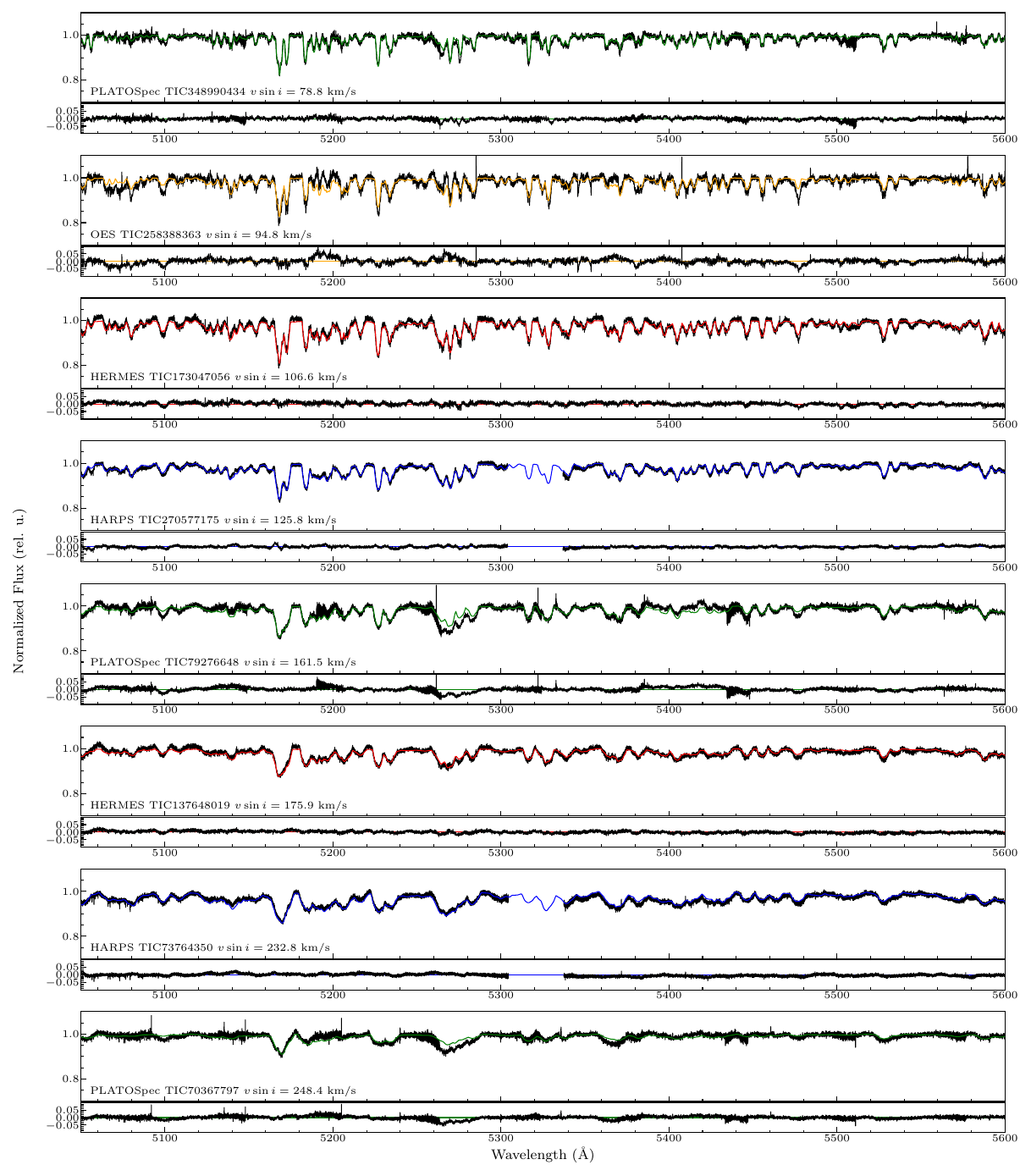}}
     \caption{Fig.~\ref{fig:Apendix:iSpec1} continued.}
      \label{fig:Apendix:iSpec2}
\end{figure*}

\begin{figure*}[ht!]
    \centering
    {\includegraphics[width=.4\linewidth]{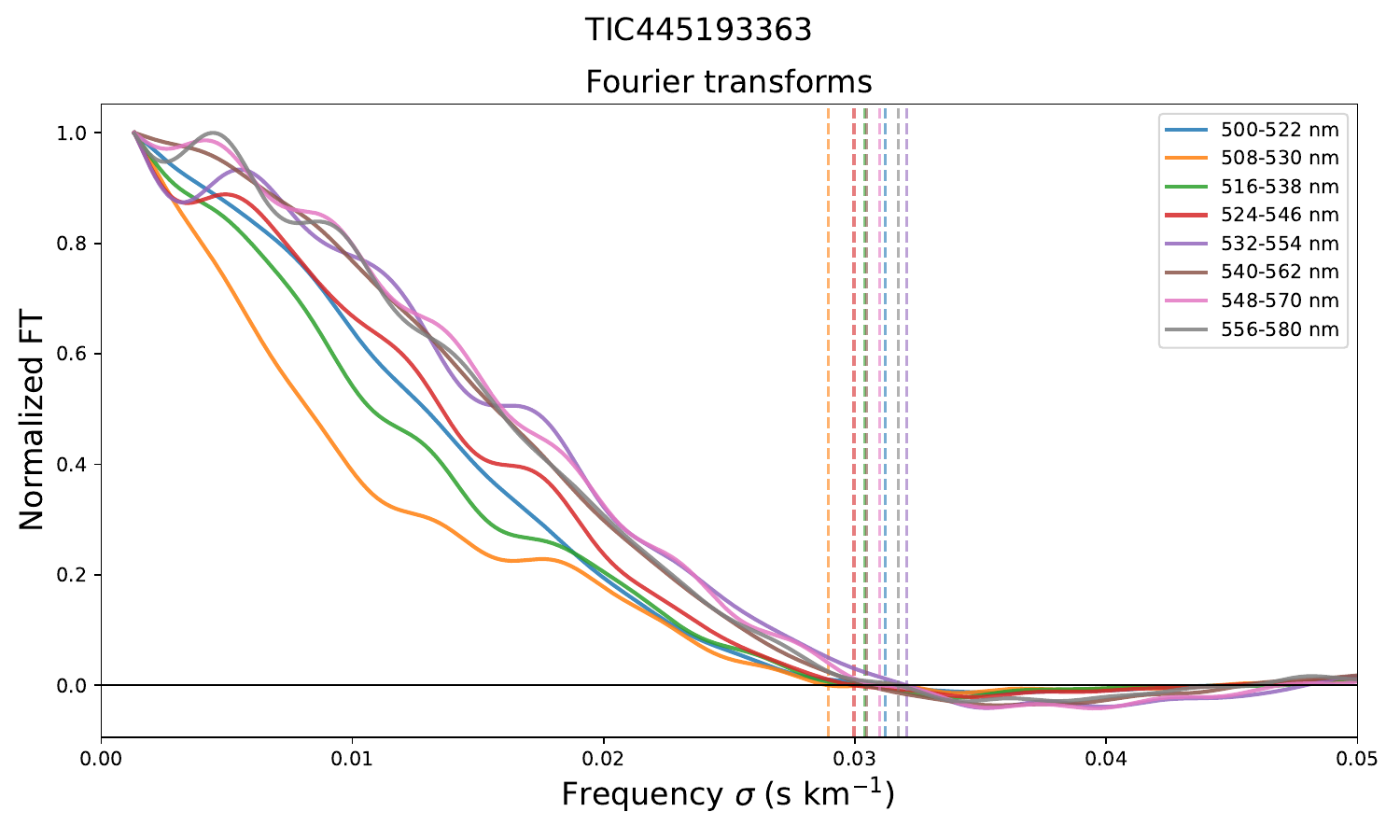}}
     {\includegraphics[width=.4\linewidth]{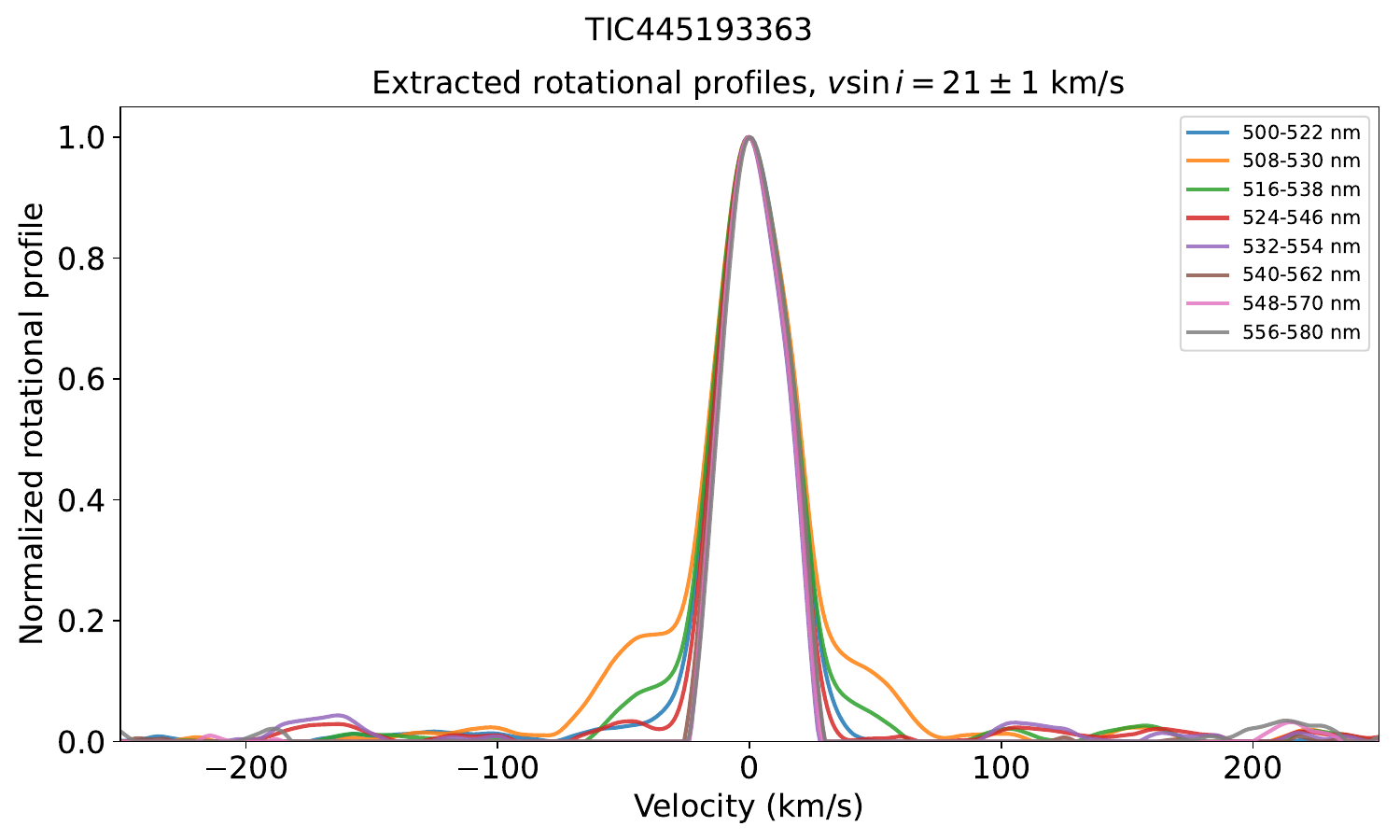}}
     {\includegraphics[width=.4\linewidth]{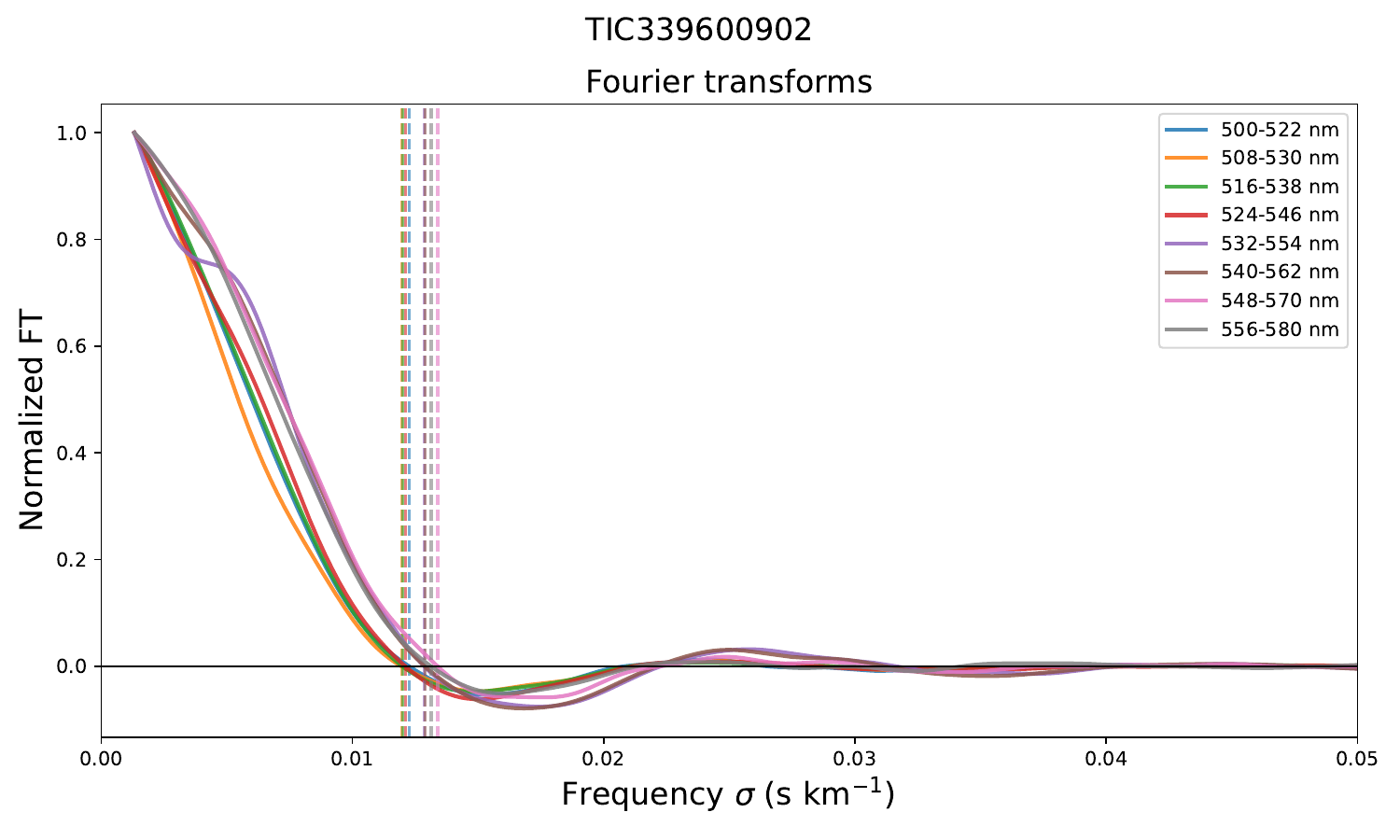}}
     {\includegraphics[width=.4\linewidth]{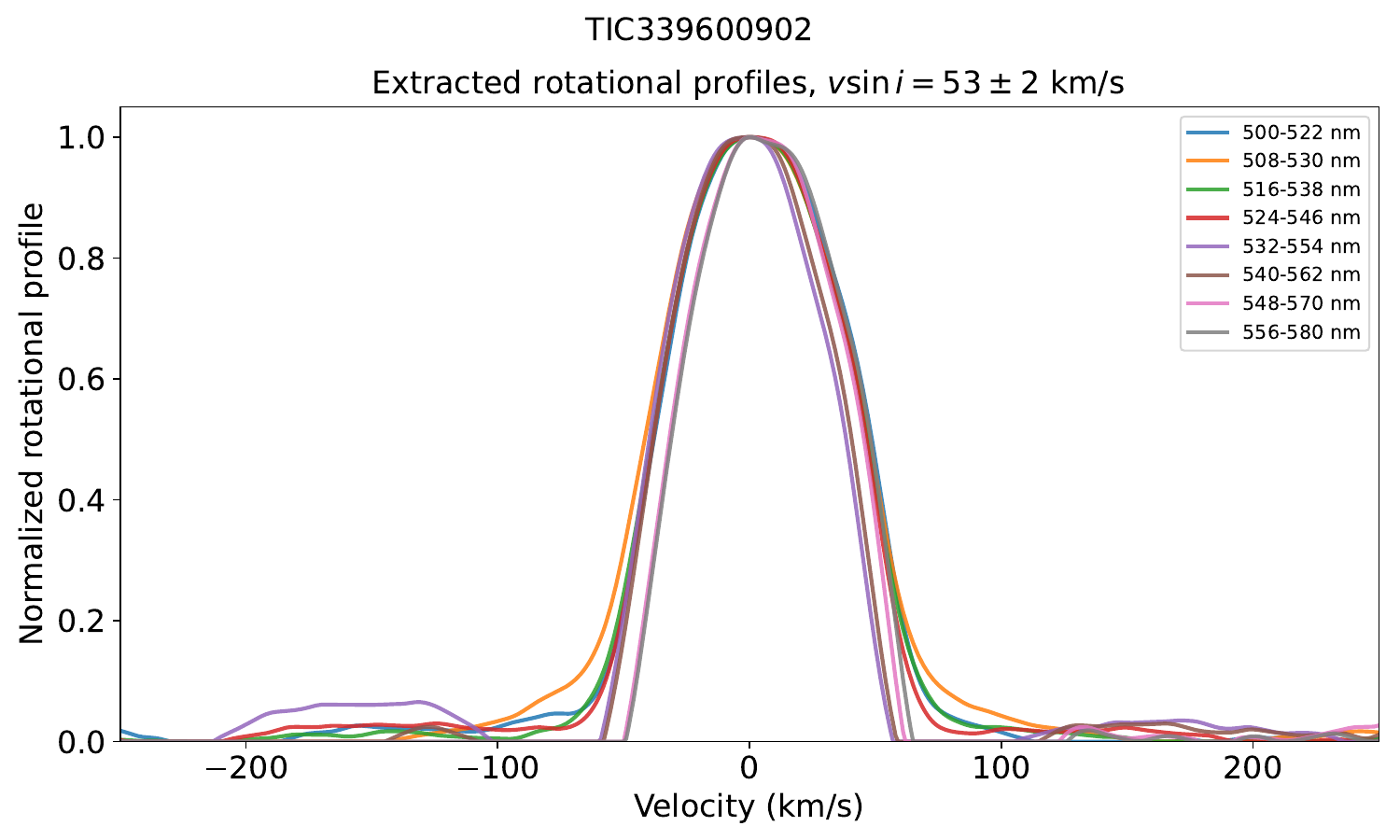}}
     {\includegraphics[width=.4\linewidth]{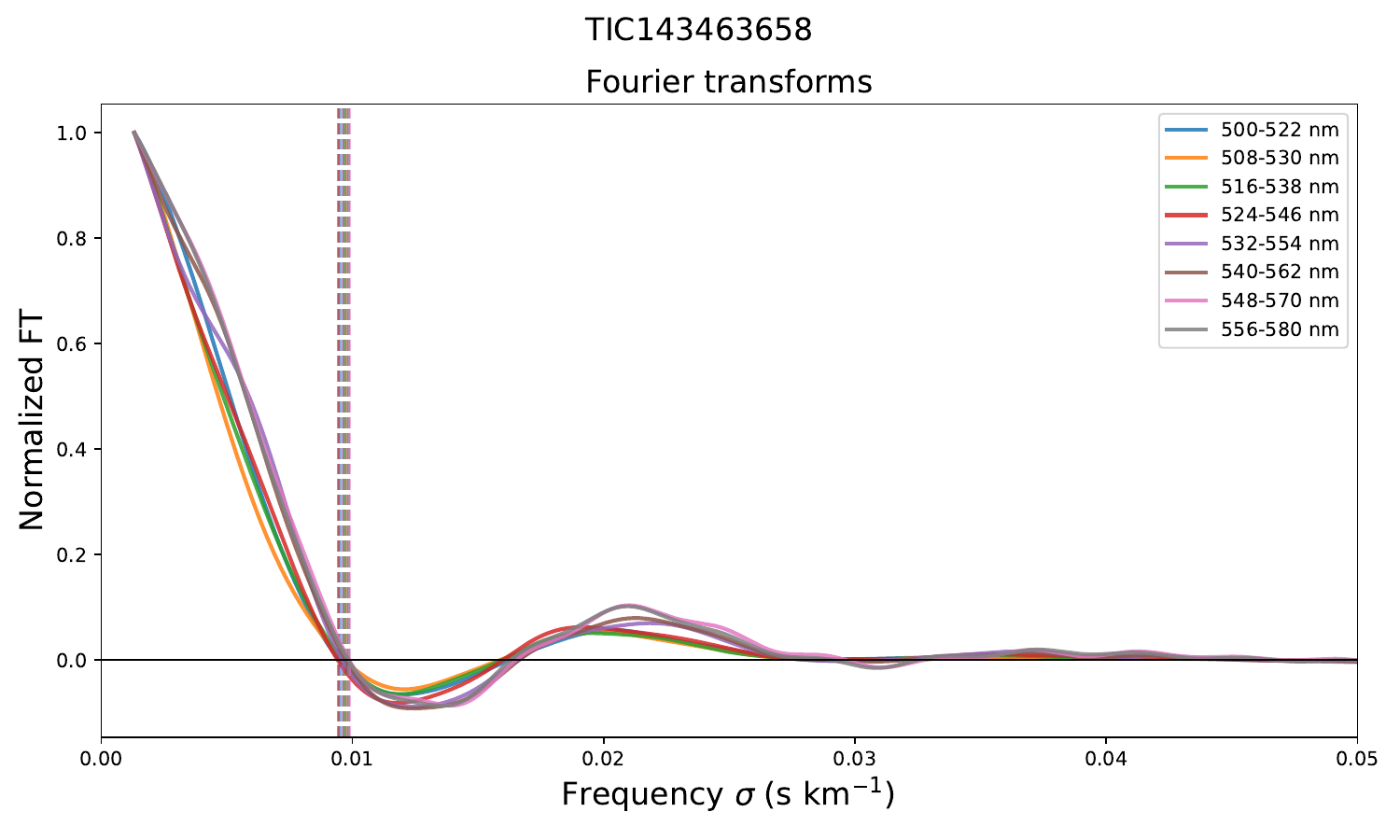}}
     {\includegraphics[width=.4\linewidth]{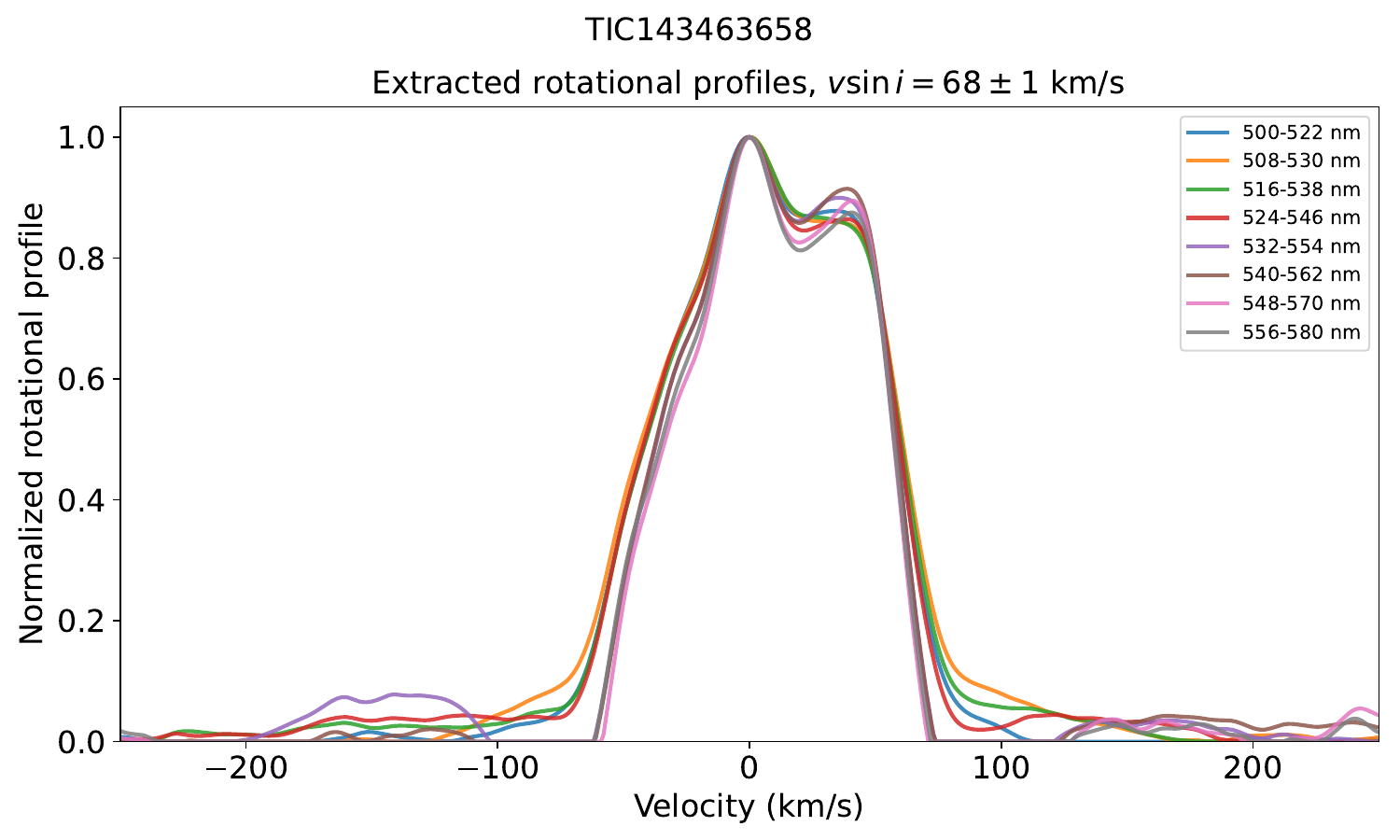}}
     {\includegraphics[width=.4\linewidth]{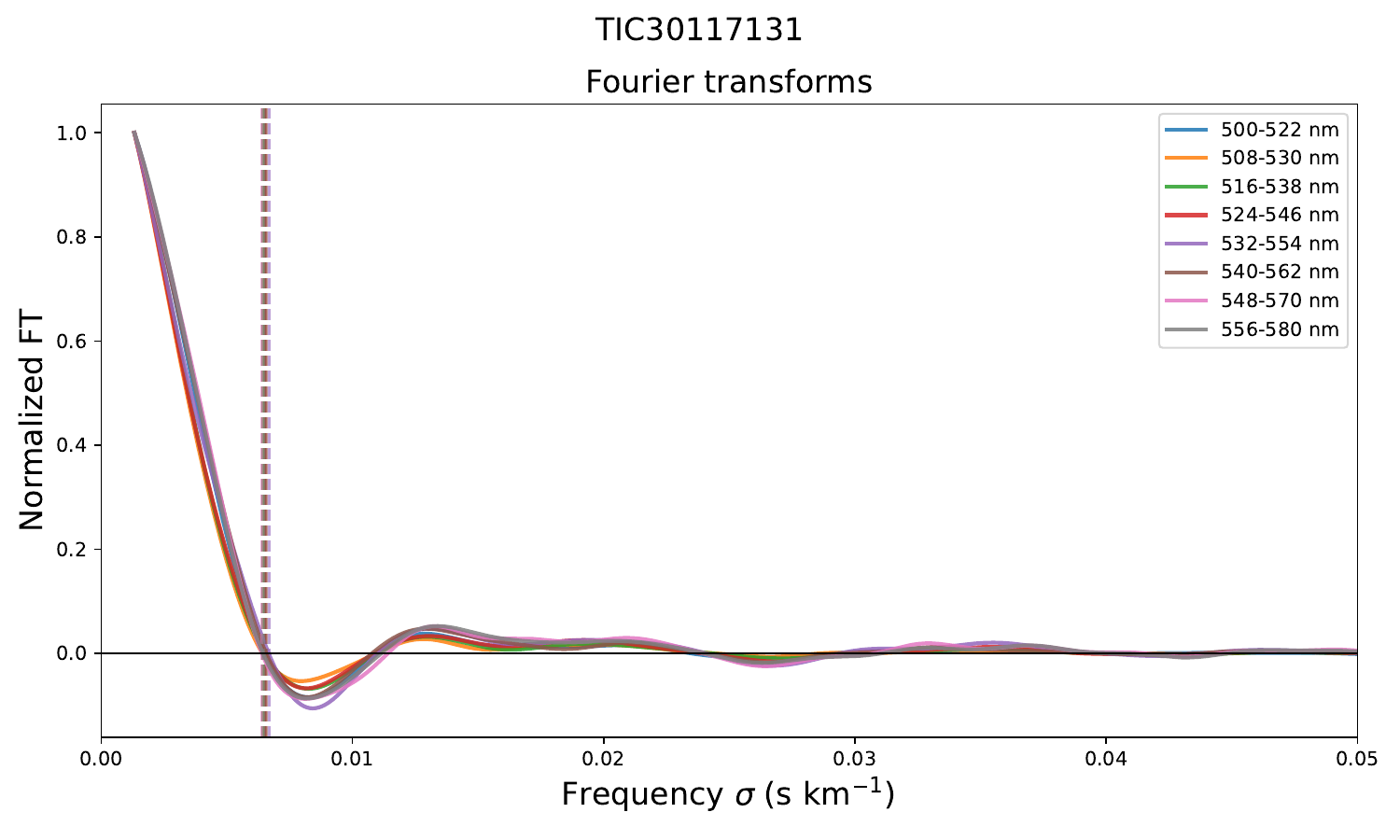}}
     {\includegraphics[width=.4\linewidth]{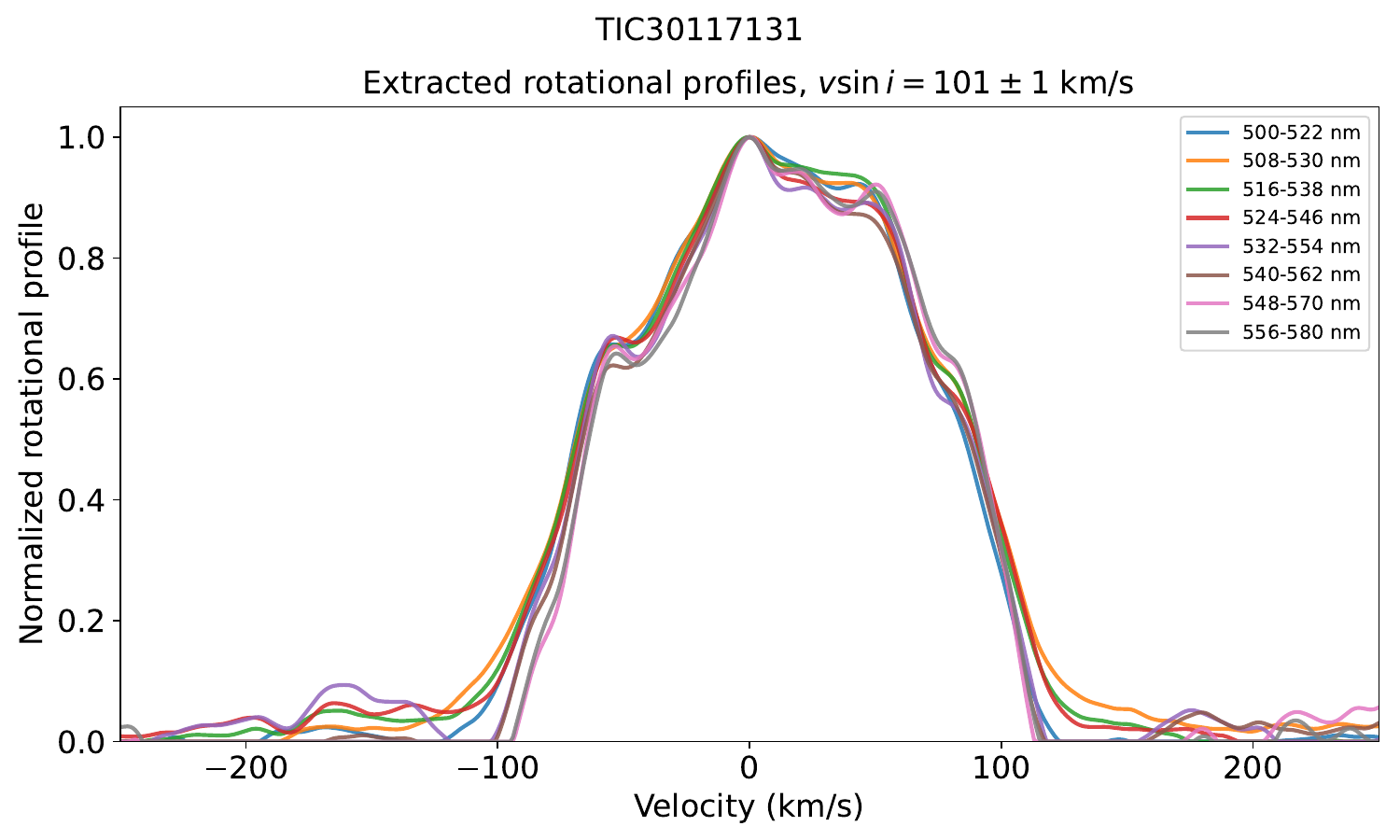}}  
     {\includegraphics[width=.4\linewidth]{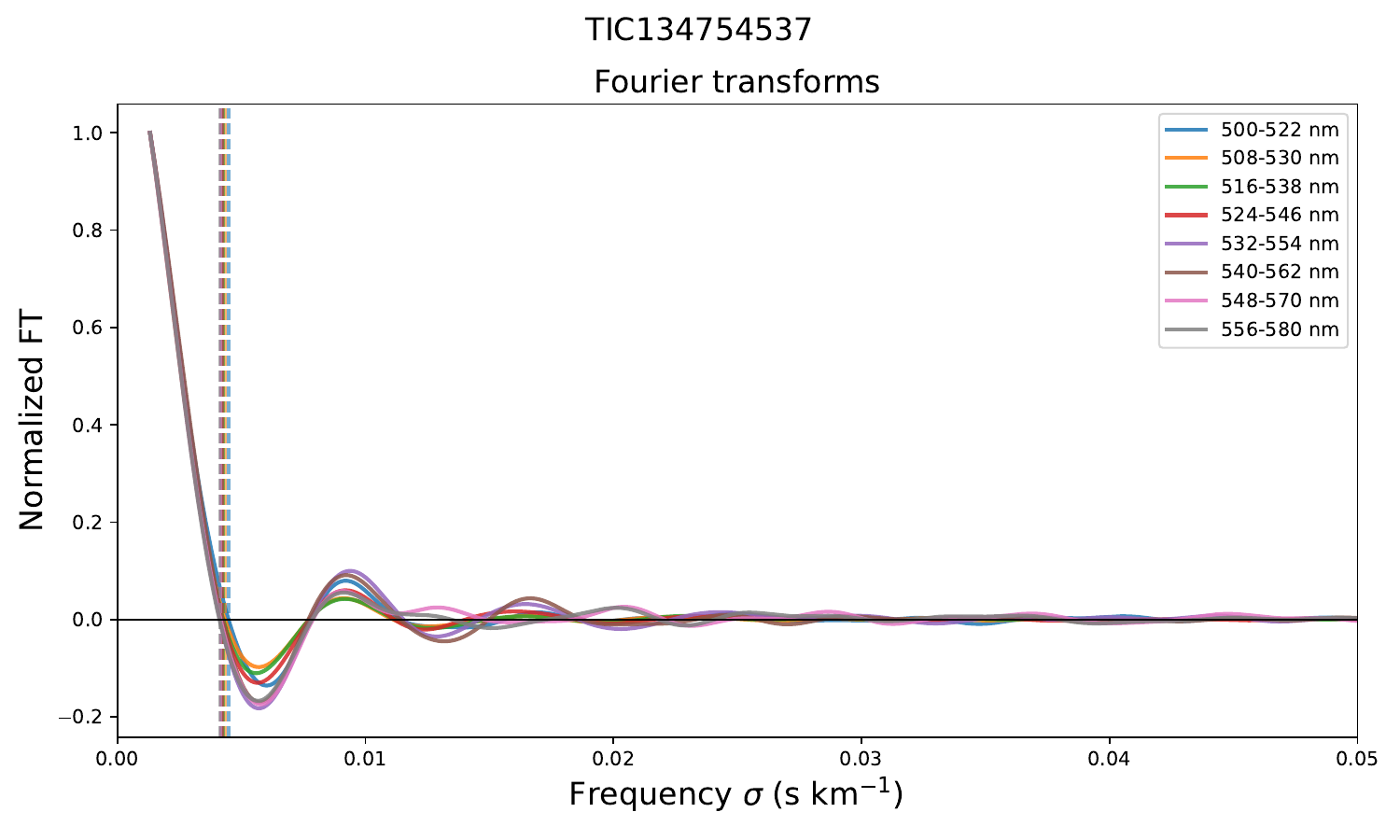}}
     {\includegraphics[width=.4\linewidth]{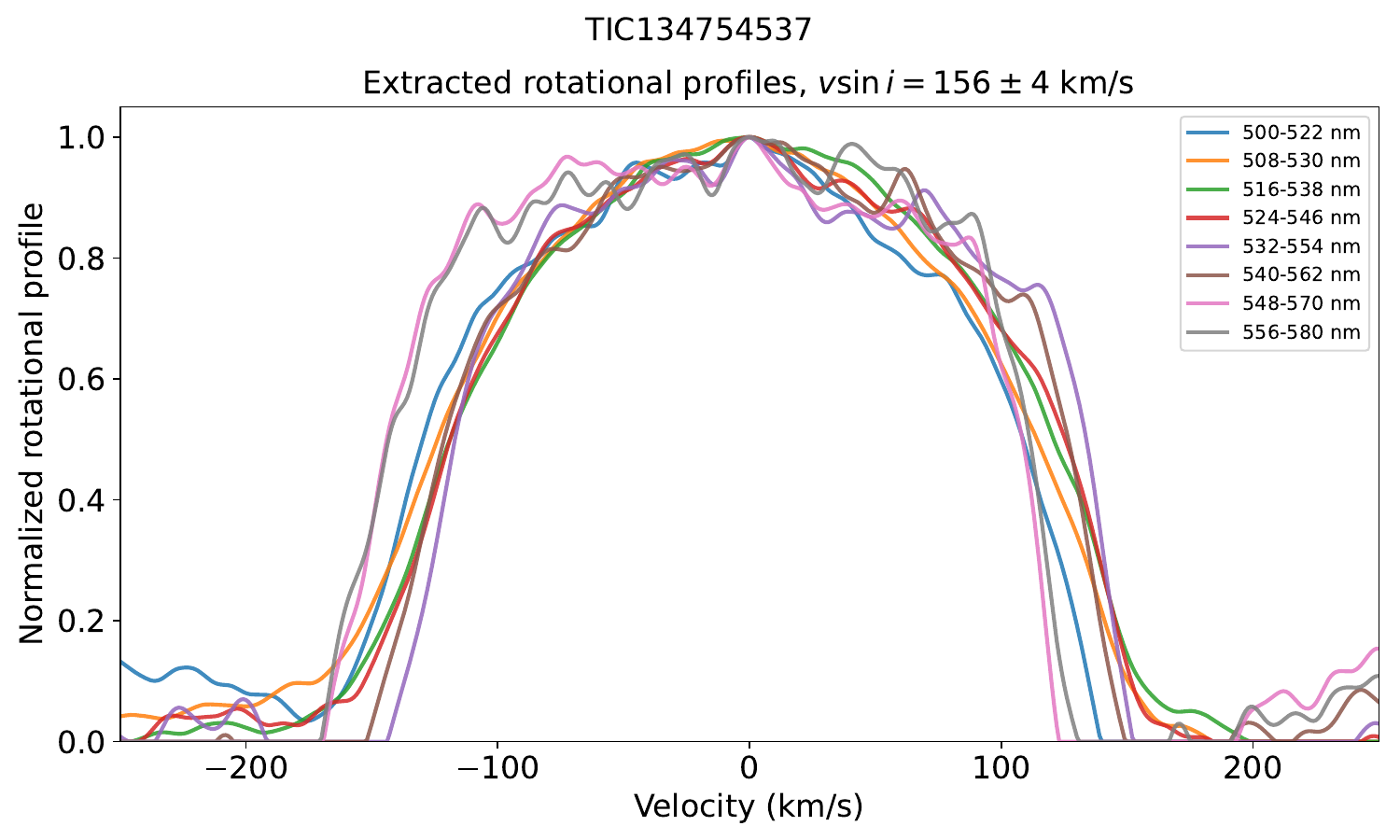}}

    \caption{Examples of \textit{method~3} ordered with increasing $v\sin i$. Left: Fourier transforms of spectral regions. First zeros of the Fourier transforms, from which the $v\sin i$ is calculated is shown using vertical dashed lines. Right: Extracted rotational profiles of spectral regions.}
      \label{fig:Apendix:FourierTransform} 
\end{figure*}

\clearpage

\section{Binary candidates}
\label{Apendix:B}

\begin{table}[ht]
\label{table:Binaries}
\caption[]{List of new binary candidates in our sample of $\delta$~Sct stars.}
\begin{tabular}{ll}
 \hline \hline
 TIC & note\\
 \hline
331893210&triple system\\
409736562\\
6029735\\
278535185\\
229284388\\
\hline
\end{tabular}
\end{table}

\begin{figure*}[ht!]
    \centering
    {\includegraphics[width=.85\linewidth]{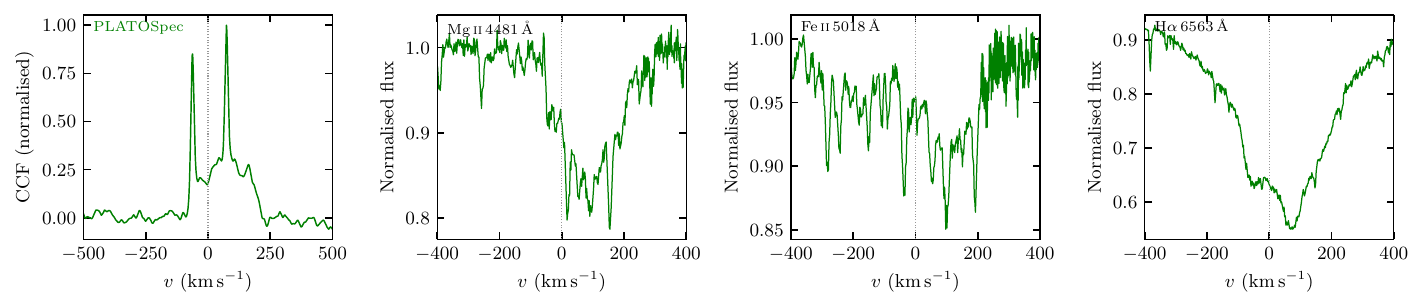}}
     {\includegraphics[width=.85\linewidth]{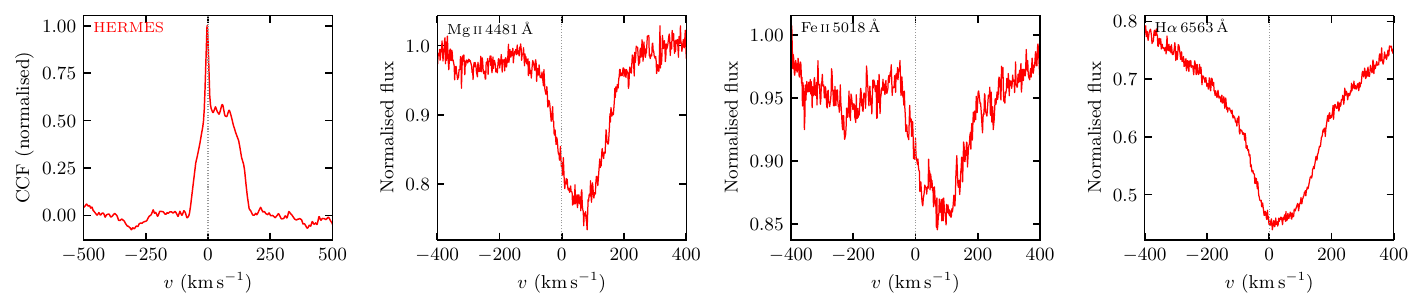}}
     {\includegraphics[width=.85\linewidth]{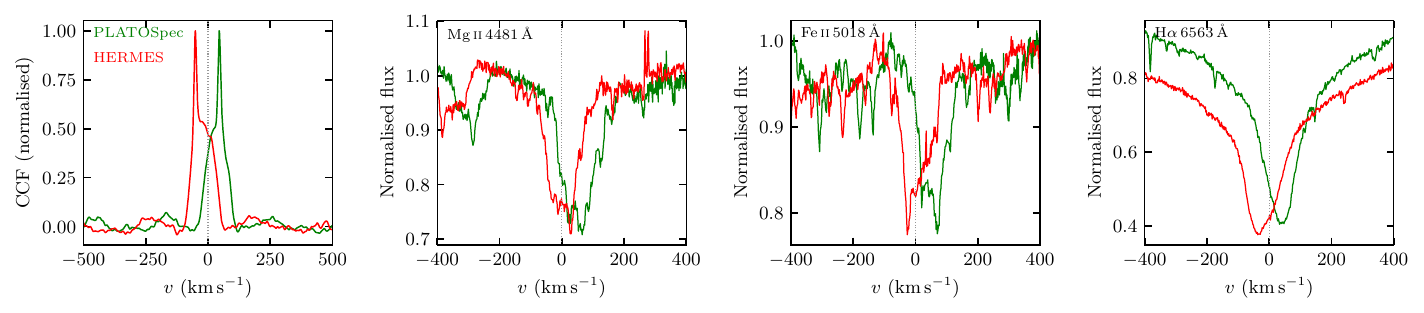}}
     {\includegraphics[width=.85\linewidth]{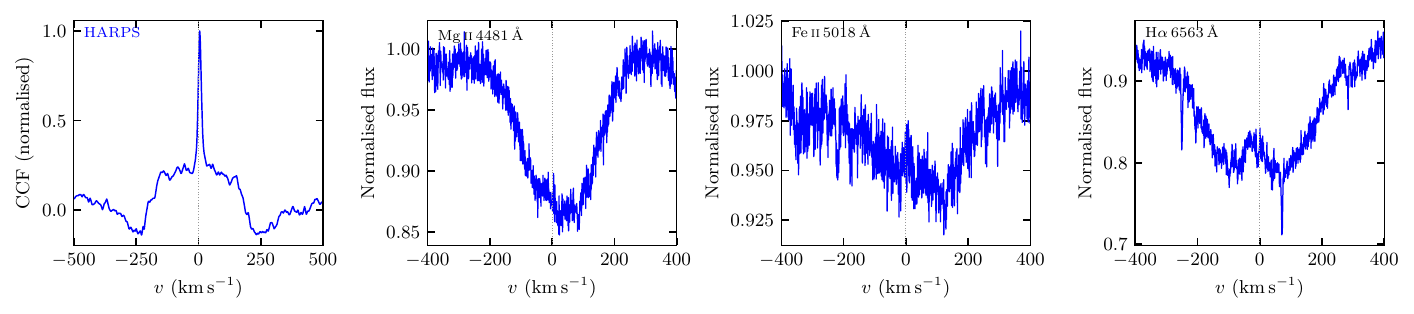}}
     {\includegraphics[width=.85\linewidth]{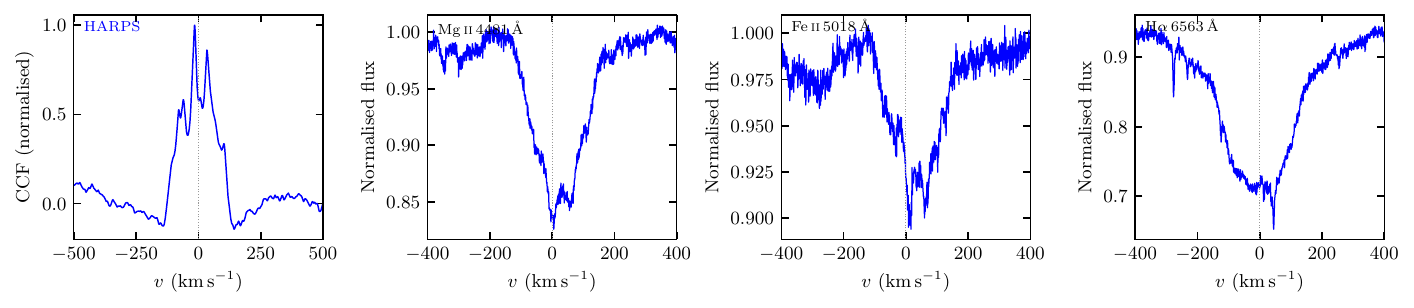}}
    \caption{Examples of new binary candidates. Top to bottom: Triple system TIC\,331893210, TIC\,409736562, TIC\,6029735, TIC\,278535185 and TIC\,229284388. Left to right: cross-correlation function of the entire observed spectrum with a slow-rotation template,  spectrum, spectral lines Mg\,\textsc{ii} \SI{4481}{\AA}, Fe\,\textsc{ii} \SI{5018}{\AA}, H$\alpha$ \SI{6563}{\AA}.}
      \label{fig:Apendix:Binaries} 
\end{figure*}

\end{appendix}
\end{document}